\documentclass[usenatbib]{mn2e}
\usepackage[pdftex]{graphicx}
\newcommand{\be}{\begin{equation}}
\newcommand{\ee}{\end{equation}}
\newcommand{\bea}{\begin{eqnarray}}
\newcommand{\eea}{\end{eqnarray}}
\newcommand{\bc}{\begin{center}}
\newcommand{\ec}{\end{center}}

\def\gsim{ \lower .75ex \hbox{$\sim$} \llap{\raise .27ex \hbox{$>$}} }
\def\lsim{ \lower .75ex \hbox{$\sim$} \llap{\raise .27ex \hbox{$<$}} }
\input amssym.def
\input amssym

\renewcommand{\thefootnote}{\fnsymbol{footnote}}

\title[Black holes in Illustris]{The Illustris simulation: the evolving population of black holes across
  cosmic time}

\author[Sijacki et al.]
       {\parbox{18cm}{Debora~Sijacki$^{1}$\footnotemark[1], Mark
           Vogelsberger$^{2}$, Shy Genel$^{3,4}$\footnotemark[2], Volker
           Springel$^{5,6}$, Paul Torrey$^{3,2,7}$, Gregory F.~Snyder$^{8}$, Dylan
           Nelson$^{3}$ and Lars Hernquist$^{3}$}\vspace{0.3cm}\\  
         $^1$ Institute of Astronomy and Kavli Institute for Cosmology,
         University of Cambridge, Madingley Road, Cambridge CB3 0HA, UK\\
         $^2$ Department of Physics, Kavli Institute for Astrophysics and
         Space Research, Massachusetts Institute of Technology, \\Cambridge, MA 02139, USA\\
         $^3$ Harvard-Smithsonian Center for Astrophysics, 60 Garden Street,
         Cambridge, MA, 02138, USA\\
         $^4$ Department of Astronomy, Columbia University, 550 West 120th Street, New York, NY 10027, USA\\
         $^5$ Heidelberg Institute for Theoretical Studies,
         Schloss-Wolfsbrunnenweg 35, 69118 Heidelberg, Germany\\ 
         $^6$ Zentrum f\"{u}r Astronomie der Universit\"{a}t Heidelberg, ARI,
         M\"onchhofstr. 12-14, 69120 Heidelberg, Germany\\
         $^7$ Caltech, TAPIR, Mailcode 350-17, California Institute of
         Technology, Pasadena, CA 91125, USA\\
         $^8$ Space Telescope Science Institute, 3700 San Martin Dr,
         Baltimore, MD 21218, USA \\}

\begin{document}

\maketitle
\begin{abstract} 

We study the properties of black holes and their host galaxies across cosmic
time in the Illustris simulation. Illustris is a large scale cosmological
hydrodynamical simulation which resolves a $(106.5\,{\rm Mpc})^3$ volume with
more than $12$ billion resolution elements and includes state-of-the-art
physical models relevant for galaxy formation. We find that the black hole
mass density for redshifts $z = 0 - 5$ and the black hole mass function at $z
= 0$ predicted by Illustris are in very good agreement with the most
recent observational constraints. We show that the bolometric and hard X-ray
luminosity functions of AGN at $z = 0$ and $1$ reproduce observational
data very well over the full dynamic range probed. Unless the bolometric
  corrections 
  are largely underestimated, this requires radiative efficiencies to be
on average low, $\epsilon_r \lesssim 0.1$, noting however that in our model
  radiative efficiencies are degenerate with black hole feedback  
  efficiencies. Cosmic downsizing of the AGN population is in
broad agreement with the findings from X-ray surveys, but we predict a
  larger number density of faint AGN at high redshifts than currently
  inferred. We also study black hole -- host galaxy scaling
relations as a function of galaxy morphology, colour and specific star
formation rate. We find that black holes and galaxies co-evolve at the massive
end, but for low mass, blue and star-forming galaxies there is no tight
relation with either their central black hole masses or the nuclear AGN
activity.  
 
\end{abstract}


\section{Introduction}
\renewcommand{\thefootnote}{\fnsymbol{footnote}}
\footnotetext[1]{E-mail: deboras@ast.cam.ac.uk}
\footnotetext[2]{Hubble Fellow}
\stepcounter{footnote}

Accretion onto supermassive black holes has been identified as the most likely
mechanism powering the engines of bright quasars \citep{LyndenBell1969, Rees1984}. Quasars are one of the
most luminous sources in the entire Universe, often outshining the whole light
emitted from the galaxies hosting them. Their large radiative power
means that we can observe quasars out to very high redshifts \citep{Fan2006,
  Mortlock2011} and thus probe their evolution over more than $90 \%$
of cosmic time. Whereas for quasars at $z \sim 6 -7$ \citep{Willott2010, DeRosa2011}
the range of luminosities probed is still relatively narrow, for $z
\lesssim 5$ the consensus on quasar luminosities is more complete thanks to
both optical and X-ray surveys, such as SDSS, GOODS, COSMOS and Chandra Deep Fields
North and South. 

While it is imperative for any state-of-the-art cosmological simulation to
compare against this wealth of data, the study of supermassive black holes
involves some broader and more fundamental questions. In a series of
seminal theoretical papers \citep{Silk1998, Haehnelt1998, Fabian1999,
  King2003} principal ideas have been developed to explain the possible mutual
feedback between galaxies and their central black holes.
Observational evidence for this physical relationship has been mounting over
the years \citep{Magorrian1998, Ferrarese2000, Tremaine2002, Marconi2003,
  Haring2004, Gultekin2009, Gebhardt2011, McConnell2013, Kormendy2013}, indicating that black hole masses correlate with host galaxy stellar
properties, such as bulge luminosity, mass and velocity
dispersion. Although these scaling relations may suggest that galaxies and black
holes co-evolve, they are subject to many biases and systematic
uncertainties both at the low mass and the massive end. Indeed recent
work by \citet{McConnell2013} and \citet{Kormendy2013} revised significantly the $M_{\rm BH}$ -
$M_{\rm bulge}$ and $M_{\rm BH}$ - $\sigma$ relations \citep[see also][]{Gebhardt2011}, such that for a given bulge
mass or velocity dispersion, the best-fit black hole masses are a factor $2$ to
$3$ higher than previously thought. These studies further highlighted that galaxies with different
properties, e.g. pseudo bulges versus real bulges, or cored versus power-law
ellipticals, may correlate differently with their central black hole
masses. Uncertainties in the origin of the black hole -- galaxy scaling relations prompted some
authors \citep{Peng2007, Hirschmann2010, Jahnke2011} to consider mass
averaging in mergers as a root cause of scaling relations without the need to
invoke any feedback. In this scenario repeated galaxy -- galaxy and thus black hole -- black
hole mergers lead to the establishment of the $M_{\rm BH}$ - $M_{\rm bulge}$
relation thanks to the central limit theorem. 

One of the clues that could help shed light on the relative importance of feedback versus merger
averaging, is the redshift evolution of the black hole -- galaxy
scaling relations. While different observational pieces of evidence indicate that the
scaling relation should evolve such that, for a given host galaxy at higher
redshifts, black holes are more massive than their $z = 0$ counterparts
\citep[e.g.][for a review see \citealt{Kormendy2013}]{Treu2004, Shields2006,
  Woo2008, Merloni2010}, systematic uncertainties, selection effects and 
insufficient data do not allow yet to conclude anything secure about the redshift
evolution of the scatter. Thus from the observational point of view this
remains an unsettled point, even though there is accumulating evidence for
AGN-driven large scale outflows \citep{Cicone2012,
  Maiolino2012, Cicone2014, Genzel2014}.

It is therefore of fundamental importance to understand from a theoretical point
of view if indeed feedback from supermassive black holes affects 
their hosts significantly and if this leads to the co-evolutionary picture. The majority of
past work based on mergers of isolated galaxies found that black holes play a
crucial role in the morphological transformation of host galaxies and in the quenching
of their star formation rates \citep[for early works see e.g.][]{DiMatteo2005, Springel2005b,
  Robertson2006b} and that feedback from accreting black holes is
responsible for the existence of the black hole -- galaxy
scaling relations. Based on these simulation results \citet{Hopkins2006}
proposed a unified model for the merger-driven origin of quasars and their
host spheroids. While this picture is theoretically appealing, fully
self-consistent cosmological simulations indicate that galaxies and thus very
likely their central black holes as well assemble through a variety of physical
processes and not major mergers alone. Moreover, observations have thus far
been inconclusive in showing a clear link between enhanced star formation
rates of galaxies and AGN nuclear activity \citep[see][and references
  therein]{Azadi2014}, thus questioning the merger-driven co-evolution of the
two.    

The first cosmological simulations \citep{Sijacki2007, DiMatteo2008} to
investigate the black hole -- galaxy co-evolution confirmed that AGN-driven
outflows not only lead to black hole self-regulation but also to the
establishment of the scaling relations, as isolated galaxy merger
studies have advocated. These results were further confirmed by several
independent groups and more recent simulations \citep{Booth2009, Dubois2012,
  Hirschmann2014, Khandai2014, Schaye2014}. What however still remains unclear is whether
black hole -- galaxy co-evolution occurs 
for all galaxies hosting supermassive black holes at their core or whether
different galaxy types exhibit weaker or stronger physical links with their
black holes. The main reason why this question remained unanswered theoretically
until now stems from the difficulty to simulate representative galaxy samples
covering the observed range of morphologies. Simulated galaxies typically appeared too
centrally concentrated, formed too many stars and lacked sufficient rotational
support. This has been one of the long standing
issues in computational galaxy formation which even led some to question the 
$\Lambda$CDM cosmology \citep[e.g.][]{SommerLarsen2001}. Now we understand that this was caused by insufficient numerical
resolution, hydro-solver inaccuracies and lack of modelling of the necessary
physics. Only recently several simulation efforts \citep{Guedes2011,
  Aumer2013, Stinson2013, Hopkins2014, Marinacci2014}, mostly based on the zoom-in technique of
individual objects, have started reproducing extended,
disk-dominated galaxies which in some aspects resemble our own Milky Way. None
the less, to study the black hole -- galaxy co-evolution large scale cosmological
simulations are needed to have a sufficiently representative sample
of objects. At the same time good spatial resolution is necessary to resolve at
least the basic structural properties of galaxies hosting supermassive black holes. These
conditions pose very challenging requirements on the dynamical range
cosmological simulation should resolve.  

The Illustris simulation project \citep[][see also \citealt{Vogelsberger2014b,
    Genel2014}]{Vogelsberger2014a} is
the first cosmological simulation that is able to probe the necessary range of
spatial scales with a comprehensive set of physical processes so that we can
study black hole -- galaxy co-evolution with unprecedented detail. This means
that we not only have a statistically large and representative sample of
objects from $z \sim 4$ to $z = 0$, but that we can start to disentangle the physical
link between black holes and their host galaxies as a function of galaxy
morphology and colour. We
anticipate here that this will allow us to pin down the most likely physics which
is responsible for the establishment of the mutual feedback between galaxies and
their central black holes, but we also highlight for which types of galaxies this
feedback loop is not fully operational. 

This paper is organised as follows. In Section~\ref{Methodology} we outline
our methodology, summarising the numerical technique adopted, simulation
characteristics and physics implementation. In Section~\ref{CONVERGENCE} and
in Appendix~\ref{APP_CONV} we discuss the convergence properties of the black hole model,
while in Sections~\ref{BHAR} and \ref{MASSFUNC} we present the basic black
hole properties, namely the cosmic black hole accretion rate and mass density
as well as the mass function at $z = 0$. Section~\ref{SCALING} summarises the main results regarding
the scaling relations of galaxies and their central black holes. We further
discuss black hole Eddington ratios, AGN luminosity functions and cosmic
downsizing in Sections~\ref{EDDINGTON} and \ref{QSOLF}, while in
Section~\ref{LXSFR} we examine the link between star formation rate and
nuclear AGN triggering. We finally discuss our results and draw conclusions in
Section~\ref{Conclusions}.
   
\section{Methodology} \label{Methodology}

\subsection{Numerical method}

In this study we use a series of large scale cosmological simulations, the
so-called Illustris project\footnote{http://www.illustris-project.org}, to
investigate the link between black holes  and their host galaxies across
cosmic time. The Illustris simulations have been performed with the massively
parallel hydrodynamical code {\small AREPO} \citep{Springel2010}, which adopts
a TreePM solver for gravity and a second-order accurate unsplit Godunov method
for the hydro forces. The hydrodynamics equations are solved on an
unstructured Voronoi mesh, which is allowed to freely move with the fluid in a
quasi-Lagrangian fashion. The code has been  thoroughly tested and validated
on a number of computational problems and small scale cosmological simulations
\citep{Springel2010, Springel2011, Bauer2011, Sijacki2012, Vogelsberger2012,
  Keres2012, Torrey2012, Genel2013, Nelson2013} demonstrating excellent shock
capturing properties, proper development of fluid instabilities, low numerical
diffusivity and Galilean invariance, making it thus well posed to tackle the
problem of galaxy formation.     

\subsection{The simulation suite}

The Illustris simulation suite consists of large scale cosmological
simulations in a periodic box with $106.5 \,{\rm Mpc}$ on a side, simulated
with different physics and at different resolutions. A standard, flat
$\Lambda$CDM cosmology is assumed with $\Omega_{m,0} = 0.2726$,
$\Omega_{\Lambda,0} = 0.7274$, $\Omega_{b,0} = 0.0456$, $\sigma_8 = 0.809$,
$n_s = 0.963$ and $H_{0} = 70.4 \, {\rm km s^{-1} Mpc^{-1}}$ consistent with
the Wilkinson Microwave Anisotropy Probe 9-year data release
\citep{Hinshaw2013}. The starting redshift of the simulations is $z = 127$ and all
simulations have been evolved to $z = 0$. The physics included ranges from
dark  matter only simulations (Illustris-Dark), non-radiative hydrodynamical
simulations (Illustris-NR), to simulations with the full galaxy formation
physics module switched on (Illustris) which will be used in this study. The
simulations have been performed at 3 different resolutions: 1. Low resolution
box with $3 \times 455^3$ dark matter, gas and Monte Carlo tracer resolution
elements, a typical gas cell mass of $m_{\rm gas} = 8.05 \times 10^7 \, M_{\rm
  \odot}$, dark matter particle mass of $m_{\rm DM} = 4.01 \times 10^8 \,
M_{\rm \odot}$ and gravitational softenings\footnote{
  Note that the gravitational softenings are Plummer-equivalent.} $\epsilon_{\rm gas} = 2.84 \,{\rm
  kpc}$ and $\epsilon_{\rm DM} = 5.86 \,{\rm kpc}$; 2. Intermediate resolution
box with $3 \times 910^3$ resolution elements in total, $m_{\rm gas} = 1.01
\times 10^7 \, M_{\rm \odot}$, $m_{\rm DM} = 5.01 \times 10^7 \, M_{\rm
  \odot}$, $\epsilon_{\rm gas} = 1.42 \,{\rm kpc}$ and $\epsilon_{\rm DM} =
2.84 \,{\rm kpc}$; and 3. High resolution box with $3 \times 1820^3$
resolution elements in total, $m_{\rm gas} = 1.26 \times 10^6 \, M_{\rm
  \odot}$, $m_{\rm DM} = 6.26 \times 10^6 \, M_{\rm \odot}$, $\epsilon_{\rm
  gas} = 0.71 \,{\rm kpc}$ and $\epsilon_{\rm DM} = 1.42 \,{\rm kpc}$. For
further details of the simulations see \citet{Vogelsberger2014b} and
\citet{Genel2014}. In this study we will mainly focus on the highest
resolution box of Illustris, which we call henceforth Illustris for brevity,
while we will take advantage of the lower resolution boxes when exploring the
convergence issues. 

\subsection{The model for galaxy formation}\label{Physicsmodel}

The Illustris simulations contain a comprehensive array of modules that describe
galaxy formation physics beyond non-radiative processes. This includes: primordial
and metal-line cooling in the presence of time dependent UV background
\citep{Faucher2009} including gas self-shielding \citep{Rahmati2013}, where
the metals are naturally  advected with the fluid flow; a sub-grid model for
star formation and associated supernovae feedback as in \citet{SpringelH2003}
adopting a softer equation of state \citep{Springel2005b} with $q = 0.3$ and a Chabrier initial mass
function \citep{Chabrier2003}; a model for stellar evolution, gas recycling,
metal enrichment \citep[see also][]{Wiersma2009} and mass- and metal-loaded
galactic outflows, where the wind mass loading scales with the inverse of
  the wind velocity squared, motivated by energy conservation arguments \citep[see
    also][]{Oppenheimer2008, Okamoto2010, Puchwein2013};
and a model for black hole seeding, accretion and feedback that we will
describe in more detail in Section~\ref{BHmodel}. A full account of these
prescriptions is given in our pilot study \citep{Vogelsberger2013,
  Torrey2014} where the basic properties of galaxies are compared with
observables. Specifically, given that currently it is not possible to describe
the physics of star formation and black holes in an ab-initio manner, simple
phenomenological and empirical sub-grid models need to be employed if we are
to gain insight into the physics of galaxy formation. In the Illustris project
the free parameters of the sub-grid models are set to physically plausible
values which have been fixed after calibrating the simulations against a few
fundamental observables, such as the cosmic star formation rate history and
the stellar mass function at $z = 0$. This calibration has been performed on
smaller cosmological boxes with $35.5 \,{\rm Mpc}$ on a side and is presented
in \citet{Vogelsberger2013, Torrey2014}.

\subsection{Black hole model}\label{BHmodel}

\subsubsection{Black hole accretion} 

In the Illustris simulations collisionless black hole particles with a seed
mass of $1.42 \times 10^5 \, M_{\rm \odot}$ ($10^5 \, h^{-1}\,M_{\rm \odot}$)
are placed with the aid of the on-the-fly Friends-of-Friends (FOF) algorithm
in all halos more massive than $7.1 \times 10^{10} \, M_{\rm \odot}$ that do
not contain a black hole particle already. Thereafter, the black hole seeds
can grow in mass either through gas accretion, which we parametrise in terms
of Eddington limited Bondi-Hoyle-Lyttleton-like accretion \citep[for further
  details see][]{Springel2005b, DiMatteo2005}, or via mergers with other black
holes. At $z = 4$ our high resolution Illustris simulation already tracks
$9414$ black holes, at $z = 2$ this number more than doubles leading to
$24878$ black holes in total, while at $z = 0$ there are $32542$ black holes in total with
$3965$ black holes more massive than $10^7 M_{\rm \odot}$.  

With respect to our previous work \citep[e.g.][]{Springel2005b, Sijacki2007,
  Sijacki2009} there are a few updates in the black hole model that we list
here. First, we do not take the relative velocity of black holes with respect
to their surrounding gas into account when estimating
Bondi-Hoyle-Lyttleton-like accretion and we merge black hole pairs which are
within smoothing lengths of each other irrespective of their relative
velocity. This is motivated by the fact that we use a repositioning scheme
 to  ensure
that the black hole particles are at the gravitational potential minimum of
the host halos and do not spuriously wander around due to two body scattering
effects with massive dark matter or star particles. This leads to ill-defined black
hole velocities. Note however that our estimated sound speeds entering
  Bondi-Hoyle-Lyttleton-like accretion are typically larger than the relative
  velocity term, so that the black hole accretion rates are not affected
  significantly by this update. Moreover, during accretion
events we gradually drain the parent gas  cell of its mass rather than
stochastically swallowing one of the neighbouring gas cells. We also use the
parent gas cell to estimate the gas density instead of performing a kernel
weighted average over gas neighbours, as we have done in the past. Finally, we
introduce a black hole ``pressure criterion'' whereby the accretion rate estimate
is lowered in cases where the gas pressure of the ambient medium cannot
compress gas to a density exceeding the star-formation
threshold in the vicinity of an accreting black hole. Here
the $\alpha = 100$ pre-factor in the Bondi prescription needed to compensate
for the unresolved cold and hot clouds of our sub-grid ISM model becomes superfluous and could 
lead to the formation of an un-physically large and hot gas bubble around
massive black holes accreting from a low density medium. For further details
on the ``pressure criterion'' see \citet{Vogelsberger2013}. Note that
self-regulated growth of black holes is largely unaffected by this change. 

\subsubsection{Black hole feedback}\label{FEEDBACK}

\begin{figure*}\centerline{
\hbox{
\includegraphics[width=8truecm,height=7.5truecm]{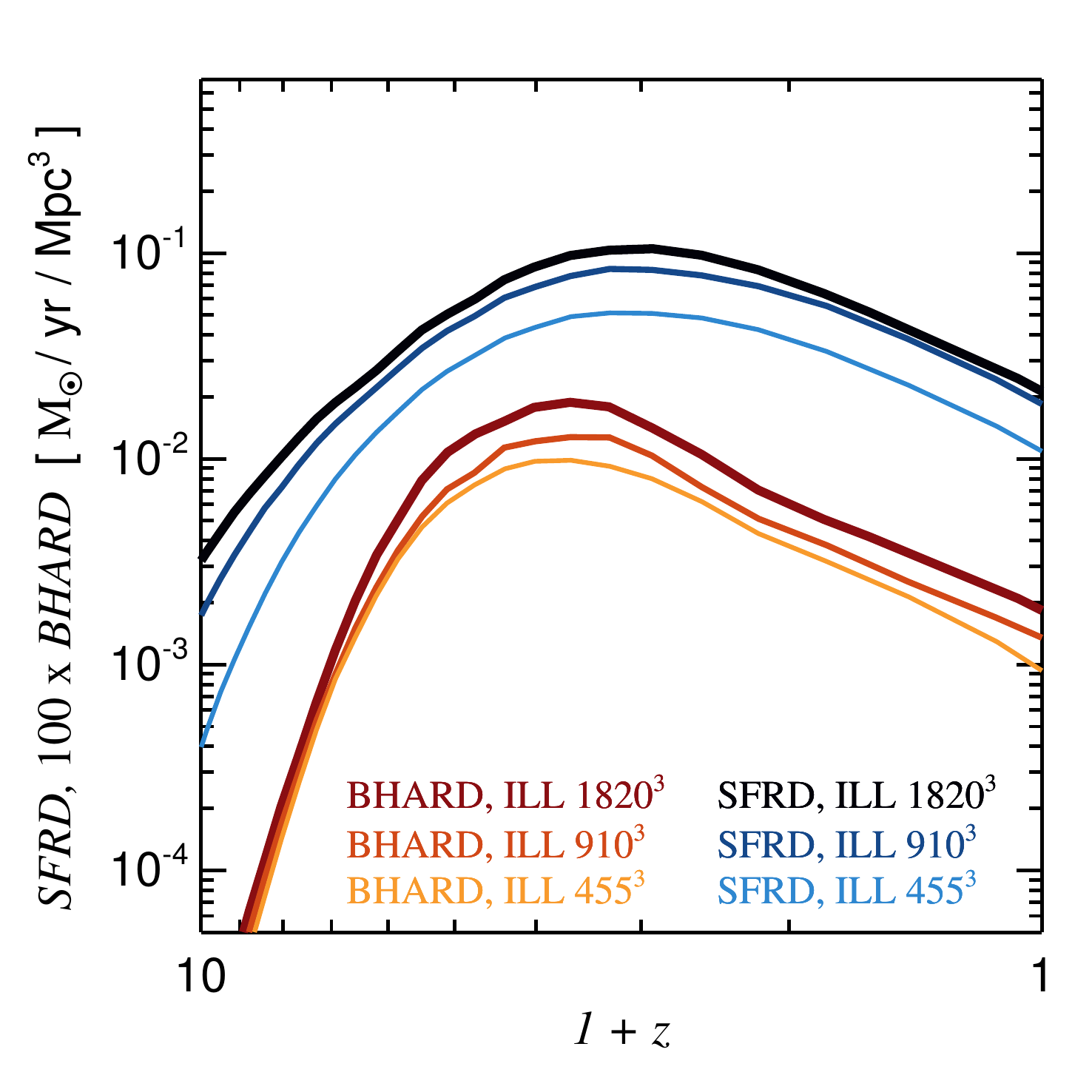}
\includegraphics[width=8truecm,height=7.5truecm]{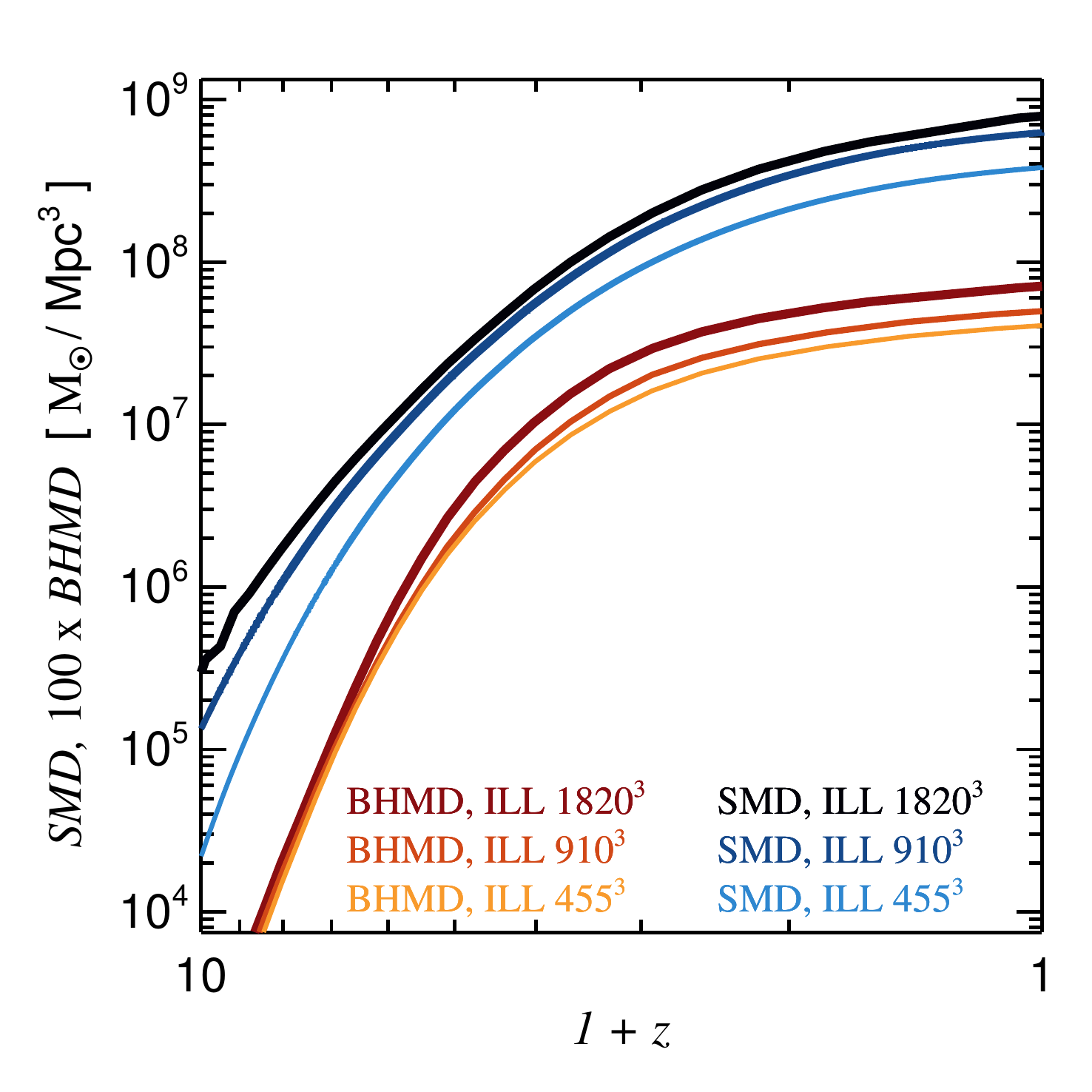}}
}
\caption{Left: time evolution of the star formation rate density (blue curves)
  and of the black hole  accretion rate density (red curves; rescaled by a
  factor of a 100) for three different  resolutions, as indicated on the
  legend. While the numerical convergence in the star  formation rate density
  is very good, the black hole accretion rate density is not yet converged  at the
  highest resolution. Note that the overall shape of the black hole accretion rate
  density  is somewhat different than that of the star formation rate density,
  i.e. it rises faster at high  redshifts and also peaks at a higher
  redshift $z \sim 2.5-3$. Right: stellar mass density  and black hole mass
  density as a function of cosmic time for the same set of simulations.}
\label{SFR_RES}
\end{figure*}

As for the black hole feedback we consider three different modes: ``quasar'',
``radio'' and ``radiative'' feedback. In the ``quasar'' mode AGN
  bolometric luminosity is computed directly from the black hole accretion
  rate assuming a given radiative efficiency. A small fraction
of the AGN bolometric luminosity is thermally coupled to the surrounding gas
with an efficiency factor of $\epsilon_f = 0.05$ thus effectively leading to
an energy-driven outflow in the case of negligible radiative
losses\footnote{Note that once gas internal energy is increased due to
    black hole feedback, gas is allowed to radiatively cool and heat, except
    for the gas within the multiphase model for star formation that is colder
    than the effective temperature of our equation of state assumed
    there. The internal energy of this cold, multiphase gas is set to the
    effective energy of the multiphase model.}. The
switch between ``quasar'' and ``radio'' mode is determined by the black hole
Eddington ratio following \citet{Sijacki2007}. In the ``radio'' mode hot
bubbles are randomly placed within a sphere around each black
hole. For all active black hole particles we estimate the local
gas density at the position of the bubble. We
then use the analytic cocoon expansion equation \citep[see equation 5 in][]
{Sijacki2007} to rescale both the radii of the bubbles and the radii of the spheres
within which the bubbles are created from the initially set default
values. The thermal energy injected into the bubbles is directly linked to the
black hole mass growth via radiative efficiency and thermal coupling
efficiency in the radio mode, $\epsilon_m$ \citep[see equation 4 in][]
{Sijacki2007}.   

With respect to the original work by \citet{Sijacki2007} we change the values of
some of the model parameters. The scaling of the radius of the sphere within
which bubbles are injected has been increased from $60 \, {\rm kpc}$ to $100
\,{\rm kpc}$. With a larger radius the energy contrast between the bubbles and
the surrounding gas can be higher and thus lead to larger feedback
effects. Note however given that for each black hole we scale this radius according
to the analytic cocoon expansion equation, thus this change in scaling is not
very significant.

We further make two more significant changes: we increase the efficiency
factor of thermal coupling, $\epsilon_m$, from $0.2$ to $0.35$ and we increase
the Eddington ratio threshold, $\chi_{radio}$ below which the ``radio'' mode
feedback kicks in from $0.01$ to $0.05$. These two changes have been motivated
by the above mentioned calibration against the observed cosmic star formation
rate history and the $z = 0$ stellar mass function. We find that a higher
$\epsilon_m$ value is needed to sufficiently suppress star formation in
massive galaxies which tends to be even higher than in previous work due to several
factors: {\it i)} stellar mass loss and metal line
cooling, which can affect the cosmic star formation rate density significantly \citep[for example, see
  Figure 15 in][]{Vogelsberger2013}; {\it ii)} more accurate 
gas cooling in {\small AREPO} with respect to standard SPH. This is due to
the much more effective gas mixing \citep[see e.g.][]{Sijacki2012, Torrey2012} which
becomes even more important when considering mixing of metal-enriched galactic
winds with the hot, diffuse halo. Also, it has been shown \citep{Bauer2011,
  Vogelsberger2012} that in {\small AREPO} there is no heating of gas due to artificial dissipation of subsonic turbulence, as is the case in the standard SPH, which may affect cooling
rates as well \citep{Nelson2013}. The higher Eddington
ratio threshold value leads to more efficient ``radio'' mode feedback being
active in somewhat lower mass galaxies and at higher redshifts as well which
helps reproducing the ``knee'' of the $z = 0$ stellar mass function. 

In addition to the ``quasar'' and ``radio'' mode, we also take into account
``radiative'' feedback where we modify the net cooling rate of gas
  (namely, photo-ionisation and photo-heating rates) in the
presence of strong ionising radiation emanating from actively accreting
black holes. Assuming a fixed spectral energy distribution we consider gas
below the density threshold for star formation to be in the optically thin
regime and compute the bolometric intensity each gas cell experiences due to the
AGN radiation field of all black holes within a given search radius which
  is set
  by a threshold in the ionisation parameter and capped to three times
  the virial radius of the parent halo. Note that
``radiative'' feedback is most effective for black holes in the ``quasar''
mode accreting close to the Eddington limit.

While the quasar efficiency factor, $\epsilon_f$, has been set by
\citet{Springel2005b, DiMatteo2005} to match the normalisation of the $M_{\rm
  BH}$ - $\sigma$ relation in the isolated galaxy mergers, note that none of the
black hole model parameters have been tuned to match any of the black hole
properties which hence can be viewed as genuine predictions of the model. For further
details on the black hole model see \citet{Springel2005b, Sijacki2007,
  Vogelsberger2013}.         

We finally note that in Illustris the radiative efficiency
has been set to $\epsilon_r = 0.2$. This change from the standardly adopted
value of $0.1$ has been motivated by the findings of \citet{Yu2002}, where it
has been shown that luminous quasars (which are the objects we are most
interested in) should have $\epsilon_r \sim 0.2$. In our model $\epsilon_r$ is
essentially unconstrained given that in all equations it is degenerate with
the values of $\epsilon_f$ and $\epsilon_m$. The only equation where
$\epsilon_r$ enters on its own is the one that determines the fraction of the
accreted mass lost to radiation, which however leads to a very small
effect. Nonetheless, as we will show in Section~\ref{QSOLF}, taking together
the results regarding black hole mass and luminosity functions we can place
interesting constraints on the average radiative efficiency of AGN.

\section{Results}\label{Results}

\subsection{Convergence issues}\label{CONVERGENCE}

We start our analysis by looking at the convergence properties of our galaxy
formation model. Here we are specifically interested in the basic black hole
properties, while the convergence of other quantities has been
discussed in \citet{Vogelsberger2013, Torrey2014, Genel2014}. In
Figure~\ref{SFR_RES} we show the cosmic star formation rate density and black hole
accretion rate density (left-hand panel), as well as stellar and black hole
mass density (right-hand panel), for the three different resolution Illustris
simulations, as indicated on the legend. With higher resolution smaller mass
galaxies are better resolved and this leads to an increase in the star
formation rate at high redshifts, which is especially pronounced between the
low and intermediate resolution simulations. However, at lower redshifts, i.e.
for $ z < 5$, the bulk of star formation occurs in sufficiently well resolved
galaxies so that the total star formation rate density does not increase much
with higher resolution. This is in particular true when we compare our
intermediate resolution simulation with the high resolution run where both
star formation rate density and stellar mass density exhibit excellent
convergence properties. 

The convergence properties of the black hole accretion rate and mass
density are however somewhat poorer. Here for $z > 5$ the black hole accretion
rate density is essentially the same in all three simulations, given that by
this time it is dominated by low mass black holes that have been relatively
recently seeded within well resolved dark matter halos. At later times there
is an approximatively constant offset between both low versus intermediate and
intermediate versus high resolution run (amounting to a factor of $\sim
1.5$). This indicates that the black hole  accretion rate density is not yet
fully converged even for our highest resolution simulation. We have investigated
whether there are any clear trends in black hole accretion rate density
convergence for different black hole mass ranges, and found that for $z > 2$
more massive black holes exhibit worse convergence, while for $z < 2$ the
convergence rate is similar regardless of the black hole mass. Clearly this is
a result that we need to keep in mind when interpreting our findings, but we
anticipate here that the convergence properties of other quantities, such as,
for example, the black hole mass function and the black hole mass -- bulge mass
relation, are still very good, as we discuss in Sections~\ref{MASSFUNC} and
\ref{SCALING} and in the Appendix~\ref{APP_CONV}.

\begin{figure*}\centerline{
\hbox{
\includegraphics[width=8truecm,height=7.5truecm]{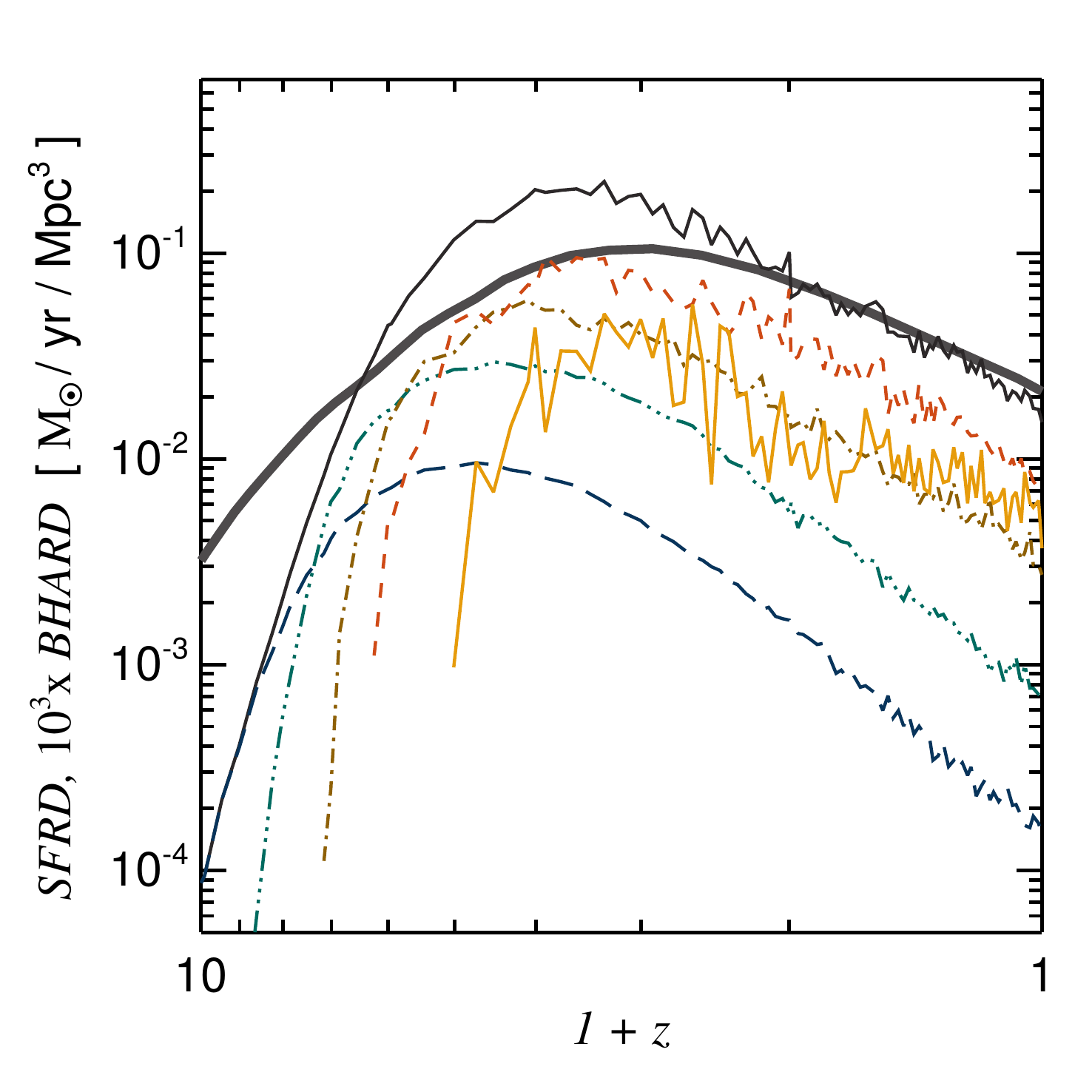}
\includegraphics[width=8truecm,height=7.5truecm]{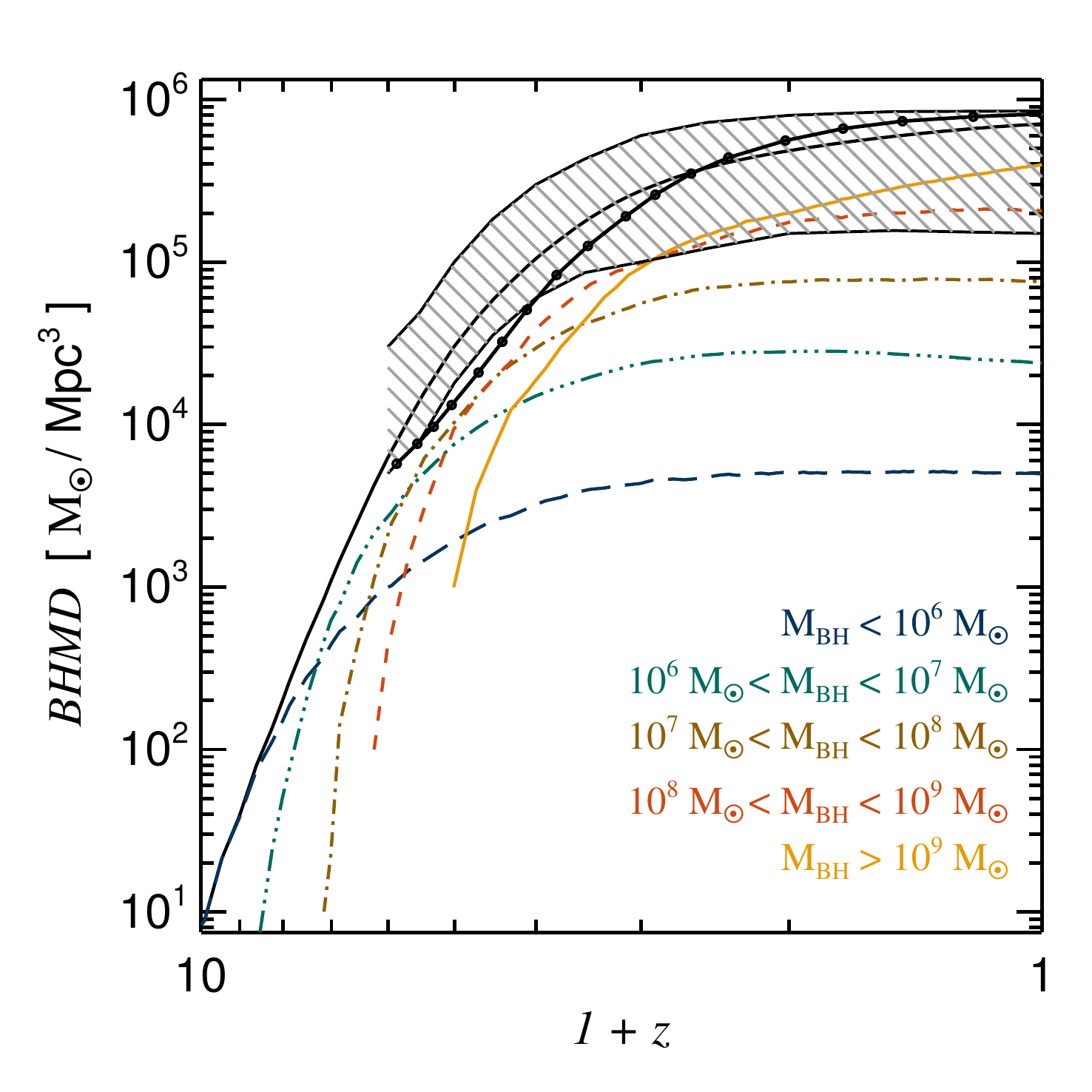}}
}
\caption{Left: time evolution of the star formation rate density (thick dark gray
  curve) and of the black hole  accretion rate density (thin black curve;
  rescaled by a factor of a 1000) for the highest resolution Illustris
  simulation. 
  Coloured lines (from blue to orange: dashed, triple-dot dashed, dot dashed,
  dashed and continuous) indicate 
  total black hole accretion rate densities for black holes in a given range
  of masses, as indicated on the legend. With decreasing redshift more
  massive black holes dominate the total black hole accretion rate density,
  except for the most massive black holes with  $M_{\rm BH} > 10^9 M_{\rm
    \odot}$. Note that for $z < 3$ and for $M_{\rm BH} > 10^8  M_{\rm \odot}$
  there is a considerable ``noise'' in the black hole accretion rate density.
  This is driven by black holes entering the ``radio'' mode feedback which is
  bursty. Right: total  black hole mass density (black thin line) for all
  black holes and split by the black hole mass bins (coloured lines from blue to
  orange with the same line styles as in the left-hand panel). The shaded
  region is the allowed range of mass densities where 
  radiative efficiency is varied from $0.057$ to $0.4$, as reported by
  \citet{Volonteri2010}. Black circles connected with a thick line are for the new
  estimate from \citet{Ueda2014}.}
\label{SFR}
\end{figure*}

\subsection{Black hole accretion rate and mass density}\label{BHAR}

Focusing now on the shape of star formation and black hole accretion rate
densities, shown in Figures~\ref{SFR_RES} and \ref{SFR}, we note that the
black hole accretion rate density rises more steeply  at high redshifts, it
has a sharper peak which occurs earlier ($z \sim 2.5-3$) and it also declines
somewhat more steeply thereafter all the way to $z = 0$. Consequently, for $z
< 2$ the total black hole mass density increases less with time than the
total stellar mass density. This  
demonstrates that while globally there is a relation between star formation
and black hole accretion rates in galaxies, these two processes are not
necessarily intimately linked \citep[see also e.g.][]{Merloni2004,
    Sijacki2007, DiMatteo2008, Somerville2008, Merloni2008}, as we will discuss more in detail in
Section~\ref{LXSFR}.

In Figure~\ref{SFR} we plot the total black hole accretion rate density
(left-hand panel) and the black hole mass density (right-hand panel) for the highest resolution
Illustris simulation, but now split by the black hole mass at a given
redshift, as indicated on the legend.  With decreasing redshift more massive
black holes start to dominate the total black hole accretion rate density,
except for the most massive black holes with $M_{\rm BH} > 10^9 M_{\rm
  \odot}$, which however dominate the total black hole mass density for $z <
2$. Also, even though the black hole accretion rate density shapes are quite
similar for the different black hole mass ranges considered, they peak at
later times for more massive black holes. Given that the distribution of
Eddington ratios is fairly flat as a function of black hole mass for $z \ge 1$
(for further details see Section~\ref{EDDINGTON}), different peaks reflect the
cosmic time when black holes of a given mass contribute most to the total
black hole accretion rate density due to combination of their number density
and accretion rate in absolute numbers. Note that for $z < 3$ and for $M_{\rm
  BH} > 10^8  M_{\rm \odot}$ black holes typically enter the ``radio'' mode
feedback which is bursty, leading to significant and rapid variations in the
black hole accretion rate density.  

In the right-hand panel of Figure~\ref{SFR} we also indicate with the shaded
region the range of possible black hole mass densities derived from
Soltan-type arguments  where radiative efficiency is varied from $0.057$ (top)
to $0.4$ (bottom), as reported by \citet{Volonteri2010}. New constraints from
a compilation of AGN X-ray luminosity surveys by \citet{Ueda2014} are shown
with black circles connected with a thick line. The Illustris result is in
excellent agreement with observational findings from \citet{Ueda2014} and
indicates that on average radiative efficiencies of accreting black holes
could be low. This is an interesting point that we will discuss more in
detail in Section~\ref{QSOLF}.

\subsection{Black hole mass function}\label{MASSFUNC}

\begin{figure*}\centerline{
\hbox{
\includegraphics[width=8truecm,height=7.5truecm]{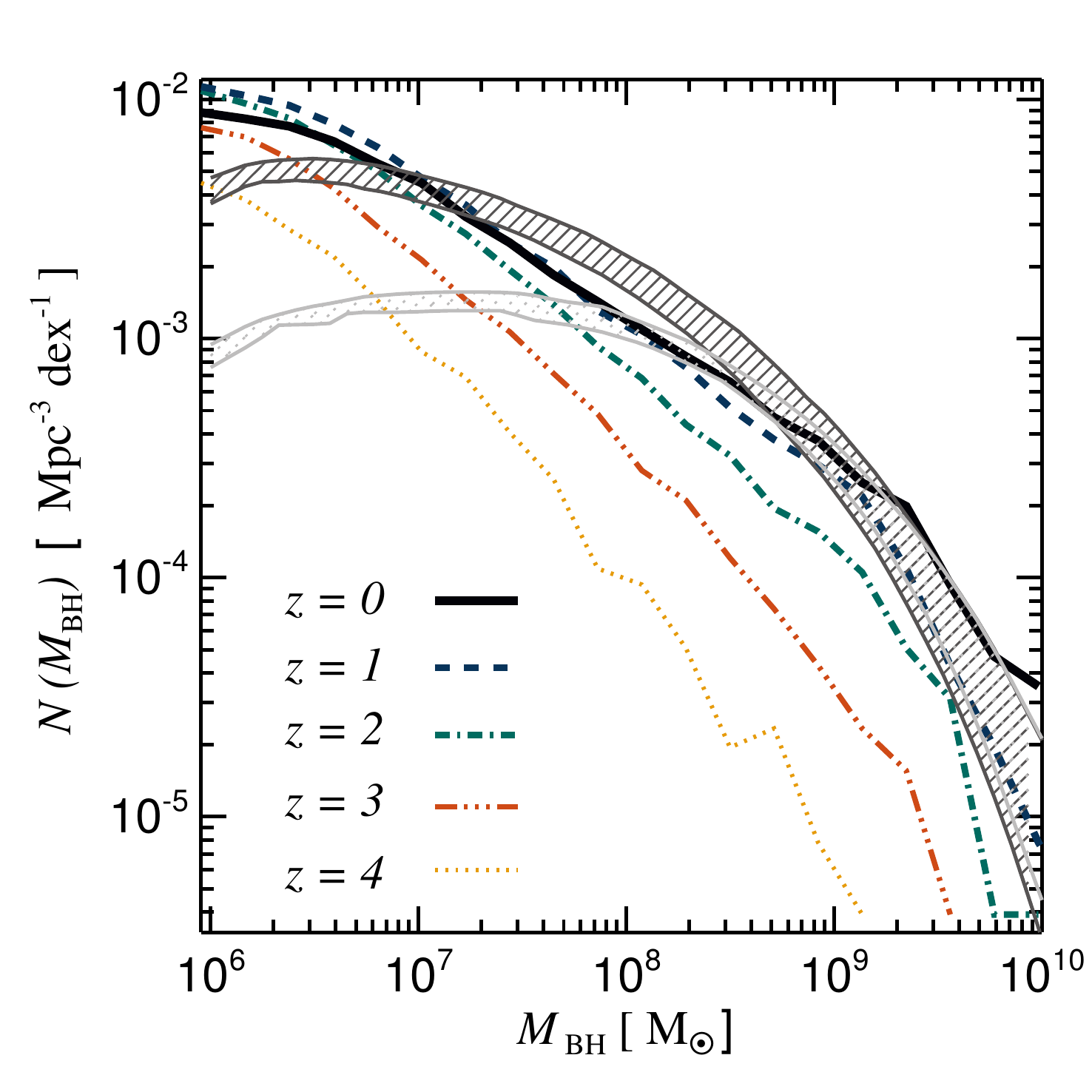}
\includegraphics[width=8truecm,height=7.5truecm]{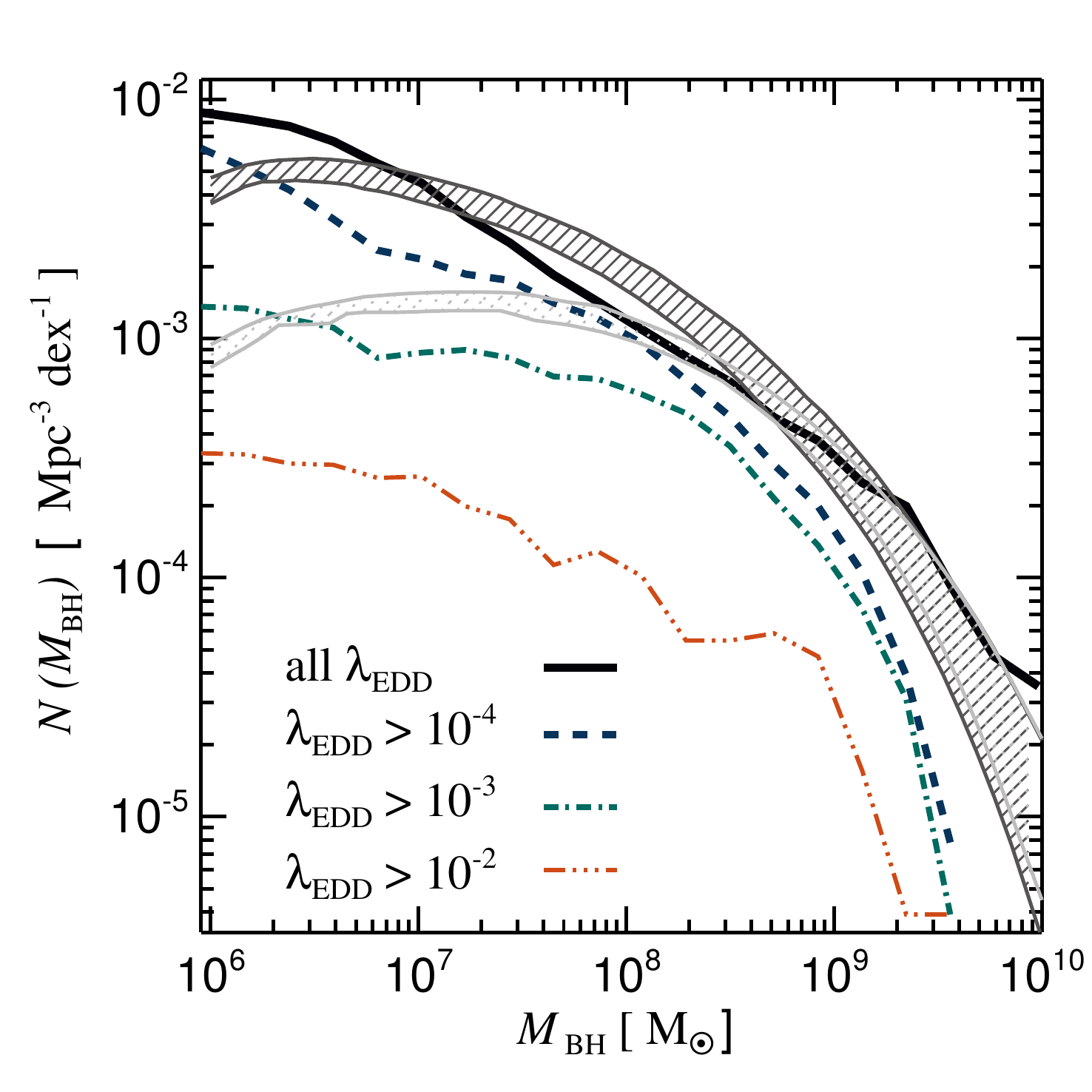}}
}
\caption{Left: black hole mass function at $z = 4, 3, 2, 1$ and $0$ for all
  black holes in the simulated volume. The hatched region is the mass function
  estimate with $1 \sigma$ uncertainty from \citet{Shankar2013}, assuming the
  revised $M_{\rm BH}$ - $\sigma$ relation from \citet{McConnell2013} and
  applying it to all local galaxies. The dotted region is the same but
  assuming Sa galaxies do not host any black holes. In Illustris the black hole
  mass function gradually builds up with cosmic time: at the low mass end
  i.e. for $M_{\rm BH} < 10^7 M_{\rm \odot}$ the mass function does not change
  much for $z < 2$, for $M_{\rm BH} < 2 \times 10^9 M_{\rm \odot}$ the mass
  function does not change much for $z < 1$, while at the massive end there is
  always evolution due to the residual ``hot mode'' accretion and black
  hole-black hole mergers. Right: black hole mass function at $z = 0$, split
  by the Eddington ratios of black holes, as indicated on the legend. It is
  clear that in the mean the Eddington ratios are moderate, but they are
  especially low 
  at the massive end where black holes are in the radiatively inefficient
  accretion regime.}
\label{BHMF}
\end{figure*}

In Figure~\ref{BHMF} we show the redshift evolution of the black hole mass
function at $z = 0, 1, 2, 3$ and $4$ (left-hand panel) and the black hole mass
function at $z =0$ split by the Eddington ratios of black holes, as indicated
on the legend. In both panels we include all black holes irrespective of their
mass or accretion rate. The hatched region marks the mass function estimate with the
$1 \sigma$ uncertainty from \citet{Shankar2013}, assuming the revised $M_{\rm
  BH}$ - $\sigma$ relation from \citet{McConnell2013} and applying it to all
local galaxies. The dotted region is the same but assuming Sa galaxies do not host
any black holes \citep{Shankar2013}. Note that this likely represents
  a lower bound on the black hole mass function. The Illustris black hole
mass function at $z = 0$ agrees quite 
well with the estimate from \citet{Shankar2013}, except for the lowest mass
black holes with $M_{\rm BH} < 10^7 M_{\rm \odot}$. This agreement is
particularly encouraging 
given that recently black hole scaling relations have been significantly
revised \citep[for further details see][]{McConnell2013, Kormendy2013}, where
for a given e.g. bulge mass the best-fit black hole mass is about a factor of
$2$ to $3$  higher with respect to the estimates by \citet{Haring2004} (we
will discuss this further in Section~\ref{SCALING}). Note that, as we show in
Appendix~\ref{APP_CONV}, the uncertainty due to the convergence of the black
hole mass function for our different resolution runs is smaller than the
observational uncertainty calculated by
\citet{Shankar2013}. Moreover, for black holes with Eddington ratios
$\lambda_{\rm EDD} > 10^{-4}$ the convergence rate in the mass function
  improves especially at the massive end (for further details see Appendix~\ref{APP_CONV}), thus indicating
that rather than some minimum black hole mass a minimum accretion rate is
  needed for the model to converge better. The disagreement between
the \citet{Shankar2013} results and the Illustris predictions at the low mass
end, 
i.e. for $M_{\rm BH} < 10^7 M_{\rm \odot}$, could be caused by a number of
reasons. Observational uncertainties increase for low mass black holes,
and at the same time our simulation results are also least reliable at the low
mass end. Here, additionally to numerical convergence issues, the black hole
number densities and masses are most dependent on our rather simplistic seeding
prescriptions and on the initial growth before the self-regulation is
achieved. Regarding the seeding prescription three issues arise: {\it i)} the
  choice of the black hole seed mass which in our model is fairly large,
  i.e. $M_{\rm BH,seed} = 10^5 h^{-1}\, M_{\rm \odot}$; {\it ii)} the fact
  that we seed all halos above a certain mass regardless of
  redshift or any other halo property, for example, such as the gas
  metallicity; and {\it iii)} the fact that due to the black hole
  repositioning, halos
  that temporarily become 
part of a larger FOF group are likely to lose their central black
holes prematurely and if above mass threshold will be re-seeded with a new black
hole.  

The left-hand panel of Figure~\ref{BHMF} shows how the black hole mass
function gradually builds up with cosmic time. The first $10^9 \, M_{\rm \odot}$
black holes are already in place before $z = 4$, while ultramassive black
holes with $\sim 10^{10} \, M_{\rm \odot}$ form  at $z < 2$. Note that the
Illustris volume of $(106.5 \,{\rm Mpc})^3$ is too small, by a factor of $\sim
300$ at least, to contain very massive black holes at $z = 6$  which are
thought to be powering high redshift quasars \citep{Sijacki2009, Costa2014}
and is thus unsuitable for studying these rare objects. At the low mass end
i.e. for $M_{\rm BH} < 10^7 M_{\rm  \odot}$ the mass function does not evolve
significantly after $z \sim 2$, while for $M_{\rm BH} < 2 \times 10^9 \,
M_{\rm \odot}$ the mass function does not change much for $z < 1$. At the
massive end, i.e. for $M_{\rm BH} > 2 \times 10^9 \, M_{\rm \odot}$ there is
always evolution due to the residual ``hot mode'' accretion and black
hole-black hole mergers. To study the downsizing properties of the black
  hole population it is thus much more straightforward to analyze (hard
  X--ray) luminosity functions, as we discuss in Section~\ref{QSOLF}.
 
In the right-hand panel of Figure~\ref{BHMF} we split the Illustris black hole
mass function by the Eddington ratios of black holes, which demonstrates that
in the mean Eddington ratios are moderate (see also
Section~\ref{EDDINGTON}). At the massive end, i.e. for $M_{\rm BH} > 10^9 \,
M_{\rm \odot}$ Eddington ratios plummet indicating that these black holes are
in a radiatively inefficient accretion regime, as expected from the observed
scarcity of luminous quasars in the local Universe. For $M_{\rm BH} \sim 10^8
\, M_{\rm \odot}$ almost all black holes have $\lambda_{\rm EDD} > 10^{-4}$,
while towards lower masses there is an increasing number of black holes with
very low Eddington ratios.

\subsection{Scaling relations with host galaxies and their evolution}\label{SCALING}

\begin{figure*}\centerline{
\includegraphics[width=18truecm]{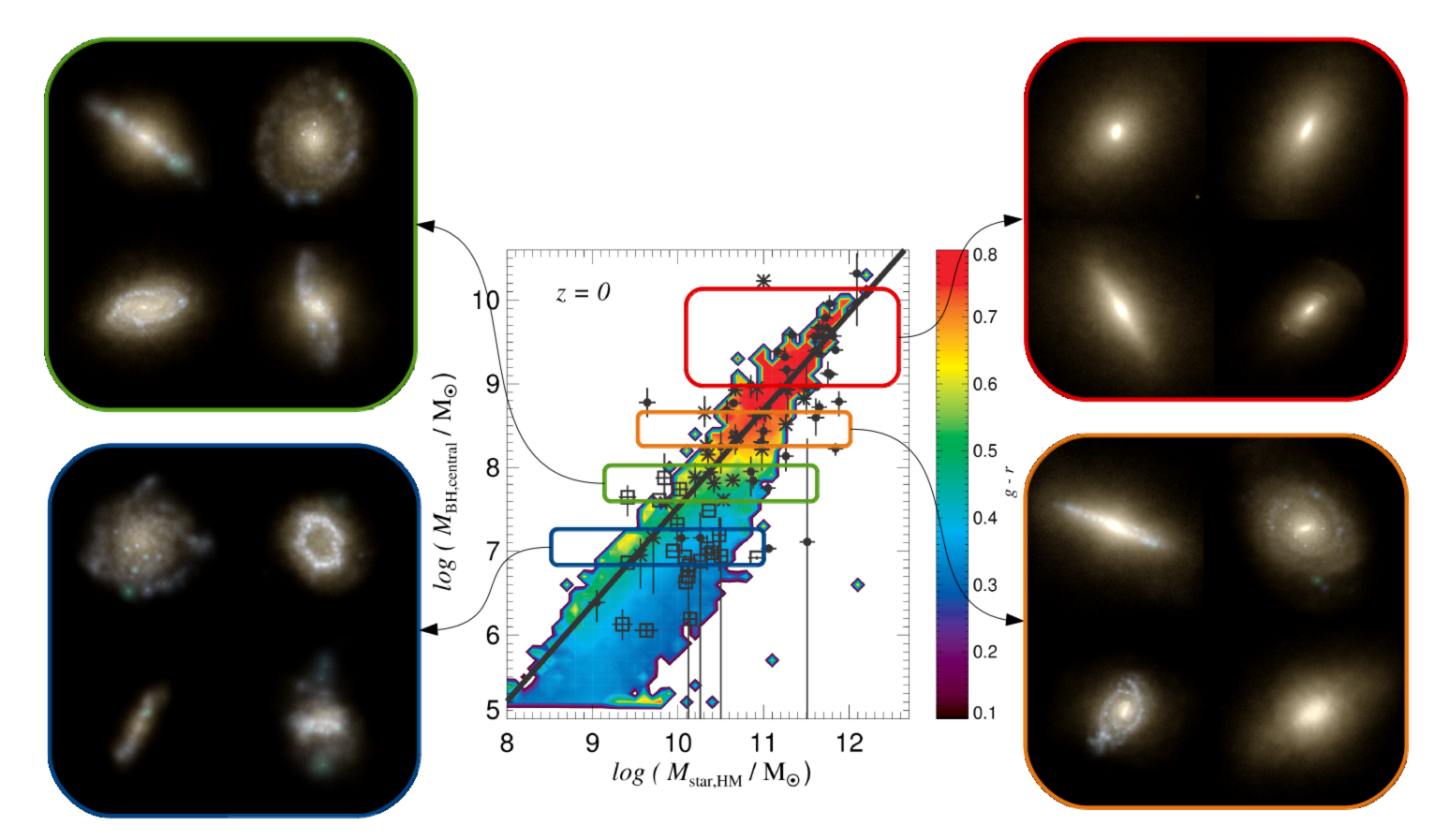} }
\caption{Central panel: stellar half-mass of all galaxies at $z =0$ versus
  their central black hole mass. Colour-coding is according to the $g-r$ colours
  of galaxies. The thick black line denotes the best-fit $M_{\rm BH}$ - $M_{\rm
    star,HM}$ relation from \citet{Kormendy2013} fitted to ellipticals and
  galaxies with bulges only. Symbols with error bars are from
  \citet{Kormendy2013} as well, where circles are for ellipticals, stars are
  for spirals with a bulge and squares are for pseudo bulges. Overall, our
  simulation reproduces the observed findings very well. Note that for $M_{\rm
    star,HM} \lesssim 10^{11} M_{\rm \odot}$ the simulated black holes which are
  above the best-fit observed relation live in redder galaxies, indicating
  feedback from these black holes is quenching their hosts more
  efficiently. Four side panels: stellar morphologies of galaxies visualised
  using SDSS g, r and i bands \citep{Torrey2015} selected
  within a range of black hole masses, as indicated by the coloured
  boxes. While for all four black hole mass ranges there is a morphological
  mix of host galaxies, lower mass black holes are preferentially hosted in
  bluer star-forming and ``diskier'' galaxies.}
\label{MBHMGAL_IMAGE}
\end{figure*}

\subsubsection{$M_{\rm BH}$ - $M_{\rm bulge}$ relation at $z = 0$}

In Figure~\ref{MBHMGAL_IMAGE} we show the Illustris prediction for the black
hole mass -- stellar bulge mass relation. Here the total stellar mass within the
stellar half-mass radius has been adopted as a proxy for the bulge mass. Note
that we do not morphologically distinguish between the real bulges and pseudo
bulges but we do split galaxies into different categories based on their
colours\footnote{Morphologically or kinematically based definitions of bulge
  masses might lead to somewhat different results if, for example, the bulge
  mass fractions 
  depend strongly on the stellar mass, but this analysis is beyond the scope
  of this paper.}. Furthermore, from now on we take into account all galaxies hosting supermassive
black holes with stellar half-mass greater than $10^8 \, M_{\rm \odot}$
and we refer to the $M_{\rm BH}$ - $M_{\rm bulge}$ relation of the whole
population, even though many of these galaxies might not contain a real bulge
or might be effectively bulgeless. The
colour-coding in Figure~\ref{MBHMGAL_IMAGE} is according to the $g-r$ colours of host galaxies and we
consider only the central galaxies of each FOF halo (i.e. the main subhalo of
each FOF halo that contains at least one black hole particle) thus excluding
the  satellites from this analysis. For each subhalo, in case it contains
multiple black  holes, we select the black hole that is closest to the centre
of the subhalo (defined as the position of the most bound particle). We have
also repeated the analysis selecting the most massive black hole of each
subhalo and this does not lead to any significant difference. The thick black
line in Figure~\ref{MBHMGAL_IMAGE} denotes the best-fit $M_{\rm BH}$   -
$M_{\rm bulge}$ relation from a recent compilation by \citet{Kormendy2013}
fitted to ellipticals and spirals with bulges only. Symbols with error bars
are from \citet{Kormendy2013} as well, where circles are for ellipticals,
stars are for spirals with a bulge and squares are for pseudo bulges. As
mentioned in Section~\ref{MASSFUNC}, the black hole mass -- host galaxy relations
have been recently significantly revised \citep[see e.g.][]{Gebhardt2011,
  McConnell2013, 
  Kormendy2013} and the best-fit black hole mass is a factor $2-3$ higher than
previously estimated \citep{Haring2004}, thus it is important to
compare against the newest observational findings.   

The agreement between the Illustris result and the observations is very good,
in particular taking into account that the best-fit observed relation is for
ellipticals and bulges only and that our quenched galaxies lie exactly on this
relation. The result not only reproduces the slope and the normalisation of
the observed $M_{\rm  BH}$ - $M_{\rm bulge}$ relation, but qualitatively also
matches the colours and the morphologies of galaxies on this relation in
agreement with the morphological split performed by \citet{Kormendy2013}. This
is the first time, to our knowledge, that such a wealth of properties of
galaxies hosting supermassive black holes is predicted by 
self-consistent cosmological simulations of galaxy formation. Furthermore,
this implies that with the Illustris simulations we can not only study how
black holes and galaxies co-evolve in the mean, but we can gain a much deeper
insight into which galaxy types are strongly physically linked with their
central black holes and which are much less affected by the presence of a
supermassive black hole in their centre. This seems to be the case, for
example, for the pseudo bulges which correspond to the simulated blue
star-forming galaxies below the best-fit $M_{\rm BH}$ - $M_{\rm bulge}$
relation as we discuss below.

\begin{figure*}\centerline{
\vbox{
\hbox{
\includegraphics[width=8truecm,height=7.5truecm]{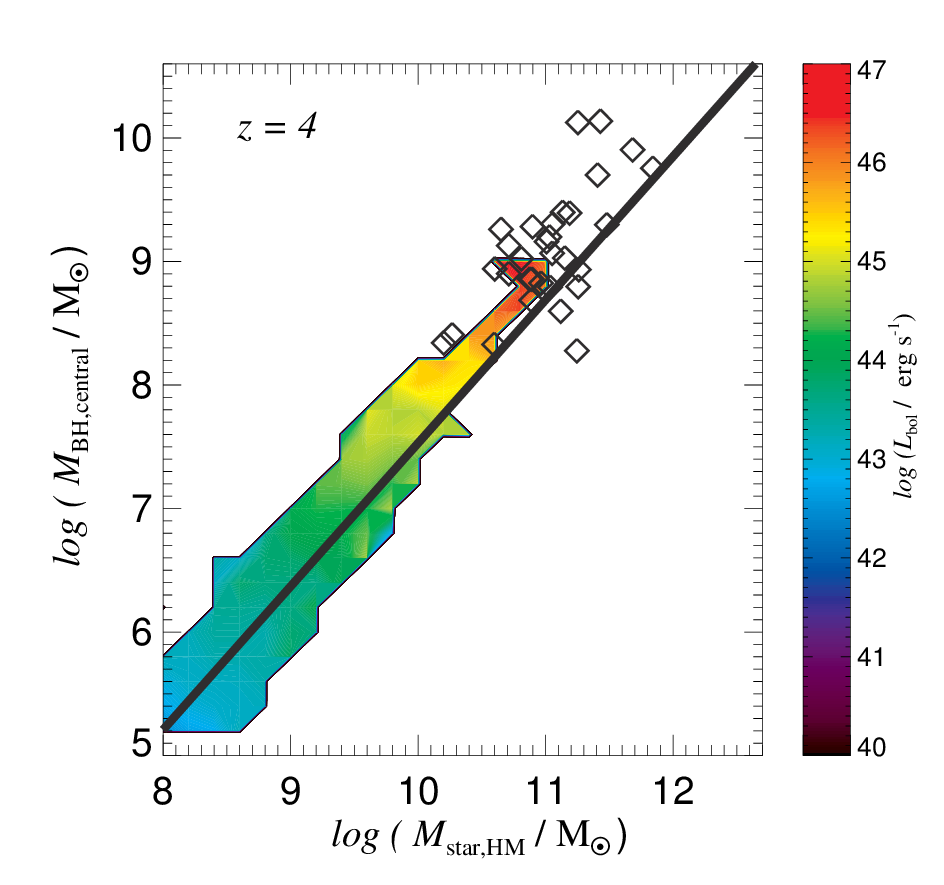}
\includegraphics[width=8truecm,height=7.5truecm]{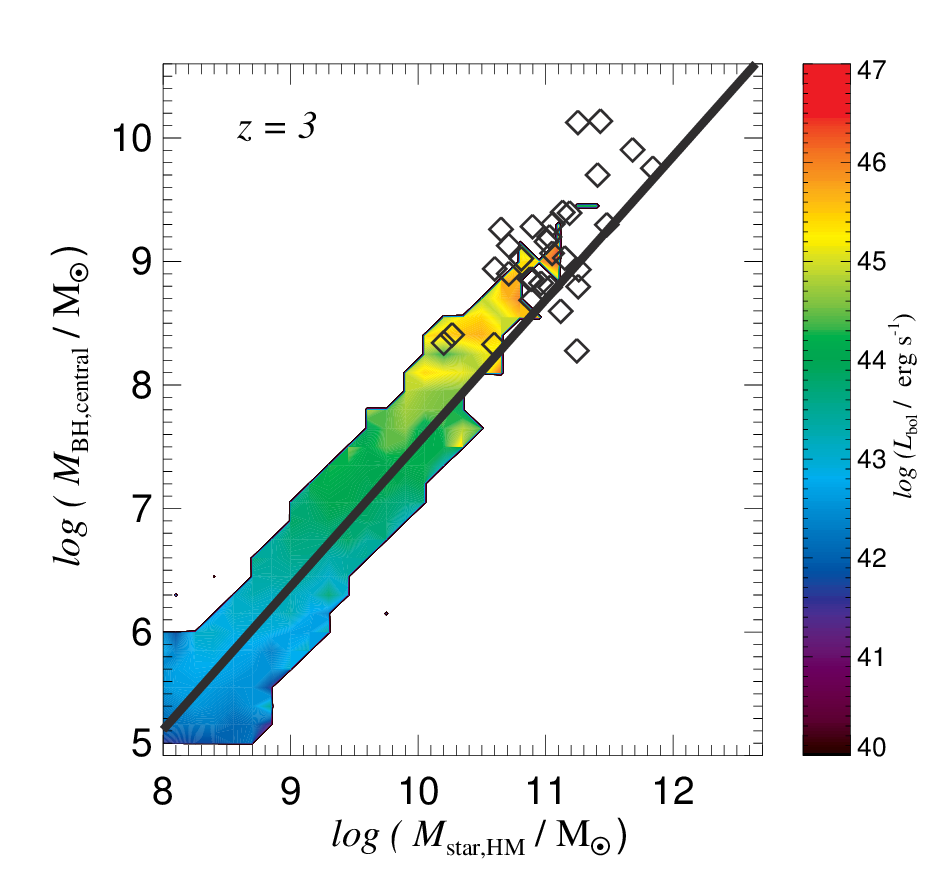}}
\hbox{
\includegraphics[width=8truecm,height=7.5truecm]{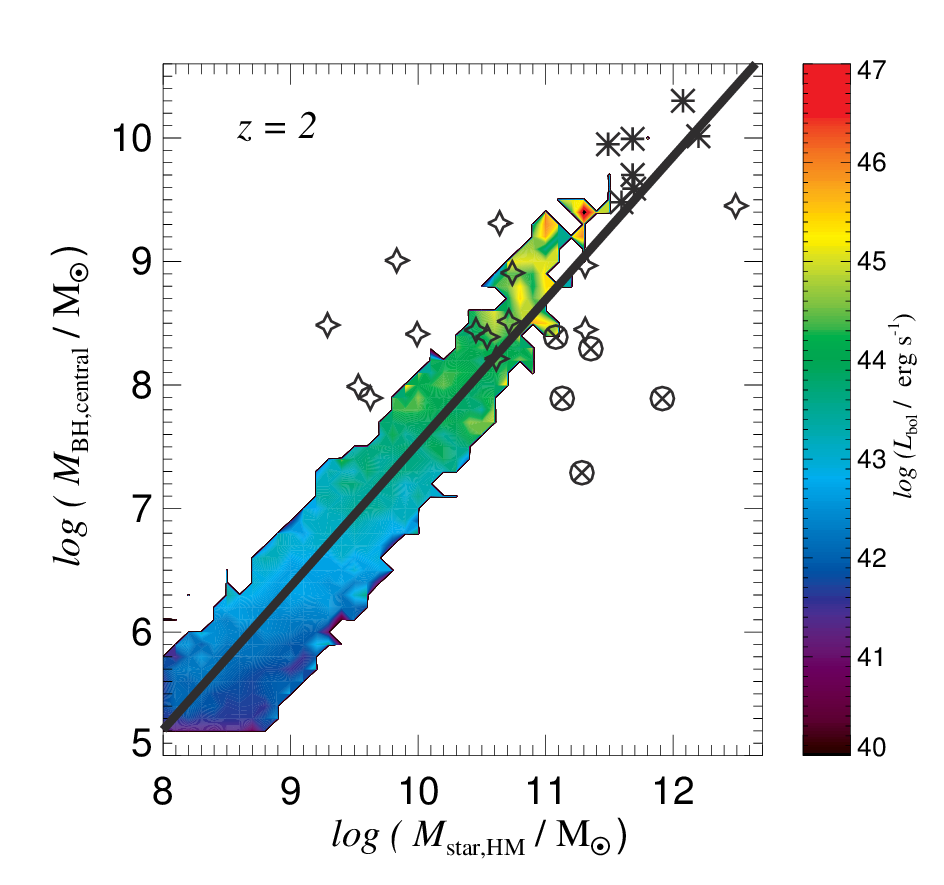}
\includegraphics[width=8truecm,height=7.5truecm]{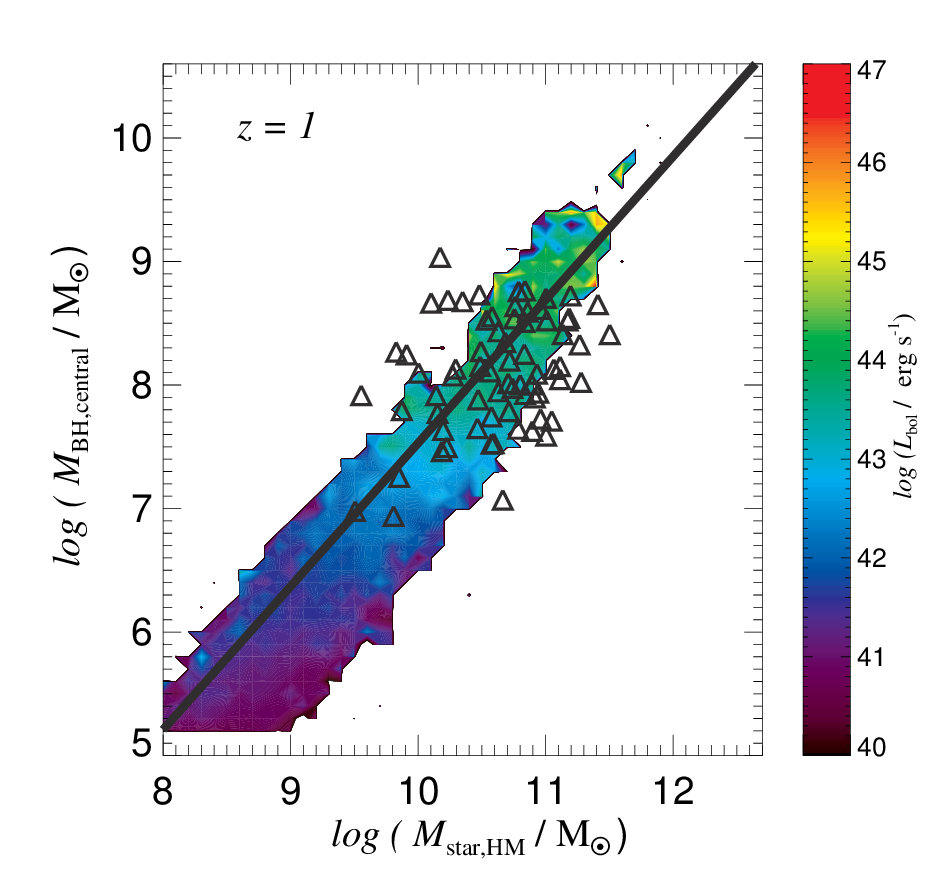}}
}}
\caption{Redshift evolution of the black hole mass -- stellar bulge mass
  relation at $z = 4, 3, 2$ and $1$. Illustris results are shown as $2$D
  histograms, where colour-coding is according to the black hole bolometric
  luminosity (contours include all black holes). Data points at different redshifts are from \citet{Kormendy2013}
  (see text for more details and their Figure 38), while the solid line is the fit from
  \citet{Kormendy2013} to $z=0$ ellipticals 
  and bulges, as in Figure~\ref{MBHMGAL_IMAGE}. Simulation results are
  consistent with data points at all redshifts and indicate
  evolution mostly in the normalisation of the best-fit relation.}
\label{MBHMGAL_ZEVO}
\end{figure*}

Specifically, by focusing on the massive black hole end, i.e. for  $M_{\rm
  BH} > 10^9 \, M_{\rm \odot}$, the simulated host galaxies are very red with $g-r$ colours
greater than 0.7 and the typical morphologies resemble ellipticals, which have
strong central light concentrations, extended red envelopes, post-merger shells
and sometimes red disks, as illustrated in the top-right side panel. As we
move along the $M_{\rm BH}$ - $M_{\rm bulge}$ relation towards black holes with
masses of a
few $10^8 \, M_{\rm \odot}$, typical $g-r$ colours are $0.65$ and the host galaxies
exhibit more of a morphological mix with some red spheroidal galaxies, quenched
extended disks, as well as blue star-forming disks but with prominent red
bulges (see bottom-right side panel). For $\sim 10^8 \, M_{\rm \odot}$ black
holes this transition is more evident with host galaxies lying in the
so-called ``green
valley'' \citep{Schawinski2014}, with mean $g-r$ colours of $0.5$ and morphologies showing both a red
quenched population as well as blue star-forming disks which are sometimes
tilted with respect to the old stellar population indicating a different
assembly history (see top-left side panel). It is interesting to note that the
black holes in ``green valley'' galaxies are the most efficient accretors (see
Section~\ref{MASSFUNC}) and this likely leads to the rapid transition between
star-forming and quenched populations. Finally, for black hole masses
$\le 10^7 \, M_{\rm \odot}$, the majority of hosts are blue and star-forming
with $g-r$ colours of $0.3-0.4$ and irregular and perturbed morphologies (see
bottom-left side panel). 
  
While in general host galaxy colours change from blue to red with increasing
black hole mass, note that for $M_{\rm
  star,HM} \lesssim 10^{11} M_{\rm \odot}$ simulated black holes which are
above the best-fit observed relation live in redder galaxies. This indicates
that the feedback from these black holes, which are more massive than the
average $M_{\rm BH}$ at a given $M_{\rm bulge}$, is quenching their hosts more
efficiently. Conversely, black holes that are under-massive for their host
bulge mass (again for $M_{\rm star,HM} \lesssim 10^{11} M_{\rm \odot}$) tend
to live in the bluest galaxies, which are least affected by their
feedback. The same conclusion has been reached in a recent paper by
  \citet{Snyder2015}, where it has been also shown that at a fixed halo mass
  galaxies with above-average black hole masses have below-average stellar
  masses and earlier-than-average morphological types.

 We finally note that the scatter in the simulated $M_{\rm BH}$ - $M_{\rm bulge}$
 relation becomes smaller with higher black hole mass. This trend is also in
 accordance with observational findings. The reasons for this are twofold: {\it
   i)} as the black holes become more massive and reach certain critical mass
 \citep{King2003, Springel2005b, Costa2014b} their feedback is sufficiently
 strong to self-regulate not only the black hole mass itself but also the
 properties of their host galaxy; {\it ii)} for higher black hole masses more
 galaxy-galaxy dry mergers and thus black hole -- black hole mergers happen and
 due to the central limit theorem \citep{Peng2007, Hirschmann2010,
   Jahnke2011}, the scatter in $M_{\rm BH}$ - $M_{\rm bulge}$ tightens. We 
 emphasise here that black hole feedback is still a necessary and crucial
 ingredient to reproduce the $M_{\rm BH}$ - $M_{\rm bulge}$ relation and that this
 establishes a physical link between the black holes and their central
 galaxies.  

\subsubsection{Redshift evolution of the $M_{\rm BH}$ - $M_{\rm bulge}$ relation}

In Figure~\ref{MBHMGAL_ZEVO} we show the simulated $M_{\rm BH}$ - $M_{\rm
  bulge}$ relation at $z = 1, 2, 3$ and $4$. Colour-coding is now according to
the black hole bolometric luminosity. The solid line is the fit from
\citet{Kormendy2013} to $z=0$ ellipticals and bulges, as in
Figure~\ref{MBHMGAL_IMAGE}, which we plot to emphasise the evolution of the
simulated $M_{\rm BH}$ - $M_{\rm bulge}$ relation. Data points at different
redshifts are from \citet{Kormendy2013}, with triangles corresponding to AGN
($z = 0.1 -1$), stars to radio galaxies (RGs) ($z \sim 2$), circles with a
cross to sub-millimeter galaxies (SMGs) ($z \sim 2$), starred diamonds to
low-redshift quasars (QSOs) ($z = 1- 2$) and diamonds to high-redshift QSOs
($z = 2- 4$)\footnote{See also Figure 38 of \citet{Kormendy2013}.}. Simulation
results are in generally good agreement with the observations. Nonetheless,
due to the limited box size we can probe neither the very luminous nor the
very rare objects, so we cannot conclude much about the evolution of the
$M_{\rm BH}$ - $M_{\rm bulge}$ relation at the very massive end or about
whether the relatively large scatter seen in observations especially for the
SMGs is reproduced (but see also \citealt{Sparre2015} who find a too low
fraction of starbursts in Illustris). We do however find a trend for the
simulated black holes to be more massive for their bulge host mass at higher
redshift (see also recent observational work by \citealt{Bongiorno2014}
  who find a similar trend in agreement with \citealt{Merloni2010}, and
  \citealt{Schulze2014} who instead find no evidence for
  evolution). This is quantified in Table~\ref{TABLE_MBHMBULGE}, where we
list the 
best-fit slope and normalisation of the simulated $M_{\rm BH}$ - $M_{\rm
  bulge}$ relation from $z = 0$ to $z= 4$.\footnote{Note that this best fit
  relation should not be directly compared with the best fit relation from
  \citet{Kormendy2013} as the two fits have been performed on very different
  samples of galaxies.} This is in agreement with the works by
  \citet{Hopkins2007b} and 
  \citet{DiMatteo2008} who also found that at fixed stellar mass black
  holes are more massive at higher redshifts. Note that the trend found by
  \citet{DiMatteo2008} is more evident for galaxies with stellar masses larger
  than $6 \times 10^{10} \, M_{\rm \odot}$.

\begin{table}
\centering
\begin{tabular}{lcc}
\hline
redshift & slope & normalisation \\
\hline
$ z = 0$ & $1.21$ & $-5.29$ \\
$ z = 1$ & $1.23$ & $-5.07$ \\
$ z = 2$ & $1.23$ & $-4.85$ \\
$ z = 3$ & $1.25$ & $-4.91$ \\
$ z = 4$ & $1.28$ & $-5.04$ \\
\hline
\end{tabular}
\caption{The best-fit simulated $\log(M_{\rm BH} / M_{\rm \odot}) = A \times
  \log(M_{\rm bulge} / M_{\rm \odot}) + B$ relation from $z = 0$ to $z= 4$,
  where $A$ is the slope and $B$ the normalisation. Here we take into  account
  only black holes hosted by galaxies with stellar half-mass greater than
  $10^8 \, M_{\rm \odot}$.}
\label{TABLE_MBHMBULGE} 
\end{table}

We finally note that, as expected, there is a very strong trend along the
simulated $M_{\rm BH}$ - $M_{\rm bulge}$ relation for more massive black
holes to have higher bolometric luminosities. There are, however, some very
interesting further trends with cosmic time, namely: {\it i)} for low mass
black holes bolometric luminosities are highest at early times and decrease
thereafter; {\it ii)} the same holds at the massive end where $10^9 \, M_{\rm
  \odot}$ black holes are powering QSOs with $10^{47} {\rm erg \,s^{-1}}$
luminosities preferentially at high redshifts; {\it iii)} the engines of $10^{44} {\rm erg
 \, s^{-1}}$ QSOs (corresponding to green colours on the histograms)
systematically shift from $\sim 10^7 \, M_{\rm \odot}$ to $\sim 10^9 \, M_{\rm
  \odot}$ black holes over the redshift interval considered. These are the
tell-tale signs of cosmic downsizing of the whole AGN population, which we will investigate
in more detail in Sections~\ref{EDDINGTON} and \ref{QSOLF}.

\subsubsection{$M_{\rm BH}$ - $\sigma$ relation at $z = 0$}

\begin{figure*}\centerline{
\includegraphics[width=8truecm,height=7.5truecm]{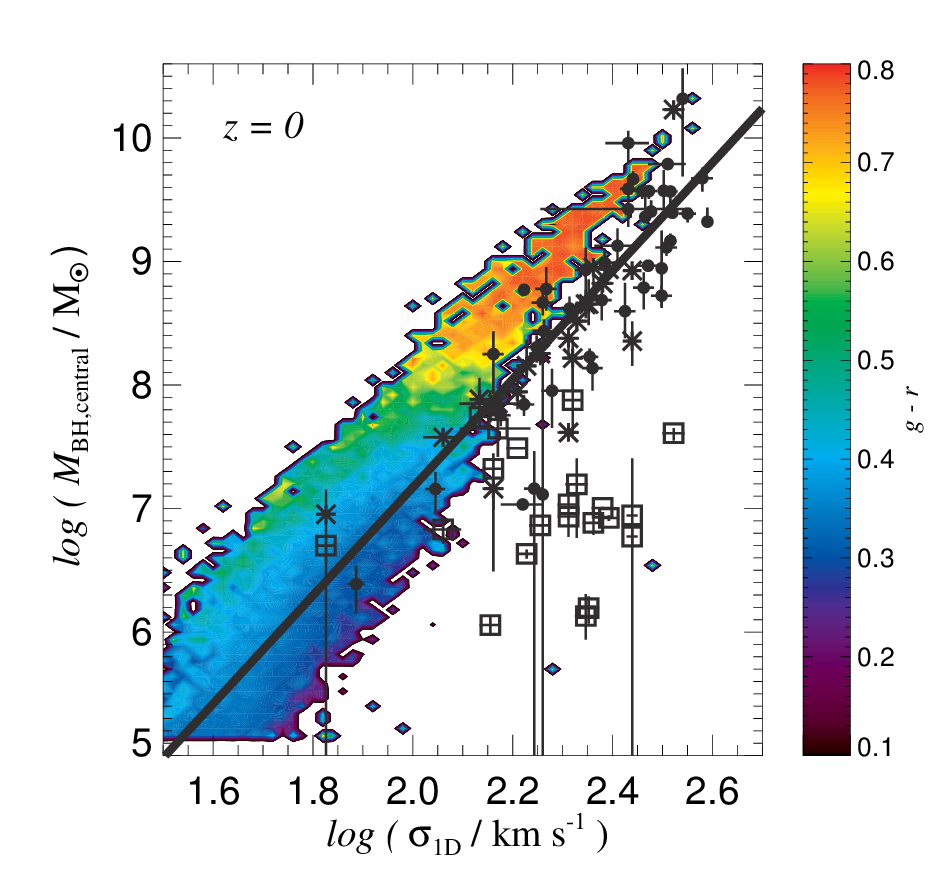}
\includegraphics[width=8truecm,height=7.5truecm]{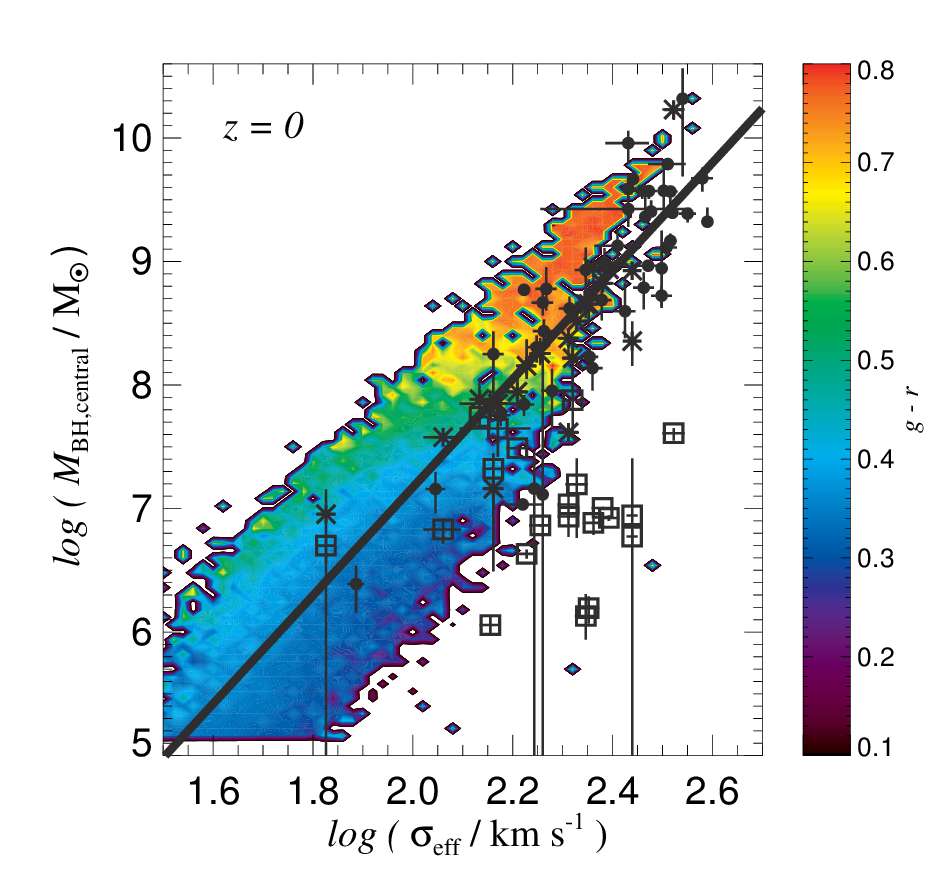}}
\caption{Black hole mass -- stellar velocity dispersion relation at $z =
  0$. Illustris results are shown as $2$D histograms 
  colour-coded according to the host galaxies $g-r$ colours. In the left-hand
  panel we compute the $1$D velocity dispersion of stars, $\sigma_{\rm 1D}$, from
  the mass-weighted $3$D velocity dispersion within the stellar half-mass
  radius. In the right panel we instead compute $\sigma_{\rm eff} =
  (\sigma_{\rm 1D}^2 + V_{\rm rot}^2)^{0.5}$ within the stellar half-mass
  radius (see text for more details). On both panels the thick black line denotes
  the best-fit $M_{\rm BH}$ - $\sigma$ relation at $z = 0$ from \citet{Kormendy2013}
  fitted to ellipticals and galaxies with bulges only. Symbols with error bars
  are from \citet{Kormendy2013} as well, where circles are for ellipticals,
  stars are for spirals with a bulge and squares are for pseudo
  bulges. Illustris agrees well with the observational findings, especially
  if, like for the observations,
  $\sigma_{\rm eff}$ is used as a proxy of stellar velocity dispersion.}
\label{MBHSIGMA}
\end{figure*}

In Figure~\ref{MBHSIGMA} we show the simulated $M_{\rm BH}$ - $\sigma$
relation at $z = 0$. The two panels are for two different ways to compute the
velocity dispersion of the stars. In the left-hand panel we calculate
mass-weighted $1$D velocity dispersions of the stars, $\sigma_{\rm 1D}$, within the stellar
half-mass radii with respect to the mean mass-weighted stellar velocity within the same
radius where we average the full $3$D
velocity dispersion.  In the right-hand panel we compute the rotational
velocity within the stellar half-mass radius, by calculating the mass-weighted
angular momentum of the stars divided by the mass-weighted mean stellar radius
within the stellar half-mass radius. We then take the RMS average of
this rotational velocity and the previously computed $\sigma_{\rm 1D}$ which
we denote as $\sigma_{\rm eff}$. Observationally velocity dispersion of the stars is
typically calculated as 
\begin{equation}
\sigma^2 = \frac{\int_{r_{\rm min}}^{r_{\rm eff}} (\sigma^2(r) + V_{\rm
  rot}^2(r))\, I(r) dr}{\int_{r_{\rm min}}^{r_{\rm eff}} I(r) dr} \,, 
\end{equation}
where
$r_{\rm eff}$ is the effective radius\footnote{Note that in practice different
  authors average over a different range of radii.}, $I(r)$ is the stellar surface
brightness profile, $\sigma(r)$ is the line-of-sight velocity dispersion and
$V_{\rm rot}$ is the rotational velocity \citep{Gultekin2009, McConnell2013,
  Kormendy2013}. While \citet{Gultekin2009} argue that $V_{\rm rot}$ is
typically small compared to $\sigma$ for a subset of their galaxies that have
central stellar velocity dispersion measurement available from HyperLEDA,
\citet{Harris2012} show that many of their galaxies hosting type-1 AGN exhibit
a significant rotational component. For these reasons, we show the simulated
$M_{\rm BH}$ - $\sigma$ relation both with $\sigma_{\rm 1D}$ and $\sigma_{\rm
  eff}$. Furthermore, we have verified that if we compute  $\sigma_{\rm 1D}$
within half (or twice) the stellar half-mass radius the $M_{\rm BH}$ -
$\sigma$ relation is essentially unchanged. Thick black lines denote
  the best-fit $M_{\rm BH}$ - $\sigma$ relation at $z = 0$ from \citet{Kormendy2013}
  fitted to ellipticals and galaxies with bulges only. Symbols with error bars
  are from \citet{Kormendy2013} as well, where circles are for ellipticals,
  stars are for spirals with a bulge and squares are for pseudo
  bulges.

\begin{table}
\centering
\begin{tabular}{lcc}
\hline
redshift & slope & normalisation \\
\hline
for $\sigma_{\rm 1D}$ & &  \\
$ z = 0$ & $5.04$ & $-2.69$ \\
$ z = 1$ & $4.97$ & $-2.62$ \\
$ z = 2$ & $4.81$ & $-2.37$ \\
$ z = 3$ & $4.64$ & $-2.16$ \\
$ z = 4$ & $4.26$ & $-1.60$ \\
\hline
for $\sigma_{\rm eff}$ & &  \\
$ z = 0$ & $4.42$ & $-1.83$ \\
$ z = 1$ & $4.24$ & $-1.48$ \\
\hline
\end{tabular}
\caption{The best-fit simulated
  $\log(M_{\rm BH} / M_{\rm \odot}) = A \times \log(\sigma / {\rm km\,
    s^{-1}}) + B$ 
  relation from $z = 0$ to $z =
  4$, where A is the slope and B the normalisation. Here we take into 
  account only black holes hosted by galaxies with stellar half-masses greater
  than  
  $10^8 \, M_{\rm \odot}$, as we did for the $M_{\rm BH}$ - $M_{\rm bulge}$
  relation as well. Top $5$ rows are for $\sigma_{\rm 1D}$, while the bottom two rows are for
  $\sigma_{\rm eff}$ (see text for more details.)}
\label{TABLE_MBHSIGMA} 
\end{table}

\begin{figure*}\centerline{
\includegraphics[width=8truecm,height=7.5truecm]{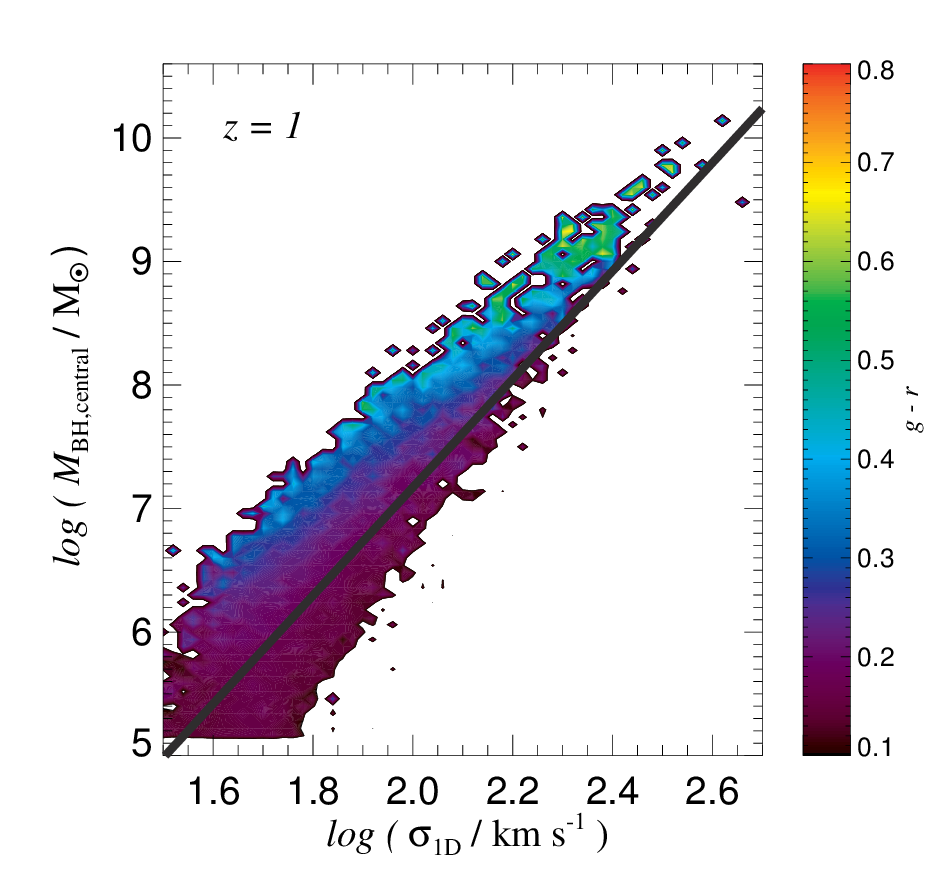}
\includegraphics[width=8truecm,height=7.5truecm]{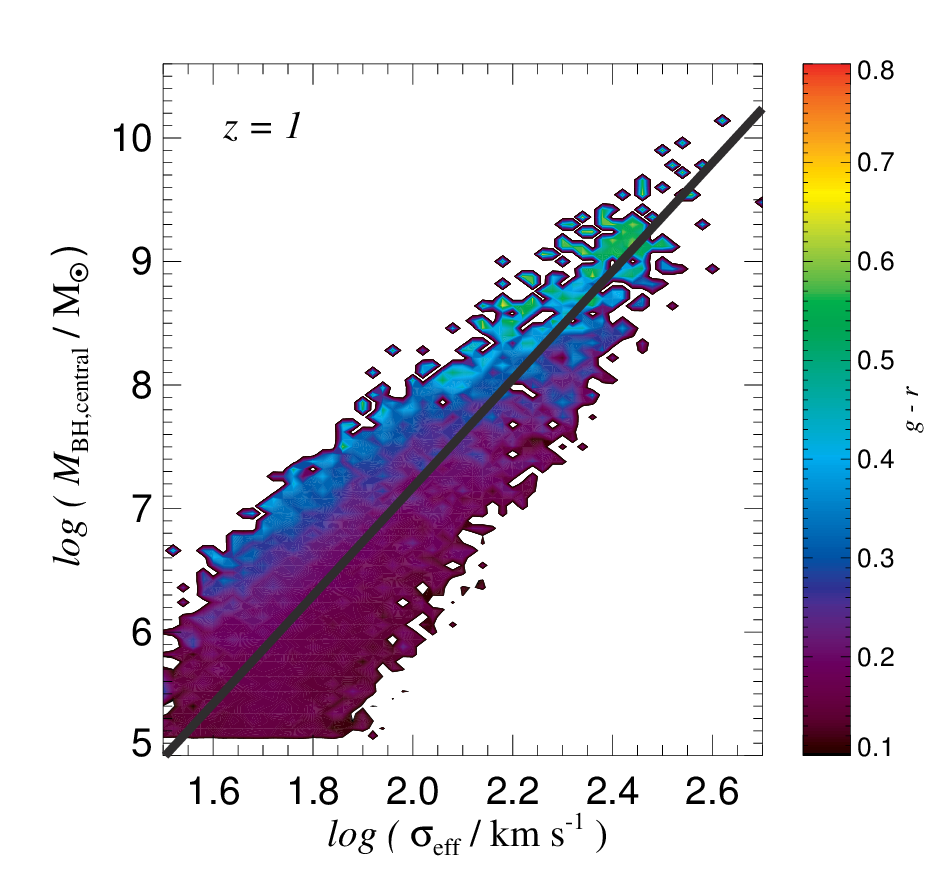}}
\caption{Black hole mass -- stellar velocity dispersion relation at $z =
  1$. Illustris results are shown as $2$D histograms 
  colour-coded according to the host galaxies $g-r$ colours (rest-frame). In
  the left  panel we compute the $1$D velocity dispersion of stars, $\sigma_{\rm
    1D}$, from the mass-weighted $3$D velocity dispersion within the stellar
  half-mass radius. In the right panel we instead compute $\sigma_{\rm eff} =
  (\sigma_{\rm 1D}^2 + v_{rot}^2)^{0.5}$ within the stellar half-mass radius
  (see text for more details). On both panels the thick black line denotes the
  best-fit $M_{\rm BH}$ - $\sigma$ relation at $z = 0$ from \citet{Kormendy2013} fitted
  to ellipticals and galaxies with bulges only. Note the considerable evolution in
  host galaxy colours from $z = 0$ (see Figure~\ref{MBHSIGMA}), and that the
  rotational velocity is now contributing more to $\sigma_{\rm eff}$ at almost
  all black hole masses.}
\label{MBHSIGMA_Z1}
\end{figure*}

\begin{figure*}\centerline{
\vbox{
\hbox{
\includegraphics[width=8truecm,height=7.5truecm]{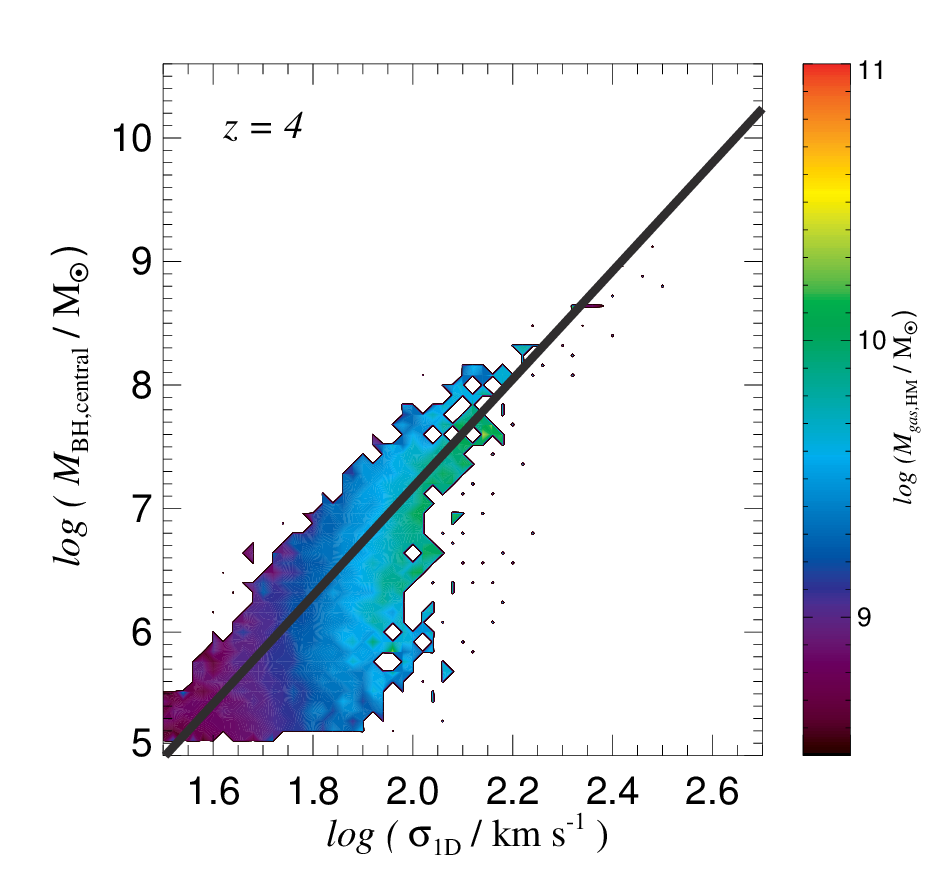}
\includegraphics[width=8truecm,height=7.5truecm]{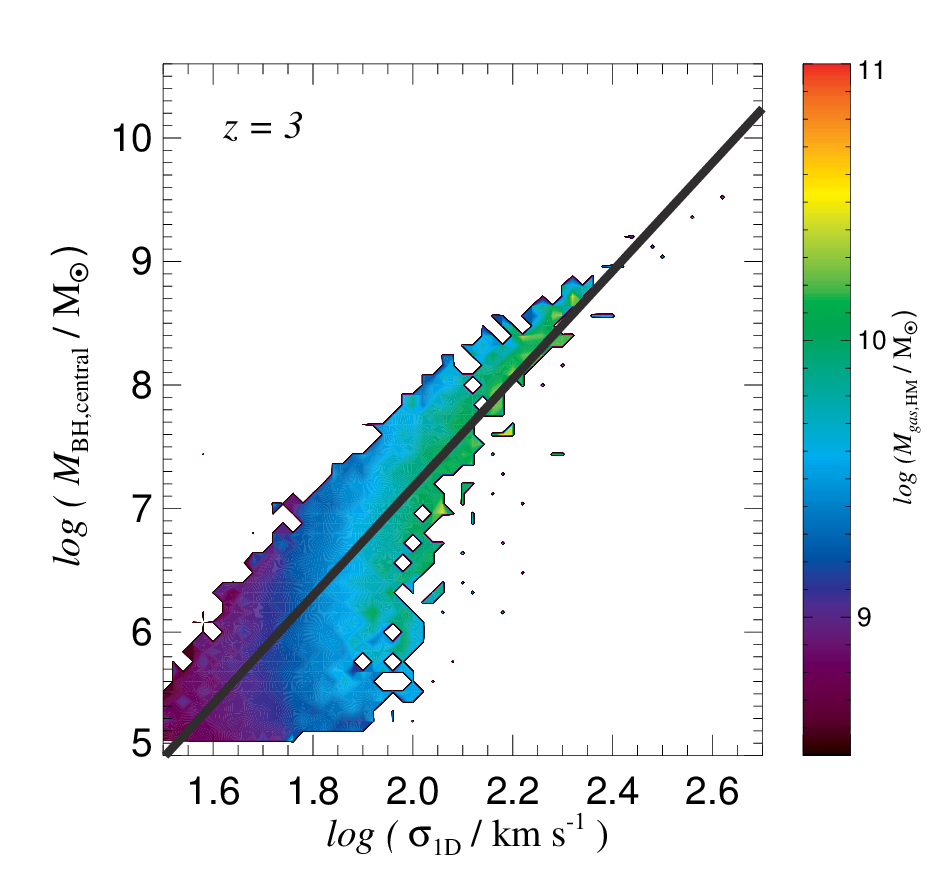}}
\hbox{
\includegraphics[width=8truecm,height=7.5truecm]{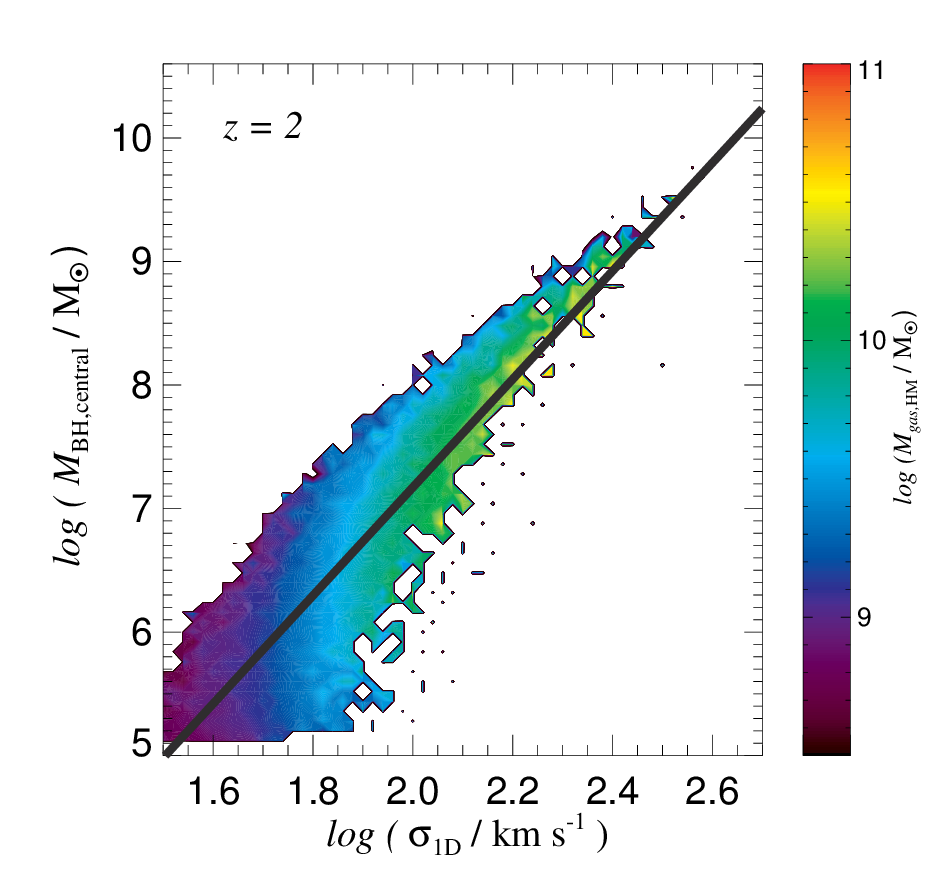}
\includegraphics[width=8truecm,height=7.5truecm]{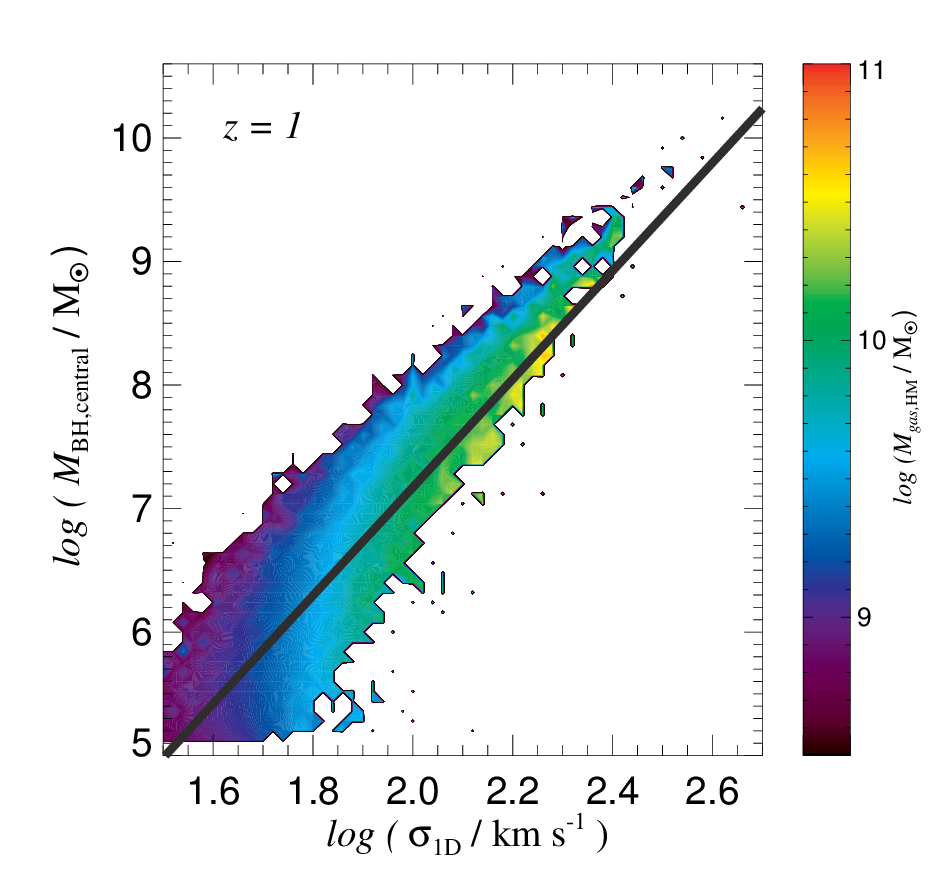}}
}}
\caption{Redshift evolution of the black hole mass -- stellar velocity
  dispersion relation at $z = 4, 3, 2$ and
  $1$. Illustris results are shown as $2$D  histograms, where colour-coding is
  according to the total gas mass within stellar half-mass radius. The thick solid
  line is the fit from \citet{Kormendy2013} to $z=0$ ellipticals and bulges,
  as in Figure~\ref{MBHSIGMA}. Simulation results indicate that the $M_{\rm BH}$ -
  $\sigma$ relation is evolving with the slope steepening from
  $\sim 4.26$ at $z = 4$ to $\sim 4.97$ at $z = 1$.}
\label{MBHSIGMA_ZEVO}
\end{figure*}

From Figure~\ref{MBHSIGMA} we draw several important conclusions: {\it i)}
the rotational velocity is very subdominant for the galaxies hosting the most
massive black holes, where $\sigma_{\rm 1D}$ and $\sigma_{\rm eff}$ give very
similar results; {\it ii)} for black hole
masses less than $10^9 \, M_{\rm 
  \odot}$ a fraction of galaxies has a non-negligible $V_{\rm rot}$ component
such that for a given black hole mass the stellar velocity dispersion can be shifted
further to the right, increasing the scatter of the $M_{\rm BH}$ - $\sigma$
relation;  {\it iii)} at the massive end the agreement between the Illustris
prediction and observational findings is very good, especially if we use
$\sigma_{\rm eff}$ (for the best-fit relation see Table~\ref{TABLE_MBHSIGMA}),
 but note that at fixed black hole mass our simulated $\sigma_{\rm eff}$
  is still on average slightly lower than the \citet{Kormendy2013} sample of
  ellipticals; {\it iv)} at the low mass end the inclusion of $V_{\rm rot}$ 
does not help us to completely explain the scatter seen in the pseudo bulges
(even though there are a few simulated galaxies which have extremely high
$\sigma$ values for their black hole mass). Future observations of low
$\sigma$ galaxies hosting supermassive black holes will be crucial to
constrain $M_{\rm BH}$ - $\sigma$ relation at the low black hole mass end. In
terms of host galaxy colours we 
see the same trends described for the $M_{\rm BH}$ - $M_{\rm bulge}$ relation,
namely a well defined sequence of increasingly red $g-r$ colours with
higher black hole mass and over-/under-massive black holes hosted by the
redder/bluer than average galaxies at a given $\sigma$.

\subsubsection{Redshift evolution of the $M_{\rm BH}$ - $\sigma$ relation}

We now explore the difference between $\sigma_{\rm 1D}$ and $\sigma_{\rm eff}$
at $z = 1$ to identify possible systematic uncertainties for the future
observational determinations of the $M_{\rm BH}$ - $\sigma$ relation, given that
currently the vast majority of $M_{\rm BH}$ - $\sigma$ measurements are
  for $z 
  \lesssim 1.0$ \citep[for high redshift measurements see
    e.g.][]{Salviander2013}. As
the fraction of rotationally supported systems is higher at $z = 1$ than at $z
= 0$ we find that the inclusion of the rotational velocity becomes a more
significant effect leading to a larger scatter and a different slope and
normalisation of the best-fit relation, as shown in
Table~\ref{TABLE_MBHSIGMA}\footnote{It is interesting to note that
  \citet{Kassin2007} also consider the combination of the rotational velocity
  and velocity dispersion of the gas when studying the redshift evolution of the
  stellar mass Tully-Fisher relation, but find that the scatter is reduced
  with respect to using gas rotational velocity alone.}. We thus conclude that
the observational 
uncertainties and systematic biases in the $M_{\rm BH}$ - $\sigma$ relation at $z
= 0$ \citep[see also][]{Bellovary2014}, and especially at $z > 0$ could be
more significant than currently assumed. At $z > 0$ there are other
  important biases stemming from samples based on the AGN luminosity
  \citep{Lauer2007b} and from the uncertainty associated with the single-epoch
  virial black hole mass estimators \citep{Shen2010}, which both lead to systematic
  overestimation of black hole masses (see also discussion about the evolution
  of the $M_{\rm BH}$ - $M_{\rm bulge}$ relation by
  \citealt{Kormendy2013}). Thus, the real evolutionary trends of black
  hole -- host galaxy relations with
  cosmic time are currently rather uncertain.

By comparing Figures~\ref{MBHSIGMA} and \ref{MBHSIGMA_Z1} for which $2$D
histograms have been both colour-coded according to the rest frame $g-r$ colours
we can quantify how colours of galaxies hosting supermassive black holes
evolve with cosmic time. We find that for a given black hole mass,
galaxies at $z = 1$ are bluer. However the overall trend of a well defined
sequence of increasingly redder $g-r$ colours along the relation and
over/under-massive black holes sitting in redder/bluer than average galaxies
at a given $\sigma$ still remains, highlighting the importance of AGN feedback
for $z > 0$.

\begin{figure*}\centerline{
\hbox{
\includegraphics[width=8truecm,height=7.5truecm]{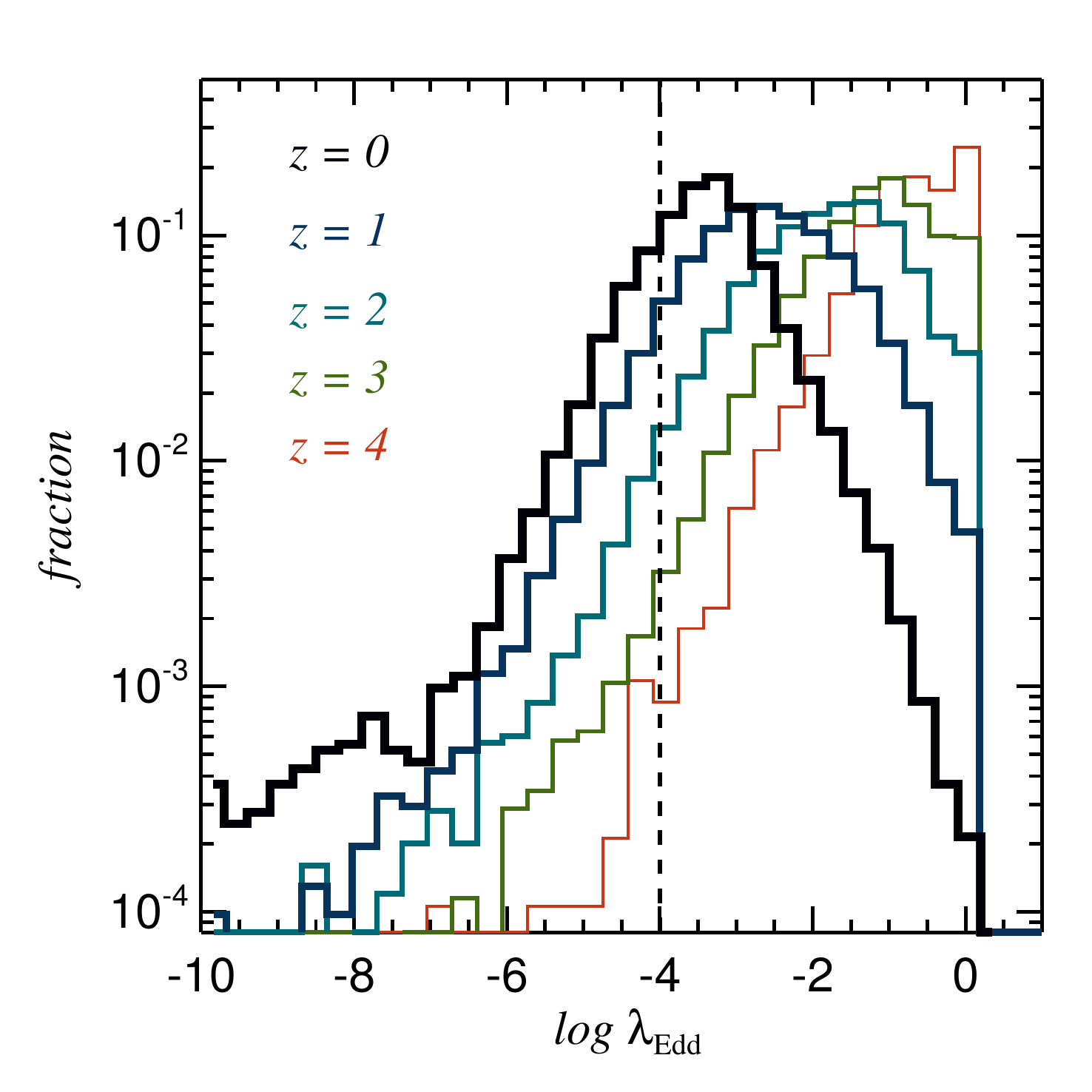}
\includegraphics[width=8truecm,height=7.5truecm]{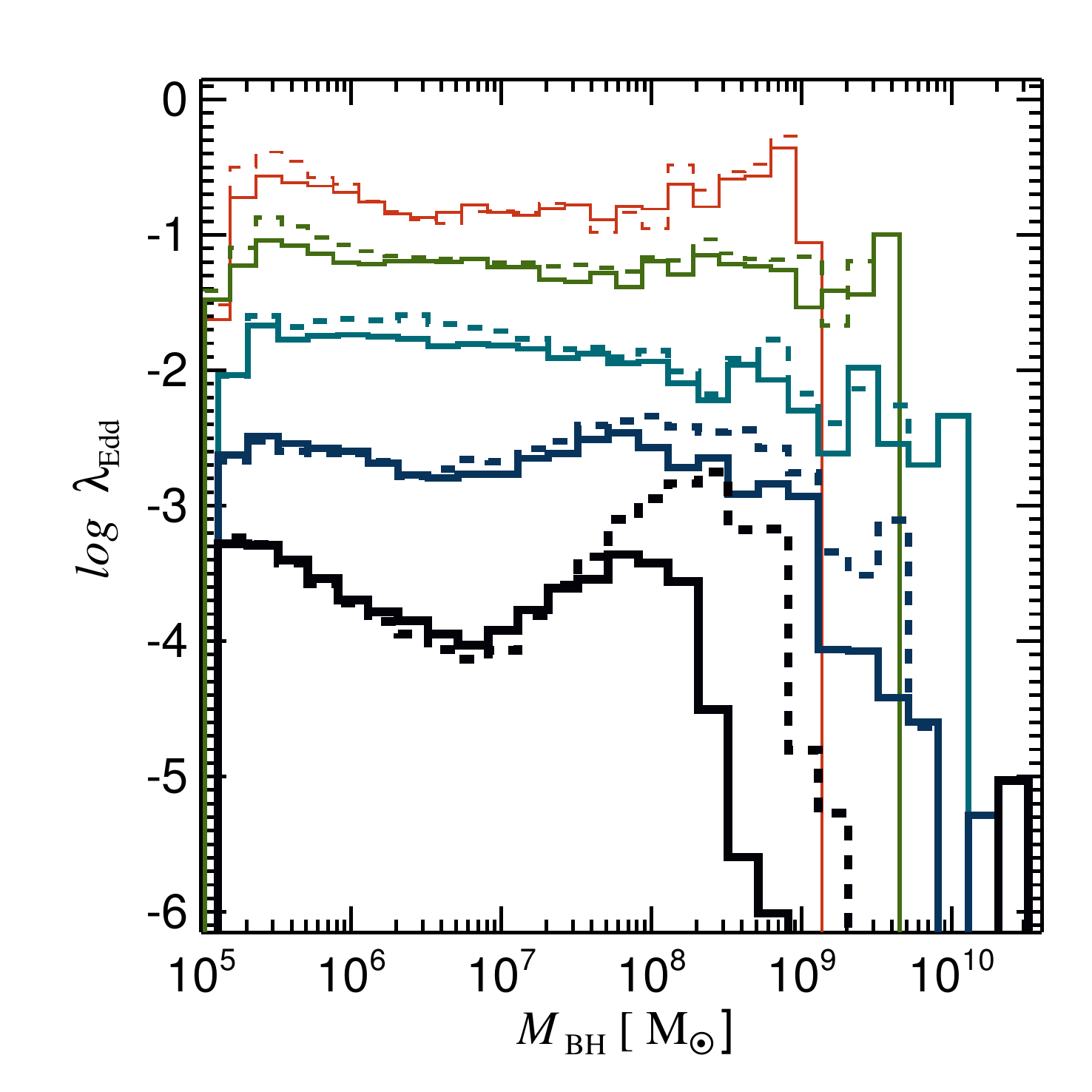}}
}
\caption{Left: distribution of black hole Eddington ratios at $z = 4, 3, 2, 1$
and $0$, as indicated on the legend. For Eddington ratios lower than
  $\lambda_{\rm Edd} = 10^{-4}$ (dashed vertical line) the model is not well
  converged. There is a clear evolution in the
Eddington ratios with cosmic time, with many black holes accreting
close to the maximal rate at $z = 4$, while for $z = 0$ the mean
  Eddington ratio is low. Right: Eddington ratios as a function of
black hole mass at $z = 4, 3, 2, 1$ and $0$ (same colour-coding as on the left
panel). Continuous lines denote the mean of the logarithm of $\lambda_{\rm
  Edd}$ in each bin, while the dashed lines show the median values. Apart
from $z = 0$ result where the average Eddington ratio is lowest for $\sim
10^{7} M_{\rm \odot}$ black holes, the distribution of $\lambda_{\rm Edd}$ is
fairly flat. Note that at the massive end, i.e $M_{\rm BH} \ge 2 \times 10^{9} M_{\rm
  \odot}$, the mean $\lambda_{\rm Edd}$ drops most with decreasing redshift, which
is a clear signature of cosmic downsizing.}
\label{LEDD}
\end{figure*}

In Figure~\ref{MBHSIGMA_ZEVO} we show the simulated $M_{\rm BH}$ - $\sigma$
relation at $z = 1, 2, 3$ and $4$, where we only show the result for
$\sigma_{\rm 1D}$. Here the colour-coding is according to the total gas mass
within the stellar half-mass radius and the thick solid line is the fit from
\citet{Kormendy2013} to $z=0$ ellipticals and bulges.  There are several 
important features to note: {\it i)} for $z > 0$ the scatter in the $M_{\rm
  BH}$ - $\sigma$ relation is larger than for the $M_{\rm BH}$ - $M_{\rm bulge}$
relation even without including the effect of rotational velocity; {\it ii)}
the scatter in the relation significantly increases at higher redshifts. We
have verified that calculating velocity dispersion in different ways does not
decrease the scatter (e.g. computing $\sigma_{\rm 1D}$ with respect to the median
velocity or the mean velocity of the whole subhalo, computing $\sigma_{\rm
  1D}$ within $2 \times R_{\rm HM}$, defining the centre as the centre of mass
of the whole subhalo or as the position of the black hole particle, changing
the bin size of $2$D histograms); {\it
  iii)} due to the finite box size we are not able to probe the redshift
evolution of black holes at the massive end, but we note that once black holes
reach the $M_{\rm BH}$ - $\sigma$ relation they tend to stay on it (or become
somewhat more massive); {\it iv)} most of
the evolution happens at the low mass end, where a large fraction of black
holes at high redshifts is below the $z = 0$ relation; {\it v)} this also
drives a strong steepening of the simulated best-fit slope with cosmic time
which evolves from $\sim 4.26$ at $z = 4$ to $\sim 5.04$ at $z = 0$ (for further
details see Table~\ref{TABLE_MBHSIGMA}); {\it vi)} as evident from
colour-coding the total gas mass within the stellar half-mass radius not only
increases with $\sigma$ (which would be a simple dependence on host halo mass), but
it also shows a trend roughly perpendicular to the best-fit relation: for a given
$\sigma$ under-massive black holes live in more gas rich environments than is
the case for the over-massive black holes and this difference is up to a factor
of $10$. This demonstrates that the strong AGN feedback not only quenches
galaxy colours but it also efficiently expels gas from the galaxies' innermost
regions.
 
The evolution of the slope of the simulated $M_{\rm BH}$ - $\sigma$ relation
is very interesting. According to the well established analytical models
\citep{Silk1998, Haehnelt1998, Fabian1999, King2003} and as recently
demonstrated by detailed numerical simulations 
\citep{Costa2014} momentum-driven AGN outflows lead to a slope of the
$M_{\rm BH}$ - $\sigma$ relation equal to $4$, whereas energy-driven outflows
correspond to steeper slopes of $5$. While we are not directly injecting
any momentum into the medium surrounding black holes, large energy injections
where the gas becomes outflowing and where at the same time considerable energy is
lost due to the radiation resemble momentum-driven flows and lead to
slopes of the $M_{\rm BH}$ - $\sigma$ relation closer to $4$. Conversely, large
energy injections where most of the energy is not radiated away should lead to
slopes closer to $5$. The evolutionary trend that we see in Illustris can thus
be interpreted in the following way: at high redshifts there are copious
amounts of very dense and cold gas due to the rapid cooling regime
\citep{White1991, Birnboim2003, Nelson2013} and thus AGN injected energy can
be easily 
radiated away, leading to a  $M_{\rm BH}$ - $\sigma$ relation with a slope of $\sim 4$;
at low redshifts and especially in massive galaxies there is less cold gas
inflow as gas is supported by quasi hydrostatic atmospheres. Thus AGN feedback
becomes more energy-driven. This is particularly the case for our ``radio''
mode heating which is more bursty, more energetic and thus less prone to the
radiative losses. Moreover, at lower redshifts, $z \lesssim 2$, AGN
  feedback starts to become very efficient at quenching galaxies and more dry
mergers take place. Dry mergers tend to increase both black hole and stellar
mass, while keeping the stellar velocity dispersion largely unchanged
\citep[see e.g.][]{Boylan2006}. Taken together, these arguments may
help explain why the slope of the $M_{\rm BH}$ - $\sigma$ steepens for the
most massive black holes, as seen in observations. 

\begin{figure*}\centerline{
\hbox{
\includegraphics[width=8truecm,height=7.5truecm]{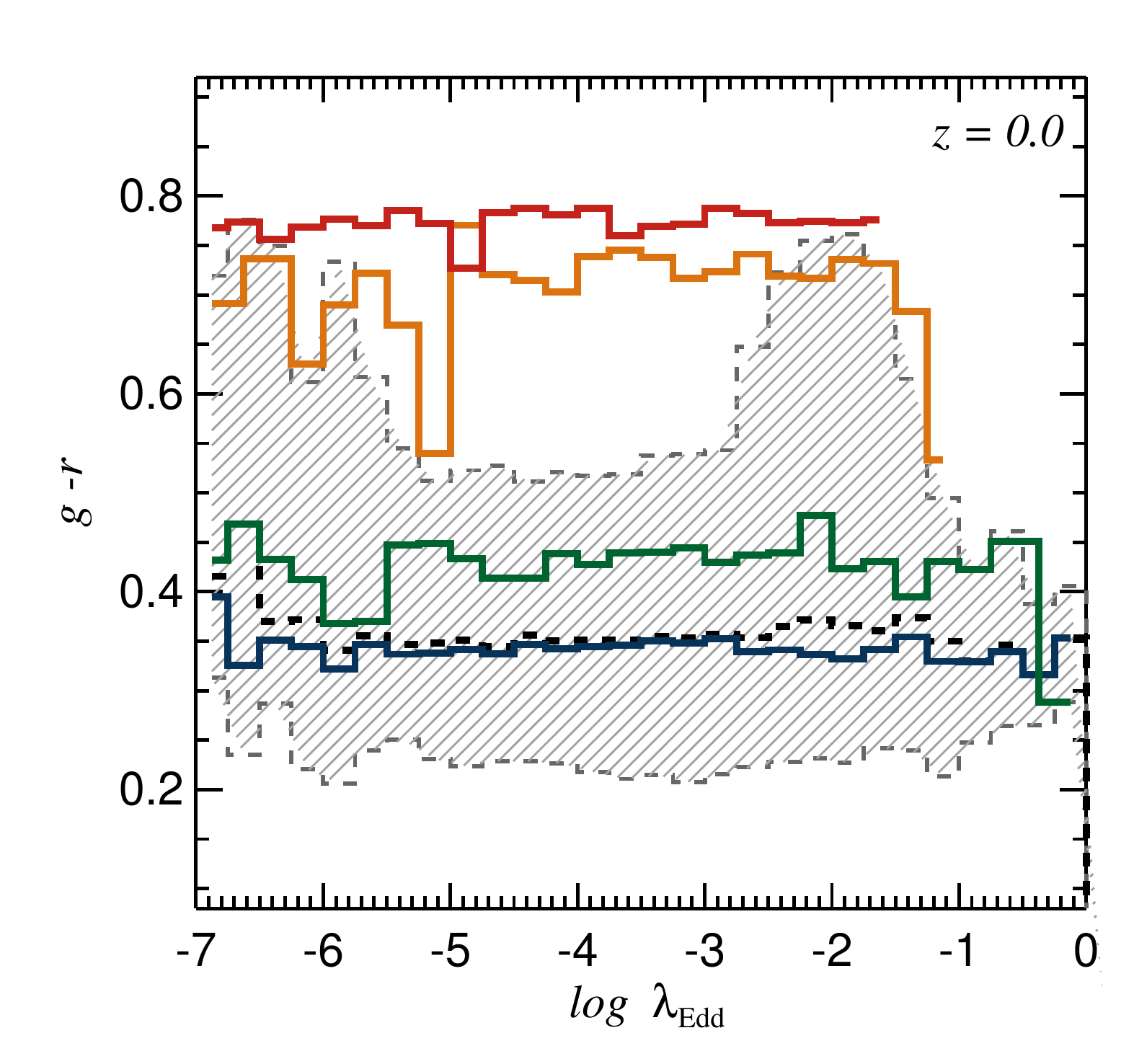}
\includegraphics[width=8truecm,height=7.5truecm]{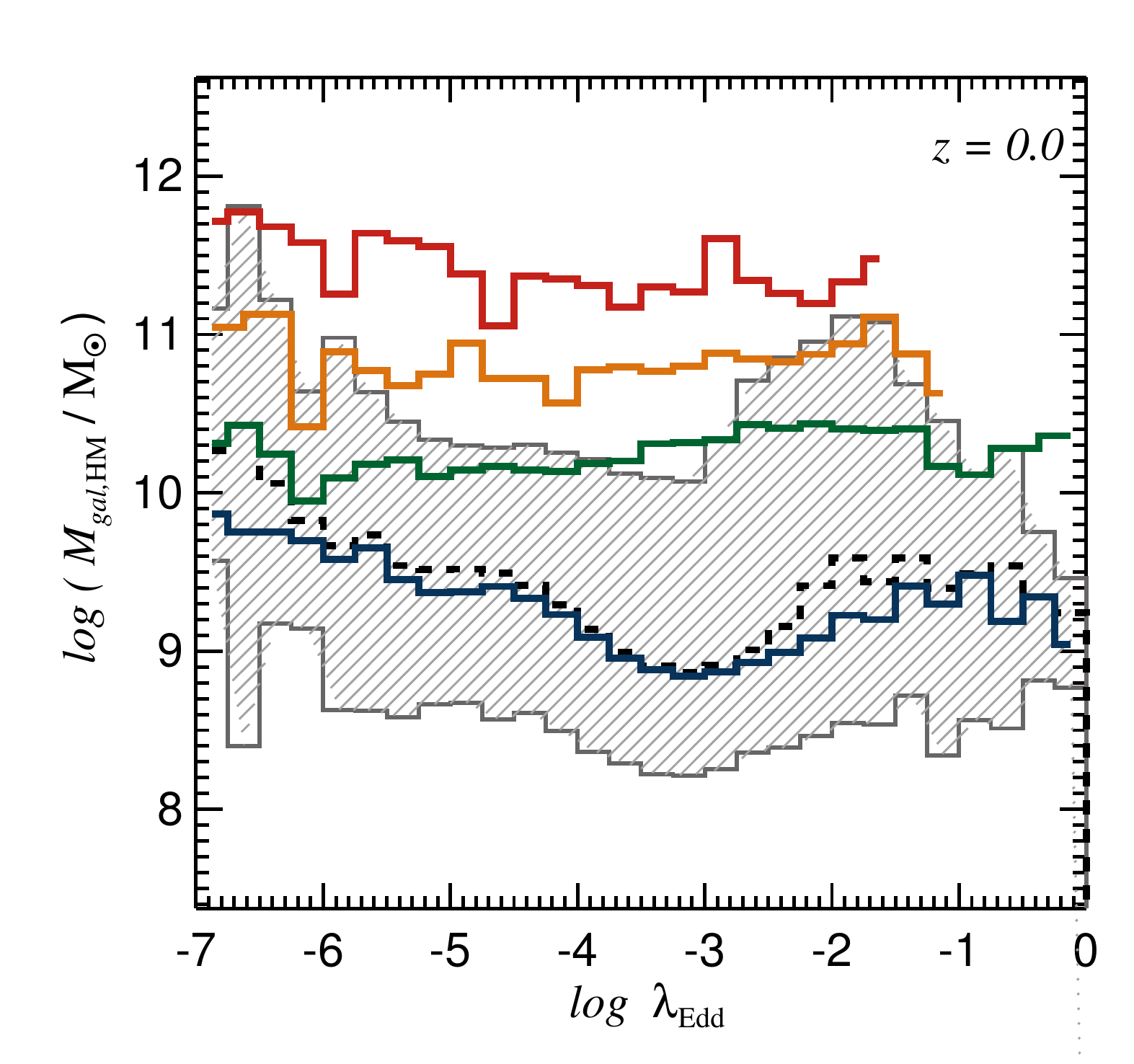}}
}
\caption{Distribution of $g-r$ colours (left) and stellar masses within the half-mass
  radii (right) of galaxies at $z = 0$ as a function of the Eddington ratios of
  their central black holes. The black dashed histogram denotes the median of
  the 
  distribution, and the shaded region encloses the $5$th to $95$th percentile
  of the 
  distribution. Coloured histograms are for galaxies
  hosting black holes in different mass ranges: $M_{\rm BH} < 10^7 M_{\rm
    \odot}$ (blue), $10^7 M_{\rm \odot} \le M_{\rm BH} < 10^8 M_{\rm \odot}$
  (green), $10^8 M_{\rm \odot} \le M_{\rm BH} < 10^9 M_{\rm \odot}$ (orange),
  $10^9 M_{\rm \odot} \le M_{\rm BH} $ (red), where the colours have been
  chosen to roughly match the colour-coding of 
Figure~\ref{MBHMGAL_IMAGE}. For any given $\lambda_{\rm Edd}$ there is a very
large spread in galaxy colours and stellar masses, while the median of the
distributions is driven by galaxies hosting low mass black holes. More massive
black holes live in redder and more massive galaxies but have a vast
range of Eddington ratios, with a clear trend only for the highest
$\lambda_{\rm Edd}$ values.}
\label{LEDD_COLOUR}
\end{figure*}

\subsection{Eddington ratios}\label{EDDINGTON}

In Figure~\ref{LEDD} we show the distribution of black hole Eddington ratios
from $z = 4$ to $z = 0$ (left-hand panel) and the mean and median Eddington
ratios as a function of 
black hole mass for the same redshift interval (right-hand panel). Note
  that for Eddington ratios lower than
  $\lambda_{\rm Edd} = 10^{-4}$ (denoted with a dashed vertical line) the model is not well
  converged. While the
distribution of Eddington ratios is quite broad there is a clear trend with
cosmic time \citep[see also e.g.][]{Sijacki2007, DiMatteo2008}. At high redshifts the majority of black holes accrete at a high rate
and the mean Eddington ratio is essentially flat as a function of black hole
mass. With cosmic time the peak of the Eddington ratio distribution
systematically shifts towards lower $\lambda_{\rm Edd}$ values, the
fraction of the population accreting at the maximal rate decreases, while at
the same time a larger tail of very low Eddington accretors builds up.
  Mean Eddington ratios at $z = 0, 1, 2, 3$ and $4$ are $< \log \lambda_{\rm
    Edd}> = -3.6, -2.6, -1.8, -1.2$ and $-0.7$, while if we consider only AGN
  with bolometric luminosities greater than $10^{42}\, {\rm erg\, s}^{-1}$
the respective values are $-2.3, -1.6, -1.2, -0.8$ and $-0.5$. These results
are in 
good qualitative agreement with a study of broad-line SDSS quasars by \citet{Shen2012}, whose mean Eddington ratios
interpolated on the same redshifts yield $-2.2, -1.5,-1.0, -0.7$ and
$-0.4$. Interestingly, while \citet{Shen2012} find a strong evolution of the
mean  
Eddington ratio with redshift, they do not find that it strongly depends on
the black hole mass, similar to our simulation results. Note, however, that the comparison with \citet{Shen2012} is not
straightforward as our sample is volume-limited and theirs is flux-limited.
Note that at the massive end, i.e $\ge 2 \times 10^{9} M_{\rm
  \odot}$, the mean $\lambda_{\rm Edd}$ drops most with decreasing redshift, which
is a clear signature of cosmic downsizing. We will explore this further in
Section~\ref{QSOLF} where we will directly link our simulation results with
the observational findings.

\begin{figure*}\centerline{
\vbox{
\hbox{
\includegraphics[width=8truecm,height=7.5truecm]{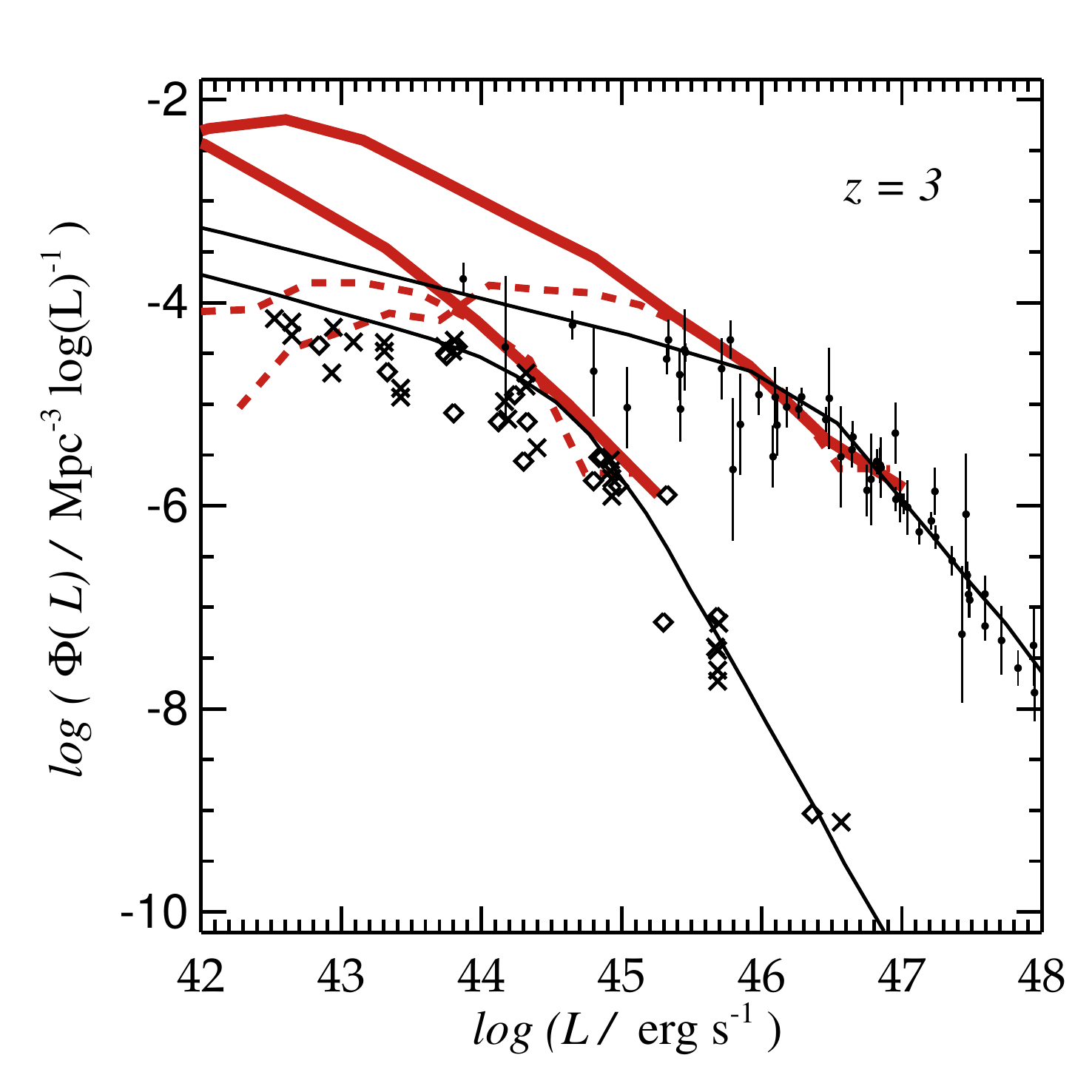}
\includegraphics[width=8truecm,height=7.5truecm]{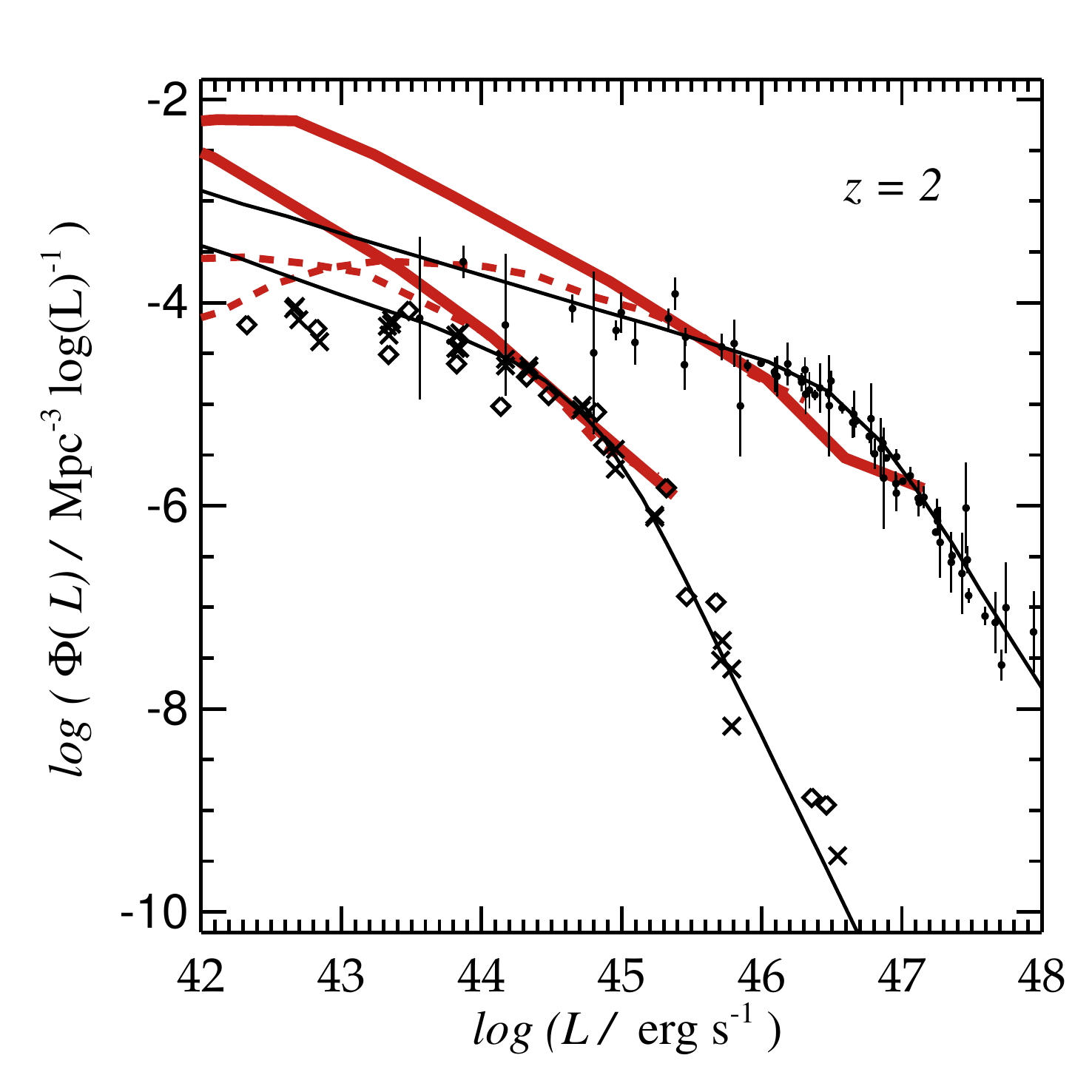}}
\hbox{
\includegraphics[width=8truecm,height=7.5truecm]{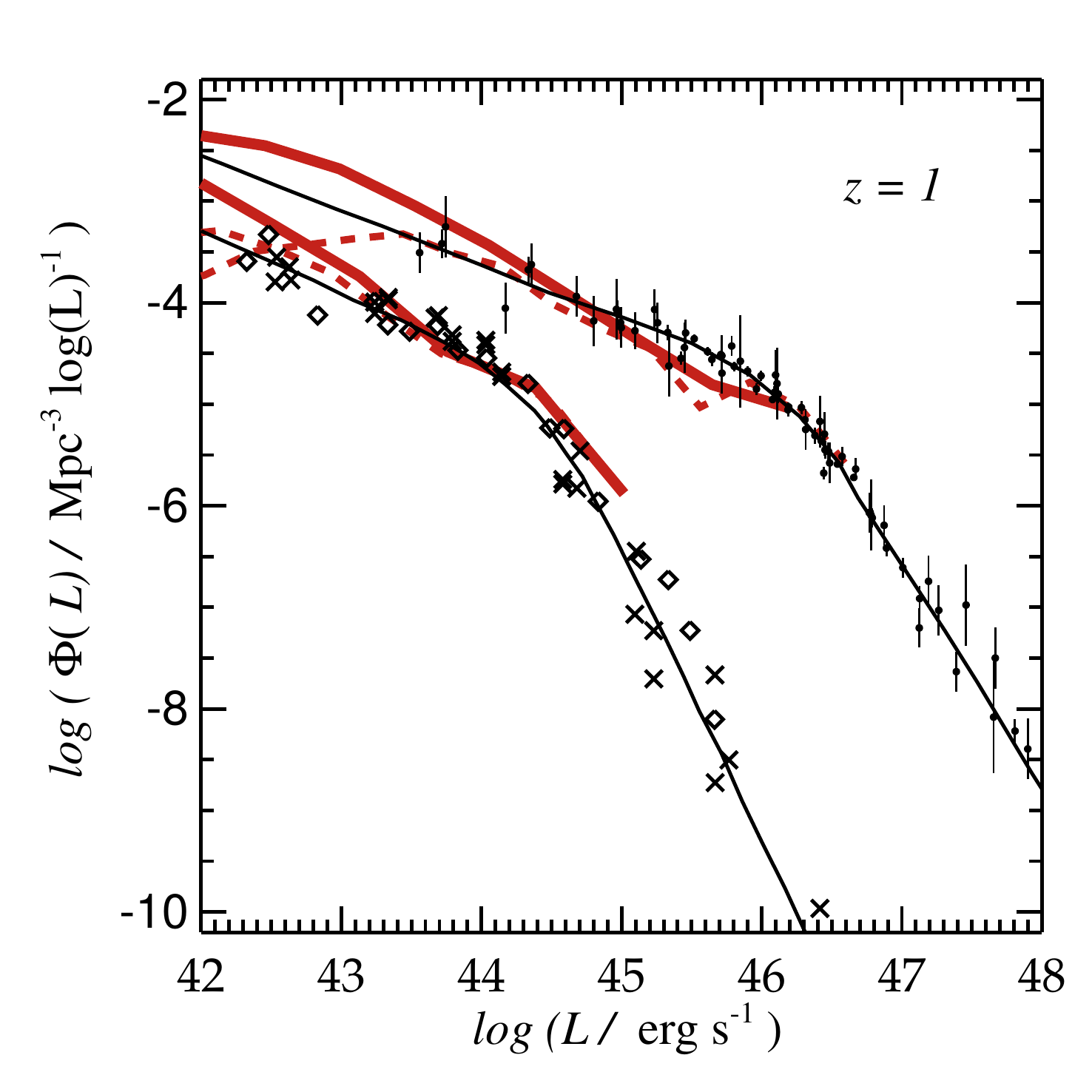}
\includegraphics[width=8truecm,height=7.5truecm]{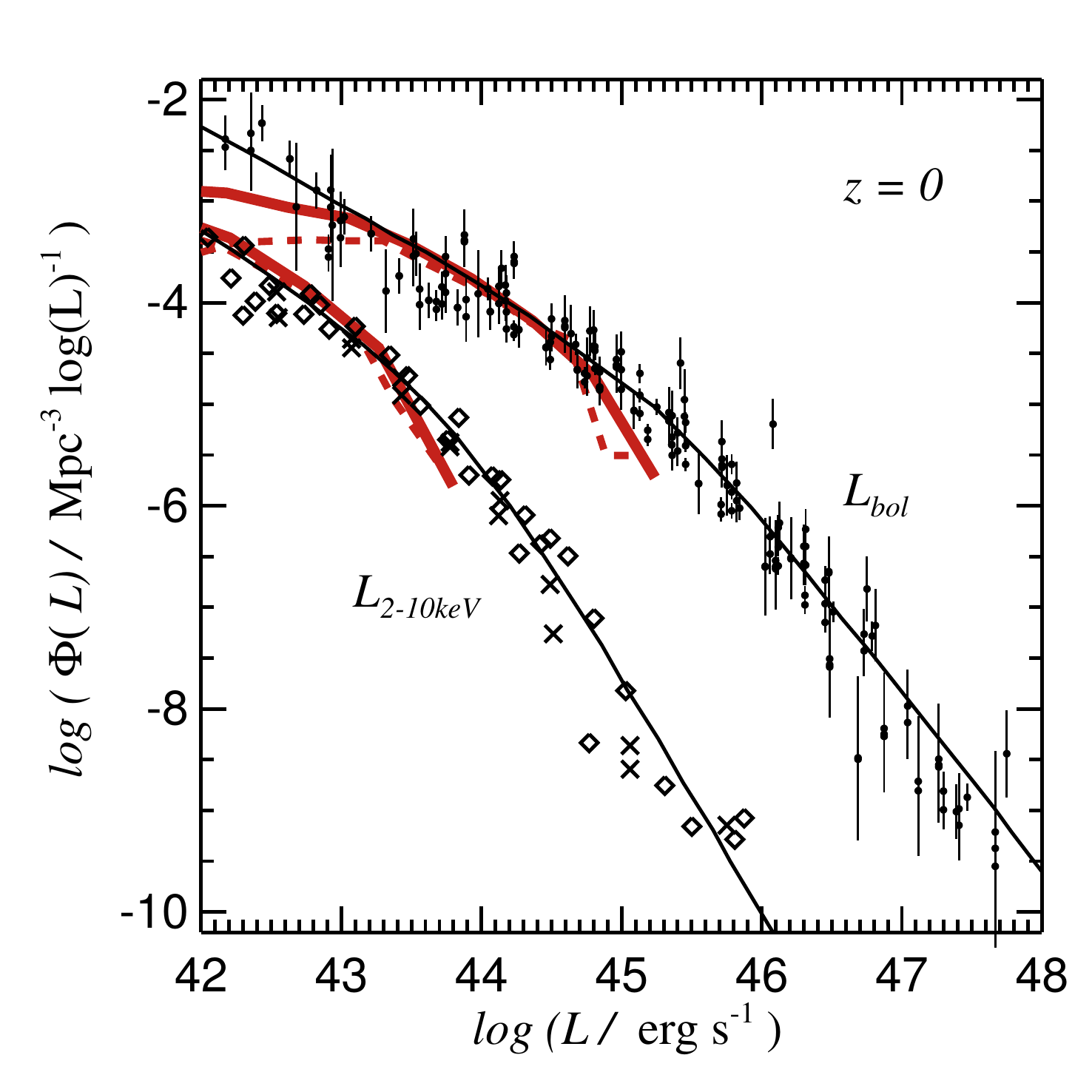}}
}}
\caption{AGN bolometric and hard X-ray luminosity functions at $z = 0,
  1, 2$ and $3$. Illustris results are shown with thick red lines. A
    constant radiative efficiency of $0.05$ is 
  assumed and black holes with Eddington ratio greater than $10^{-4}$ are
  plotted. Dashed red lines denote results for black holes that are additionally
  more massive than $5 \times 10^7 M_{\rm \odot}$. For bolometric 
  luminosity functions data points with error bars are from the
  compilation by \citet{Hopkins2007}, while for hard X-ray 
  luminosity functions additional data points are taken from
  \citet{Ueda2014}. Thin black lines are best-fit 
  evolving double power-law models to all redshifts from \citet{Hopkins2007}
  (see their Figures 6 and 7). Overall we find good agreement with the
  observed AGN luminosity functions at the bright end, while for $z > 1$ we
over-predict the number of faint AGN, unless low mass black holes are excluded.}   
\label{BHLF_FIXER}
\end{figure*}

In Figure~\ref{LEDD_COLOUR} we explore how the distribution of $g-r$ colours
(left-hand panel) and stellar masses within half-mass radii (right-hand
panel) of galaxies at $z = 0$ depends on the Eddington ratios of their central
black holes. The black dashed histogram denotes the median of the
distribution, and shaded 
region encloses the $5$th to $95$th percentile of the distribution. The
coloured histograms are for galaxies hosting black holes 
in different mass ranges, as specified in the caption. For any given
$\lambda_{\rm Edd}$ there is a very large spread in galaxy colours and stellar
masses. The median of the
distributions is driven by galaxies hosting low mass black holes, as can be
seen by comparing the black dashed and blue histograms. However, we find that
black 
holes more massive than $10^8 \, M_{\rm \odot}$ live in redder and more massive
galaxies but have a vast range of Eddington ratios (consistent with the
findings from Figure~\ref{LEDD}). Only for the highest $\lambda_{\rm Edd}$
values, i.e. greater than $0.1$ the distribution of $g-r$ colours narrows,
corresponding to galaxies that have black holes just below the $M_{\rm BH}$ -
$M_{\rm bulge}$ and $M_{\rm BH}$ - $\sigma$ relations.

We further investigate which population of galaxies and black holes is
responsible for the two distinct peaks seen in the distribution of $g-r$
colours. We find that the peak with $\lambda_{\rm Edd} < -5.7$ and $g-r > 0.55$ corresponds
to a distinct population which lies at the tip of the massive end of the simulated $M_{\rm BH}$ -
$M_{\rm bulge}$ relation. These galaxies also have low specific star formation
rates and on average low gas content (even though there is a considerable
scatter). The second peak, i.e. $-3.2 < \lambda_{\rm Edd} < -1.2$ and $g-r >
0.55$, is instead caused by galaxies with a mix of properties, which can be
roughly divided into $3$  categories: {\it i)} a large fraction of
galaxies are 
occupying the massive end of the $M_{\rm BH}$ -
$M_{\rm bulge}$ relation, but on average they have somewhat smaller black
hole masses than is the case for galaxies with $\lambda_{\rm Edd} < -5.7$ and
$g-r > 0.55$. They also typically have low specific star formation rates, but
are on average more gas rich (again with a very large scatter); {\it ii)}
the second category consists of galaxies with a very wide range of black
hole 
masses, but at a given $M_{\rm bulge}$ all of these black holes lie above the
best-fit relation; due to the ``over-massive'' black holes the host galaxies
have low specific star formation and gas mass fraction; {\it iii)} finally, the
third category consists of galaxies with low mass black holes whose feedback
is not 
strong enough to affect their hosts. Here, instead, outflows from
supernova-driven winds lead to red galaxy colours, low specific star
formation rates and low gas masses.   

\subsection{AGN luminosity functions}\label{QSOLF}

\begin{figure*}\centerline{
\hbox{
\includegraphics[width=8truecm,height=7.5truecm]{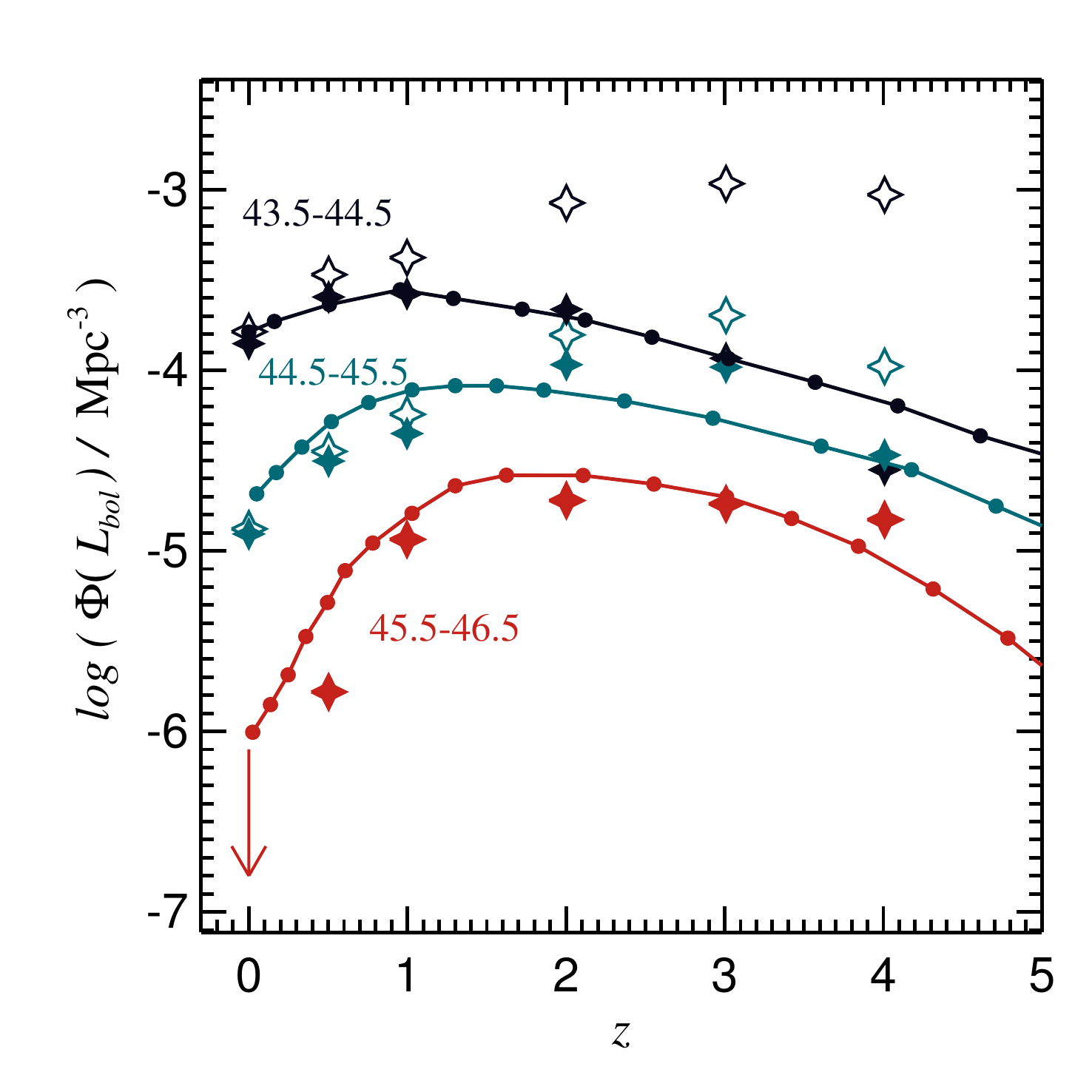}
\includegraphics[width=8truecm,height=7.5truecm]{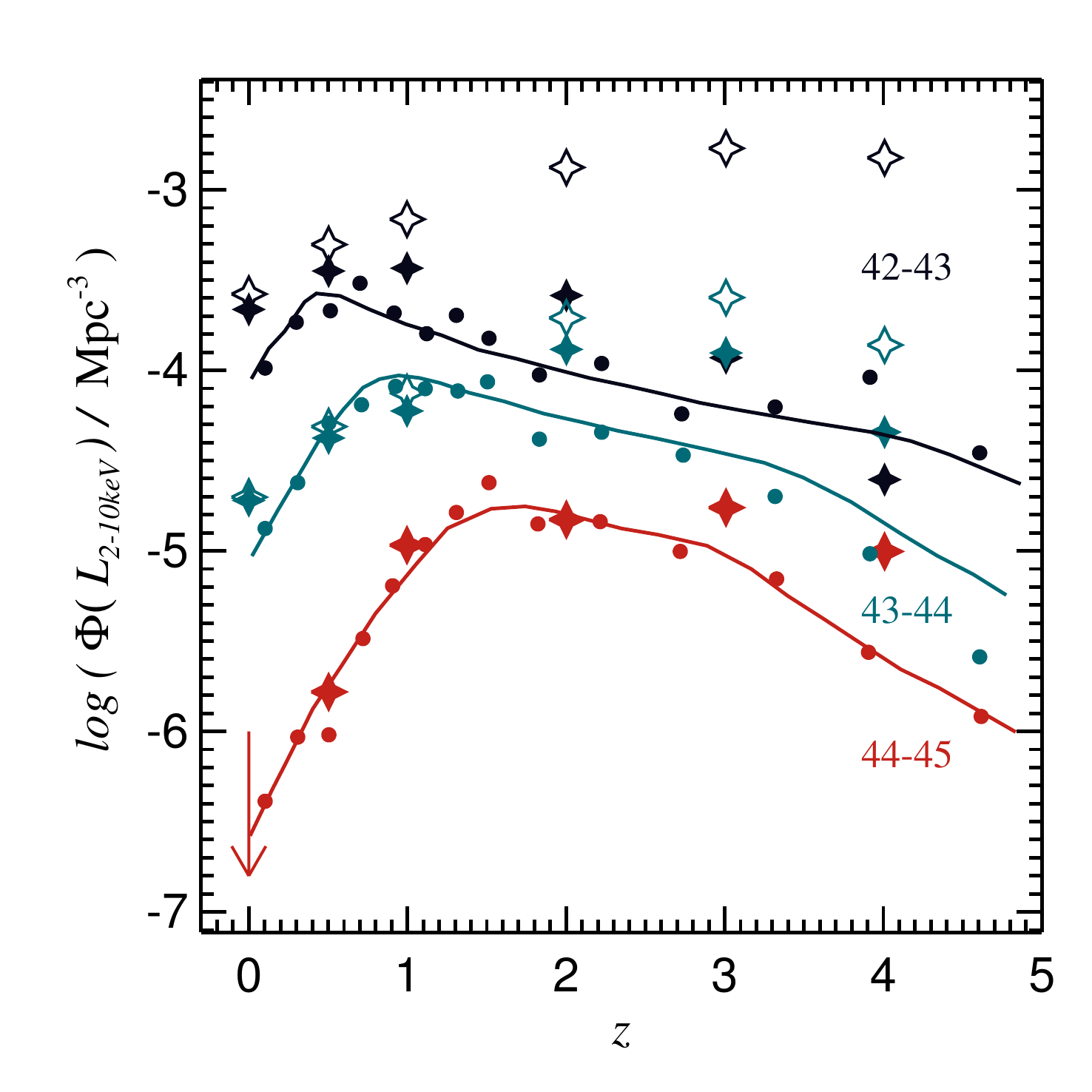}}
}
\caption{Left: redshift evolution of the comoving number density of AGN
  split into three bins based on their bolometric luminosity (in logarithmic
  units): $ 43.5 - 44.5$ (dark blue, top curve), $ 44.5 - 45.5$ (turquoise,
  middle curve) and $ 45.5 - 46.5$ (red, bottom curve). Observational
  constraints from \citet{Hopkins2007} are shown with circles joined by
  continuous lines, while the Illustris results are denoted with star
  symbols. Here we are showing the results only for black holes with
    Eddington ratios greater than $10^{-4}$ (open stars) and for
    black holes which are additionally more massive than $5 \times 10^7 M_{\rm
      \odot}$ (filled 
    stars). A constant radiative efficiency of $0.05$ is assumed. Right:
  redshift evolution of the comoving number density of AGN 
  split into three bins based on their hard X-ray luminosity (in logarithmic
  units): $42 - 43$ (dark blue, top curve), $43 - 44$ (turquoise, middle
  curve) and $ 44 - 45$ (red, bottom curve). Observational constraints from
  \citet{Ueda2014} are shown with circles and the continuous curves with
  matching colours are their best-fit model. The Illustris results are
    denoted 
  with open and filled star symbols, with the same selection criteria as in
  the left-hand panel.}
\label{BHND}
\end{figure*}

In Figure~\ref{BHLF_FIXER} we compare the AGN bolometric and hard X-ray
luminosity functions as predicted by the Illustris simulation at $z = 0, 1, 2$
and $3$ with observations. We refrain from comparing against the soft X-ray or B-band
luminosity functions because of large uncertainties in the obscuration fractions
and host galaxy contamination which could significantly bias the
interpretation of the results. Note however that even in the  
case of the bolometric and hard X-ray luminosity functions large uncertainties
remain due to poorly constrained bolometric corrections
\citep{Hopkins2007, Vasudevan2007, Vasudevan2009, Lusso2012} and the uncertain fraction of
Compton-thick sources for $z \gtrsim 0$ \citep[e.g. see recent papers
    by][]{Ueda2014, Buchner2015, Aird2015}. 

Keeping these caveats in mind, we compare
the Illustris AGN bolometric luminosity function with the bolometric luminosity
function as derived by \citet{Hopkins2007}, which is still the standard
reference in the field. When computing the bolometric luminosity function we
do not consider black holes with Eddington ratios smaller than $10^{-4}$,
which is a very conservative estimate given that these objects should be in a
radiatively inefficient regime. For
the hard X-ray luminosity function we compute the simulated X-ray luminosities
from our bolometric luminosities by adopting the bolometric corrections of
\citet{Hopkins2007}. We have also corrected the hard X-ray luminosities
assuming the obscuration fraction given by equation 4 in
\citet{Hopkins2007} which is redshift independent.

We compare our hard X-ray luminosity function
with the most recent compilation by \citet{Ueda2014}. By assuming bolometric
corrections from \citet{Hopkins2007} as well, \citet{Ueda2014} showed that to
reconcile the black hole mass function obtained from the revised $M_{\rm BH}$
- $M_{\rm bulge}$ relation \citep{Kormendy2013} with the black hole mass
function calculated from the bolometric luminosity function using Soltan-type
arguments, the mean radiative efficiencies of AGN need to be revised
downwards. Assuming an average Eddington ratio of $\sim 0.7$ that does not
depend on redshift or AGN luminosity, \citet{Ueda2014} determine
a mean radiative efficiency of $\epsilon_r \sim 0.05$. 
    
We start our analysis by fixing the radiative efficiency to $0.05$ as shown in
Figure~\ref{BHLF_FIXER}. This is not the value that yields the best match to
the observed luminosity functions, but it is merely motivated by the
considerations made by \citet{Ueda2014}. Note that any constant value of
$\epsilon_r$ simply changes the normalisation but not the shape of the
luminosity functions. For $\epsilon_r = 0.05$ we find good agreement with
observations at all redshifts, both for bolometric and hard X-ray luminosity
functions, at the bright end. For $z \ge 2$ we over-predict the number of faint
AGN with $L_{\rm bol} < 10^{45} \, {\rm erg \, s^{-1}}$ and $L_{\rm 2-10keV} < 10^{44}
\, {\rm erg \, s^{-1}}$. However, if we consider only black holes more massive
than $5 \times 10^7 M_{\rm \odot}$ (dashed red lines) we can obtain a better
agreement both in the case of the bolometric and hard X-ray luminosity
function at the faint end, while the bright end remains essentially unchanged.

There are several important conclusions to draw from this comparison: {\it i)}
for our simulated Eddington ratios, constant $\epsilon_r$ values of $ \ge 0.1$
are inconsistent with the data. This is in agreement with the
conclusions by 
\citet{Ueda2014} even though they assume a very different $\lambda_{\rm
  EDD}$. Comparison with data at higher redshifts, i.e. $z
\ge 1$, is particularly constraining given that the majority of our simulated AGN
are in the radiatively efficient regime at these epochs;
{\it ii)} a low constant value of $\epsilon_r = 0.05$ implies that our
feedback efficiency should be $0.2$ instead of $0.05$ (given that the product of
these is degenerate, see Section~\ref{FEEDBACK}) if we are to successfully
reproduce black hole mass 
function and black hole -- galaxy scaling relations; {\it iii)} there are
several lines of both theoretical and observational evidence
\citep[e.g.][]{Mahadevan1997,Ciotti2009,Ueda2014} indicating that 
radiative efficiencies might depend on black hole properties, such as their
accretion rate. By setting $\epsilon_r = 0.1$ for all black holes in the
``quasar'' mode and by computing an accretion rate dependent $\epsilon_r$ for
black holes in the ``radio'' mode (following \citealt{Mahadevan1997} or \citealt{Ciotti2009}) we can
also get a good match to the observed luminosity functions at $z = 0$. Thus,
even though we 
cannot uniquely constrain radiative efficiencies, on average they
should still be low. Alternatively, radiative efficiencies could be higher if
the bolometric corrections are currently largely underestimated. This is a
very interesting prediction of our model that 
can be verified once robust estimates of Eddington ratios for a range of
black hole masses and redshifts become available, which would break
the degeneracies between Eddington ratio and radiative efficiency
distributions.

We now discuss the possible systematic biases at the faint end of the AGN
luminosity 
function for $z \gtrsim 1$. Given that the X-ray luminosity function
determined by \citet{Ueda2014} is de-absorbed we would not need to apply any
obscuration correction, except for the contribution from Compton-thick sources
which is uncertain \citep[see also a recent paper by][which agrees with
  \citet{Ueda2014} findings]{Aird2015}. However, the \citet{Ueda2014} sample
is flux-limited, 
while our sample is volume-limited. In fact our number density of AGN is
dominated by low luminosity objects while \citet{Ueda2014} find only around $40$
sources for $z > 2$ with $L_{\rm 2-10keV} \le 10^{43} \, {\rm erg \,
  s^{-1}}$. Thus, some fraction of the discrepancy at the faint end between
our X-ray luminosity function and the observational one could be
due to this mismatch. There are other important sources of uncertainties as
well. Radiative efficiencies could be luminosity dependent, there could be
significant biases due to the uncertainties in bolometric corrections, or
the Illustris predictions could be wrong. In particular, with regard to this
last possibility, if we compute luminosity functions for black holes more
massive than $5 \times 10^7 M_{\rm \odot}$ the agreement with both bolometric
and hard X-ray luminosity function is greatly improved. This suggests that part
of the discrepancy could be attributed to our seeding prescription and to the
accretion onto low mass black holes which is not well converged and where
black holes have not yet reached the self-regulated regime. This also affects
our simulated black hole mass function at the low mass end as we have
discussed in Section~\ref{MASSFUNC}. Finally, it is interesting to note that
a recent paper by \citet{Buchner2015} advocates significantly larger
uncertainties at the faint-end of the AGN luminosity function which stem from
their non-parametric approach. Future observations of the faint
end of the AGN luminosity function and robust determination of the fraction of
Compton-thick sources as a function of redshift will be crucial to shed light
on these issues.

\begin{figure*}\centerline{
\vbox{
\hbox{
\includegraphics[width=6truecm,height=6truecm]{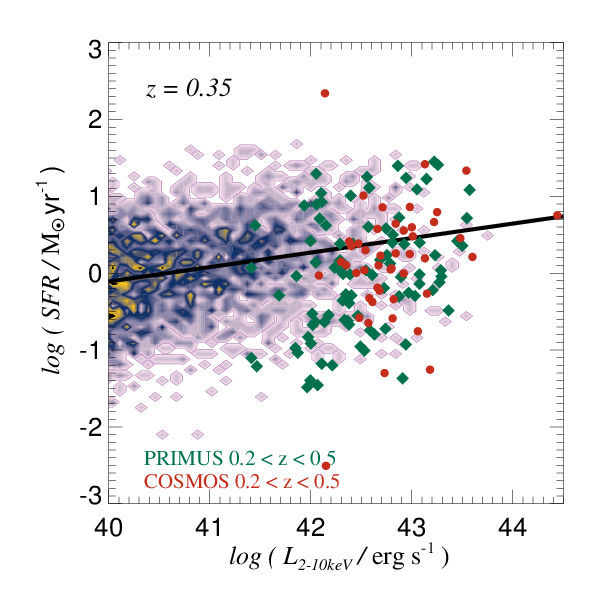}
\includegraphics[width=6truecm,height=6truecm]{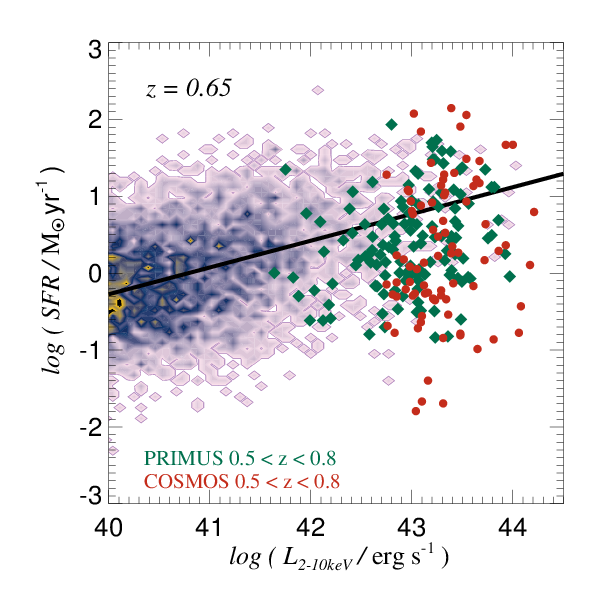}
\includegraphics[width=6truecm,height=6truecm]{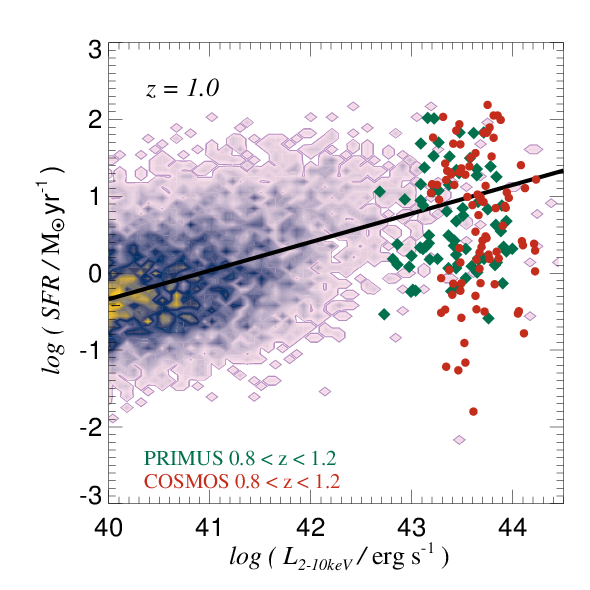}}
\hbox{
\includegraphics[width=6truecm,height=6truecm]{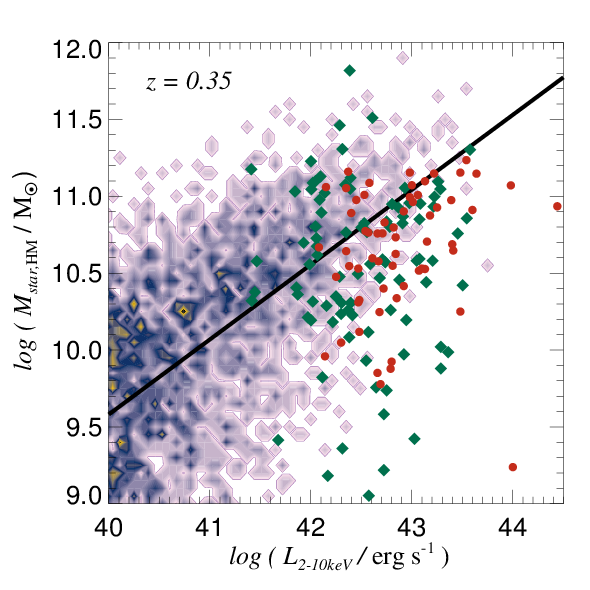}
\includegraphics[width=6truecm,height=6truecm]{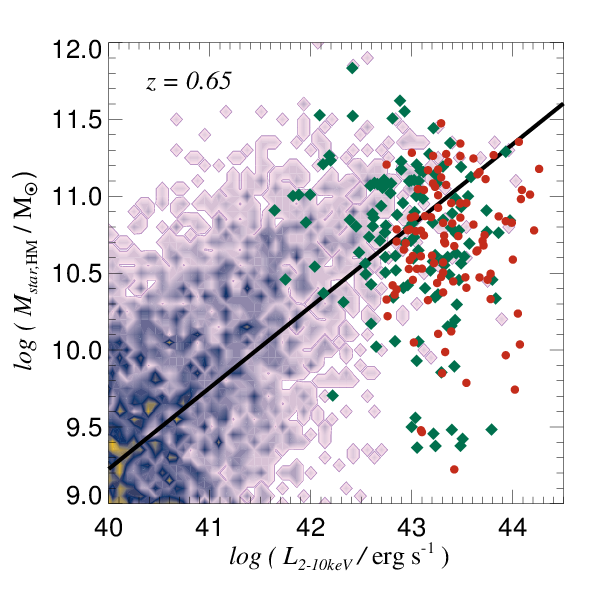}
\includegraphics[width=6truecm,height=6truecm]{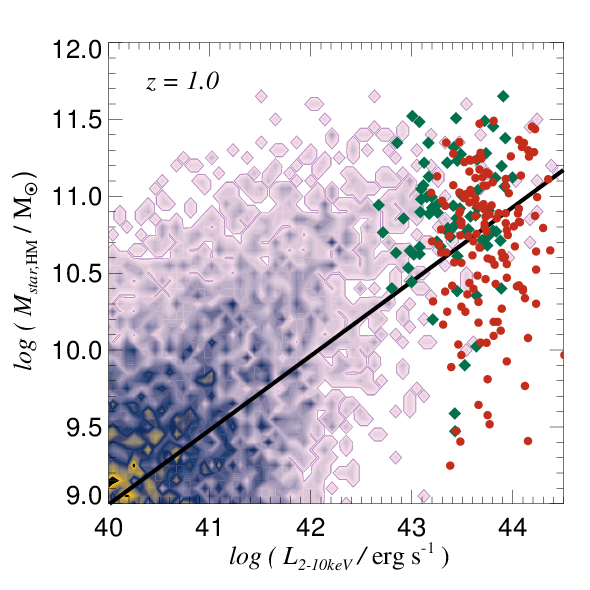}}
}}
\caption{Top panels: Star formation rate within stellar half-mass radius as a function of the
  hard 
X-ray luminosity of the central AGN. Illustris results at $z = 1, 0.65$ and
$0.35$ are shown with $2$D histograms (colours indicate number density), while
the data points (green diamonds) are for the PRIMUS galaxies from
\citet{Azadi2014} and COSMOS galaxy compilation (red circles) from
\citet{Lusso2010, Lusso2011, Brusa2010, Bongiorno2012} for the redshift ranges
$0.2 < z < 0.5$, $0.5 < z < 0.8$ and  
$0.8 < z < 1.2$, respectively. Bottom panels: Stellar mass within the stellar
half-mass radius versus the hard X-ray luminosity of the central AGN for the 
Illustris galaxies ($2$D histograms), PRIMUS galaxies from \citet{Azadi2014}
 (green diamonds), and COSMOS galaxies (red circles). In all panels the
best-fit relation (least-square linear fit) to the Illustris 
galaxies is shown with a thick black line. While we do see a correlation in
the simulated relations, there is a considerable scatter, such that for a
given $L_{2-10keV}$ value star formation rates and stellar masses can vary by up
to $2$ orders of magnitude.} 
\label{SFR_LX}
\end{figure*}

Keeping these uncertainties in mind, in Figure~\ref{BHND} we now compare the
redshift evolution of the comoving number density of AGN split into three
bins based on their bolometric (left-hand panel) and hard X-ray luminosity
(right-hand panel) with the estimates by \citet{Hopkins2007}  and
\citet{Ueda2014}. This is a more direct way of probing the observed cosmic
downsizing of the AGN population that we have discussed in
Section~\ref{EDDINGTON} in terms of Eddington ratios. We find that the
Illustris simulation 
qualitatively reproduces observations, with the agreement being
best for the highest luminosity bin and poorest for the lowest luminosity
bin, both in the case of bolometric and hard X-ray luminosities. The drop of
the AGN comoving number density in Illustris for $z < 2$ is systematically
steeper for higher luminosity objects in agreement with observations but
we do not find that the number density of lower luminosity AGN peaks at
lower redshifts. Considering only black holes more massive than $5 \times 10^7
M_{\rm \odot}$ (filled stars) significantly improves the agreement with data,
again indicating that the modelling of low mass black holes may have to be
improved.

\subsection{The link between star formation and AGN triggering}\label{LXSFR}

In Figure~\ref{SFR_LX} we show star formation rates within stellar half-mass
radius (top panels) and stellar mass within the same radius (bottom panels)
as a function of the hard X-ray luminosity of the central AGN (taking into
account black holes with $\lambda_{\rm Edd} > 10^{-4}$). We compare the
Illustris results with the recent study by \citet{Azadi2014} based on the
PRIMUS survey (green diamonds) and with the COSMOS data compilation
(\citealt{Lusso2010, Lusso2011}, with stellar masses and star formation rates
from  \citealt{Bongiorno2012} and redshifts from \citealt{Brusa2010};
red circles). The three
panels are for different redshift intervals probed by the surveys while we
plot the Illustris data at the mean redshift of each redshift bin. This
comparison reveals why it has been so hard for the present observations to
establish a clear link between star formation and AGN activity. For
  example, \citet{Mullaney2012} and \citet{Rosario2012} 
  using a combination of 
far-infrared and X-ray data found no correlation between star formation rates
and AGN X-ray luminosities, suggesting that they might be triggered by
different physical processes. However, the most luminous AGN seem to be correlated with
high star formation rates \citep[see
  e.g.][]{Lutz2008, Rosario2012} and \citet{Hickox2014} discussed the
possibility that while star formation and black hole activity are correlated
over long timescales, AGN variability introduces a significant scatter
\citep[see also discussion in][]{Alexander2012, Azadi2014}.

Regardless of
the redshift considered, Figure~\ref{SFR_LX} shows that there is considerable scatter in the simulated
relations, such that for a given $L_{2-10keV}$ star formation rates and
stellar masses can vary by up to $2$ orders of magnitude. For a given star
formation rate the variation in $L_{2-10keV}$ is even larger, spanning up to the
full X-ray luminosity range. This demonstrates that triggering of star
formation and central AGN are not necessarily always tightly linked neither in
terms of common origin nor in terms of coherent timing. For example, minor wet
mergers, cold gas inflows and local gas compression might trigger star
formation but the fresh gas supply might not 
get funnelled to the central-most region where the AGN resides (either due to
gas consumption and/or expulsion along the way or due to the residual gas
angular momentum). Moreover the timing of star formation versus AGN
triggering can be different even in the case of gas-rich major mergers which do
bring copious amounts of gas to the centre, as repeatedly shown in numerical
simulations of isolated galaxy mergers \citep[e.g.][]{Springel2005b,
  Sijacki2011, Thacker2014}. Thus, given the large intrinsic scatter in the
$SFR$ - $L_{2-10keV}$ plane as predicted by the Illustris simulation and given
that both PRIMUS and COSMOS data cover a relatively narrow range of X-ray
luminosities no correlation between the two can be observationally
inferred. Note however that regardless of the large scatter there is a
correlation between star formation rates and AGN luminosities in Illustris, as
shown by the thick black line in Figure~\ref{SFR_LX} which is the best-fit relation. This is to be
expected as the shape of the cosmic star formation rate density is similar to the
shape of the black hole accretion rate density (see Figures~\ref{SFR_RES} and
\ref{SFR}). This highlights that there is an underlying strong physical
connection between star formation and black hole growth driven by large scale
cosmological gas inflows and mergers but the details of each trigger effect
may vary. We furthermore explore whether some of the scatter seen in
  Figure~\ref{SFR_LX} is due to the AGN variability which occurs on much
  shorter timescale than the changes in the star formation rate. To test this
  idea, similarly to the recent study by \citet{Hickox2014}, we consider the
  $L_{2-10keV}$ - $SFR$ plane (essentially swapping the axis with respect to
  Figure~\ref{SFR_LX}). The best-fit relation exhibits a slightly smaller
  unreduced chi square value, which indicates that the AGN variability
  contributes to the scatter seen in Figure~\ref{SFR_LX}. We furthermore
  compute the mean $L_{2-10keV}$ in $SFR$ bins at $z = 0.35, 0.65$ and $1$ and
find an essentially redshift-independent correlation. However, the mean $SFR$
computed in bins of $L_{2-10keV}$ does change with redshift. Even though we
probe somewhat smaller luminosities than \citet{Hickox2014} this is in
qualitative agreement with their findings \citep[see also][]{Azadi2014}, further corroborating the idea that
also due to the underlying short-timescale AGN variability star formation and AGN
activity do not appear correlated, while in fact there is time-averaged
correlation.

\section{Discussion and Conclusions} \label{Conclusions}

In this work we have presented an overview of the main properties of black
holes as predicted by the Illustris simulation. Owing to its large volume
and high dynamic range we can study the properties of a representative
sample of black holes embedded within host galaxies with resolved inner
structural properties. The Illustris simulation volume is too small to
follow the formation and evolution of the most massive black holes observed in
the Universe, but it is sufficiently large that we can characterise e.g. the black
hole mass function, black hole -- host galaxy scaling relations and AGN
luminosity functions over the most important ranges. We note that while the
free parameters of the black hole model have been tuned to reproduce star
formation rate history and stellar mass function at $z = 0$
\citep{Vogelsberger2013, Torrey2014} and the quasar feedback efficiency
has been selected following \citet{DiMatteo2005, Springel2005b}, the main
properties of black holes are a genuine prediction of our model. Thus we find
it highly encouraging that the Illustris simulations 
reproduce several key observables, also allowing us to
highlight possible biases in current datasets and to make predictions for
future observational programmes. However, the successes of the model need
  to 
  be considered in view of several caveats: the black hole properties are
  not well converged, even though the convergence properties are better for
  more massive black holes with Eddington ratios greater than $10^{-4}$ (see Appendix~\ref{APP_CONV}); our
  seeding prescription is rather simplistic and uncertain, as detailed in
  Section~\ref{MASSFUNC}, which also leads to a likely over-prediction of the
  black 
  hole merger rates; unavoidably, due to resolution limitations,
  accretion onto low mass black holes is not well resolved. Keeping these
  caveats in mind our main findings are as follows:\\ 

\begin{itemize}

\item We find that the black hole mass density over the whole redshift range
  probed by observations, i.e. for $z < 5$ is consistent with the estimate
  based on the most up-to-date hard X-ray survey compilation by
  \citet{Ueda2014}. For black holes more massive than $10^7 M_{\rm
    \odot}$ the mass function at $z = 0$ is in good agreement with the
  constraints by \citet{Shankar2013}, which are based on the revised $M_{\rm
    BH}$ - $\sigma$ 
  scaling relation of \citet{McConnell2013}. These two results taken 
  together indicate that overall we have a realistic population of black
  holes formed 
  in the Illustris simulation both in terms of the total number density and 
  also in terms of the mass distribution. However, we highlight that we
    over-predict the number of low mass black holes, i.e. $M_{\rm BH} \lesssim
    10^7 M_{\rm \odot}$ with respect to the estimates by \citet{Shankar2013},
    which indicates that our seeding prescription is likely overproducing
    these object. The accretion onto these low mass black holes is also least 
  well resolved in Illustris which may contribute to the discrepancy,
  if simulated low mass black holes accrete too much gas.

\item The Illustris data set allowed us not only to construct
  the $M_{\rm BH}$ - $M_{\rm bulge}$ and $M_{\rm BH}$ - $\sigma$ scaling
  relations which are in very good agreement with the most recent estimates
  \citep{McConnell2013, Kormendy2013} but also to relate galaxy properties, in
  terms of their 
  morphologies and colours, to the position on these relations. This permits us, for the
  first time to our knowledge, to pin down for which galaxy types
  co-evolution with their central black holes is driven by a physical link,
  rather than arising as a statistical byproduct. 

\item Specifically, we find that observed pseudo bulges
  coincide with blue star-forming Illustris galaxies having under-massive
  black holes which do not significantly affect either their colours or
  their gas content. While some of the black hole and stellar mass assembly in
  these objects has a common origin, the feedback loop is not fully
  established, and other physical processes, such as supernova-driven winds
  may be prevailing. This explains why at the low mass end the scatter in the
  observed 
  $M_{\rm BH}$ - $M_{\rm bulge}$ and $M_{\rm BH}$ - $\sigma$ increases.

\item Interestingly, the most efficient accretors at $z = 0$ typically
  correspond to black holes just under the observed $M_{\rm BH}$ - $M_{\rm
    bulge}$ relation indicating that these objects, due to a sufficient gas supply,
  can transform galaxy properties from blue star-forming
  to red and quenched on short timescales. In fact, $10^8 M_{\rm \odot}$ black holes have the highest Eddington ratios on
  average and reside within galaxies with colours
  $g-r \sim 0.5$ and with a mix of morphologies, resembling the so-called ``green valley''
  objects \citep{Schawinski2014}.  

\item Black holes which are above the best-fit $M_{\rm BH}$ - $M_{\rm
    bulge}$ and $M_{\rm BH}$ - $\sigma$ relations or reside at the massive end, are hosted by galaxies which
  have red colours, low gas fractions and low specific star formation
  rates. This directly demonstrates that for these systems there is a strong
  physical link between galaxy properties and their central black holes and
  co-evolution does take place.

\item By examining the redshift evolution of the $M_{\rm BH}$ - $M_{\rm
    bulge}$ and $M_{\rm BH}$ - $\sigma$ relations, we find that black hole
  growth precedes galaxy assembly, where for a given bulge mass 
  black holes for $z > 0$ are more massive than their $z = 0$ counterparts. This is also
  in line with the shape of the black hole accretion rate density which rises
  more steeply than the cosmic star formation rate density for $z > 2$. The
  redshift evolution of the slope of the $M_{\rm BH}$ - $\sigma$ relation
  indicates that at high redshifts there are significant radiative losses in
  the AGN-driven outflows while at low $z$ and especially in massive objects
  radiation losses are subdominant so that the energy-driven flow is
  established. Additionally, at low redshifts more dry mergers take place,
  as AGN feedback becomes more efficient at quenching galaxies, which has been
  shown \citep[see e.g.][]{Boylan2006} to lead to a steepening of the
  $M_{\rm BH}$ - $\sigma$ relation. This slope evolution seen in Illustris
  also naturally explains 
  the slope steepening found in local massive ellipticals and brightest
  cluster galaxies. However we caution that the detailed comparison with
  observations can be 
  systematically biased depending on how stellar velocity dispersion is
  measured, for example, and that these biases are likely more severe at $z >
  0$. For future observations it will be of prime importance to disentangle these
  effects from a genuine redshift evolution of the scaling relations to understand how AGN
  feedback operates as a function of cosmic time.   

\item Comparison of the AGN luminosity function predicted by the Illustris
  simulations with observations reveals that on average AGN radiative
  efficiencies need to be low if we are to simultaneously match the black hole
  mass density, mass function and the normalisation of the $M_{\rm BH}$ -
  $M_{\rm bulge}$ and $M_{\rm BH}$ - $\sigma$ relations, unless the bolometric
  corrections are currently largely underestimated. This result is in line
  with the conclusions drawn by \citet{Ueda2014} based on the hard X-ray data
  and is driven by the revised black hole -- host galaxy scaling relations
  \citep{McConnell2013, Kormendy2013} with respect to the past findings
  \citep{Yu2002, Haring2004}. While we cannot uniquely predict radiative
  efficiencies as they are degenerate with the black hole feedback 
  efficiencies in our model, on average low $\epsilon_r$ values indicate that
a larger fraction of AGN luminosity needs to couple efficiently with the
surrounding gas. Given that for the Illustris simulation we have adopted
  $\epsilon_r = 0.2$ and $\epsilon_f = 0.05$, the inferred low radiative
  efficiencies imply that the feedback efficiency needs to be a factor $2-4$
  higher (as the product of the two is degenerate in our model). While such
  high feedback efficiencies are not ruled out observationally yet, different
  accretion models than the one assumed in Illustris and/or a possibility of
  super-Eddington accretion may alleviate the need for very high feedback
  efficiencies. Future observations of black hole duty cycles and Eddington
ratio distributions as a function of redshift and black hole mass will help
to shed light on these issues.

\item While the shape of the bolometric and hard X--ray luminosity
  functions is in very good agreement with the data at $z = 0$ and $1$, at higher
  redshifts we over-predict the number of faint AGN. By restricting our sample
  to black holes more massive than $5 \times 10^7 M_{\rm \odot}$ we can get a
  much better match to the data, indicating again that our seeding prescription
  and poorly resolved accretion onto low mass black holes could be responsible
  for this discrepancy. We furthermore caution that comparison of volume-
  versus flux-limited samples, bolometric corrections and cosmic evolution of
  the fraction of Compton-thick sources are additional sources of uncertainty. 

\item We find that in the Illustris AGN population there is evidence for cosmic
  downsizing 
  \citep{Barger2005, Hasinger2008, Ueda2014}. Not only does the distribution of
  Eddington ratios evolve with redshift in broad agreement with cosmic
  downsizing, 
  but we directly show that the simulated number densities of AGN, split
  into different hard X-ray luminosity bins, exhibit systematically steeper
  drops with redshift for more luminous objects. We do not find however that the
  number density of lower luminosity AGN peaks at lower redshifts in
  Illustris, unless low mass black holes are excluded from the analysis.  

\item We finally explore the physical link between star formation and black
  hole accretion triggering. Current observations have struggled to find
  clear evidence of such a link \citep[e.g. see][and the references
    therein]{Alexander2012, Azadi2014, Hickox2014}, thus questioning the
  standard lore 
  where due to 
  mergers galaxies and black holes grow hand-in-hand. We find that the black
  hole X-ray luminosities (direct proxies of the accretion rates) are correlated
  with the host galaxy star formation rate -- in accordance with the similar
  shapes of the cosmic star formation and black hole accretion rate density -- albeit with a large
  scatter. Current observations probe a too narrow dynamic range in X-ray luminosities to see this
  correlation, even though it can be inferred if the spectroscopically
  confirmed COSMOS data 
  \citep{Lusso2010, Lusso2011, Brusa2010, Bongiorno2012}  at all
  redshifts are combined together. Large scatter seen in the simulated $SFR$ -
  $L_{2-10keV}$ relation demonstrates that the physical link between star
  formation and black hole accretion triggering is more complex than
  previously envisaged. Gas-rich major mergers are responsible for the starburst
  -- AGN connection but relative timing offsets \citep[see also][]{Wild2010},
  whereby luminous quasars 
  light-up with a delay, contribute to the scatter. Moreover, in the case of
  large scale gas inflows and minor mergers star formation events might not be
  followed by black hole accretion because of gas consumption, expulsion or
  a residual angular momentum barrier. Finally, the much shorter timescale
    of AGN variability with respect to star formation can also introduce
    scatter as, for example, discussed in a recent paper by
    \citet{Hickox2014}, which is in line with our findings.        

\end{itemize}   

Large scale cosmological simulations such as Illustris where thousands of
galaxies are sufficiently well resolved to study their morphological
properties are a unique tool to dissect the cosmological co-evolution 
(or the lack thereof) for representative samples of galaxies and their central
black holes in the Universe. In fact, the results of our black hole model
  presented in this work should be viewed in a wider context of reproducing a number of key features of
  a representative sample of galaxies \citep{Vogelsberger2014a, Genel2014}, and
specifically in quenching and morphologically transforming massive galaxies
thanks to the AGN feedback. Studying black hole
growth and feedback with future improved simulations will be more timely than
ever 
and promises to give us an ever more
precise understanding of the astrophysical role these fascinating objects
play.

\section*{Acknowledgements} 
We would like to thank the referee for a very constructive referee report
which improved this paper. DS would like to thank Martin Haehnelt, Manda Banerji, Roberto Maiolino, Matt Auger, Ranjan
Vasudevan, Andy Fabian and Richard McMahon for very useful discussions. DS
also would like to thank Manda Banerji for kindly providing the data compilation
used in Figure~\ref{SFR_LX}. Simulations were run on the Harvard
Odyssey and CfA/ITC clusters, the Ranger and Stampede supercomputers
at the Texas Advanced Computing Center as part of
XSEDE, the Kraken supercomputer at Oak Ridge National Laboratory
as part of XSEDE, the CURIE supercomputer at CEA/France
as part of PRACE project RA0844, and the SuperMUC computer at
the Leibniz Computing Centre, Germany, as part of project pr85je.
VS acknowledges support by the DFG Research Centre SFB-881 ``The
Milky Way System'' through project A1, and by the European Research
Council under ERC-StG EXAGAL-308037. SG acknowledges support provided by NASA
through Hubble Fellowship grant HST-HF2-51341.001-A awarded by the STScI,
which is operated by the Association of Universities for Research in
Astronomy, Inc., for NASA, under contract NAS5-26555. GS acknowledges
support from the HST grants program, numbers HST-AR-
12856.01-A and HST-AR-13887.004A. Support for programs \#12856 and \#13887 was provided
by NASA through a grant from the Space Telescope Science Institute,
which is operated by the Association of Universities for Research
in Astronomy, Inc., under NASA contract NAS 5-26555. LH
acknowledges support from NASA grant NNX12AC67G and NSF
grant AST-1312095.

\bibliographystyle{mn2e}

\bibliography{paper}

\appendix
\section{Convergence issues}\label{APP_CONV}
\begin{figure}\centerline{
\includegraphics[width=8truecm,height=7.5truecm]{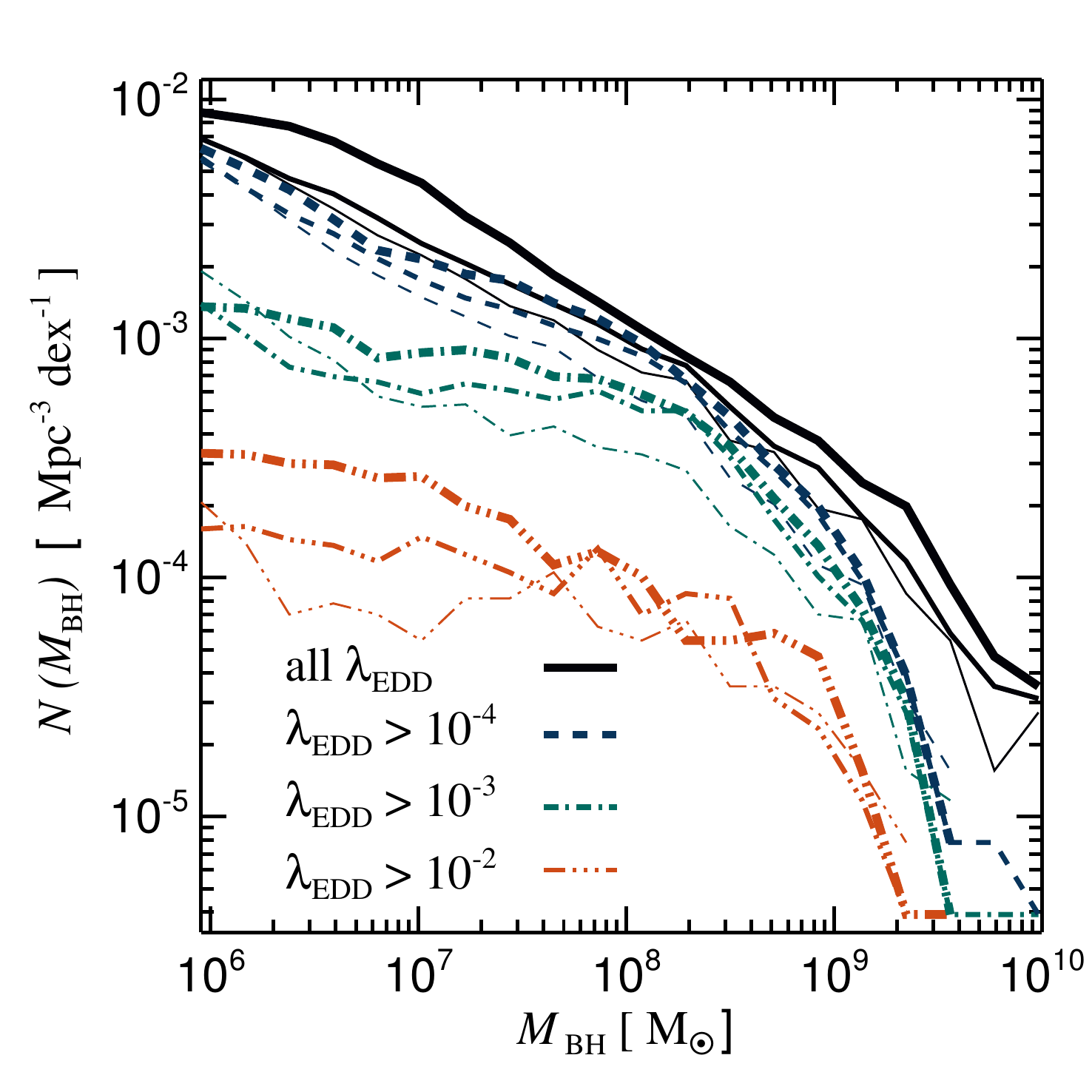}}
\caption{Black hole mass function at $z = 0$, split by the Eddington ratios of black holes, as
  indicated on the legend. For each colour thin to thick lines are for Illustris
  simulations with three different resolutions: $3 \times 455^3$, $3 \times
  910^3$ and $3 \times 1820^3$, respectively.}
\label{BHMF_CONV}
\end{figure}

Having discussed convergence properties of the black hole accretion rate and
mass density in Section~\ref{CONVERGENCE}, here we focus on the convergence
of the black hole mass function and $M_{\rm BH}$ - $M_{\rm bulge}$ relation
which we have presented for the highest resolution Illustris simulation in
Sections~\ref{MASSFUNC} and \ref{SCALING}.

\begin{figure*}\centerline{
\hbox{
\includegraphics[width=8truecm,height=7.5truecm]{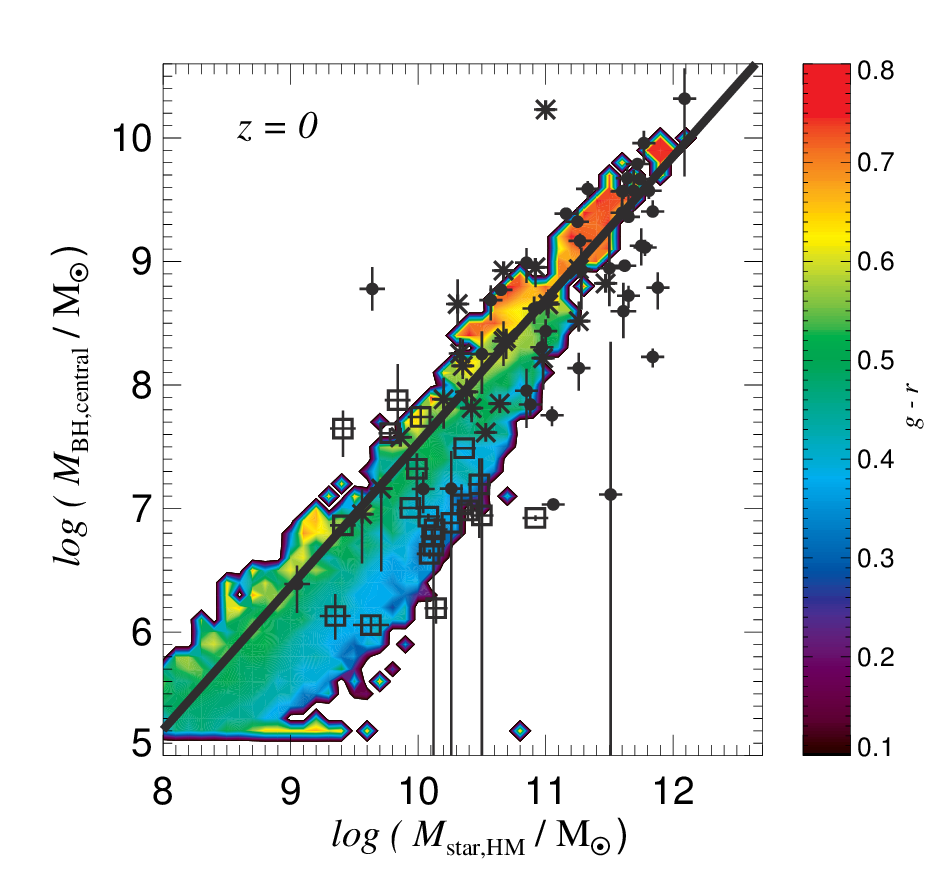}
\includegraphics[width=8truecm,height=7.5truecm]{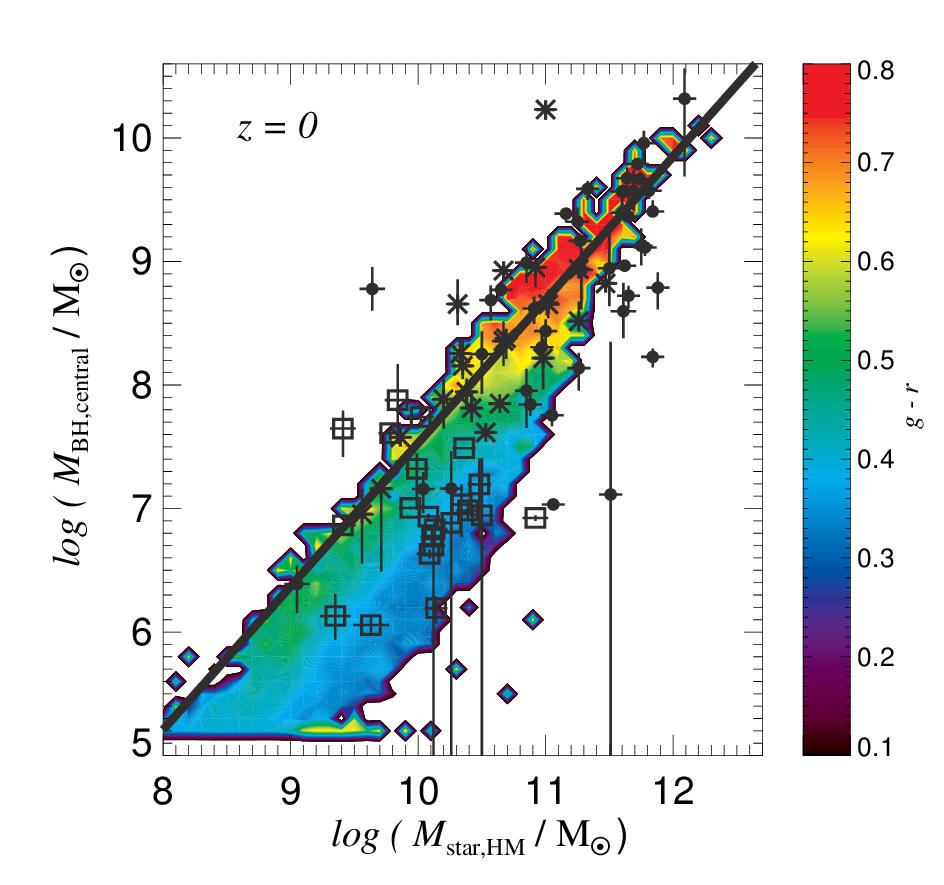}}}
\caption{Stellar half-mass of all galaxies at $z =0$ versus
  their central black hole mass for the low (left) and intermediate (right)
  resolution Illustris simulation. Colour-coding is according to the $g-r$
  colours of galaxies. The thick black line denotes the best-fit $M_{\rm BH}$
  - $M_{\rm bulge}$ relation from \citet{Kormendy2013} fitted to ellipticals and
  galaxies with bulges only. Symbols with error bars are from
  \citet{Kormendy2013} as well,
  where circles are for ellipticals, stars are for spirals with a bulge and
  squares are for pseudo bulges.}
\label{MBHMGAL_CONV}
\end{figure*}

In Figure~\ref{BHMF_CONV} we show the black hole mass function at $z = 0$,
split by the Eddington ratios of black holes, as 
  indicated in the legend. For each colour thin to thick lines are for Illustris
  simulations with three different resolutions: $3 \times 455^3$, $3 \times
  910^3$ and $3 \times 1820^3$, respectively. There are two important features
  to notice. While the convergence rate for the total mass function is not
  very good (similarly to what we found for the black hole accretion rate
  density), the uncertainty due to this is smaller than the observational
  uncertainty as estimated by \citet{Shankar2013}. Furthermore, the convergence rate
  significantly improves for black holes with Eddington ratios
  $\lambda_{\rm Edd} > 10^{-4}$, in particular at the massive end. This is
  very encouraging given that black 
  holes with higher $\lambda_{\rm Edd}$ will likely influence their host galaxies
  more than the very radiatively inefficient accretors and also considering that
  most of the black hole mass is accreted during the radiatively efficient
  accretion phase. Note also that the black hole accretion model is more
    robust (and less dependent on sub-grid physics details) for the black
    holes accreting closer to the Eddington rate.

In Figure~\ref{MBHMGAL_CONV} we show the stellar half-mass of all galaxies at $z =0$ versus
  their central black hole mass for the low (left) and intermediate (right)
  resolution Illustris simulation. The same plot for the high resolution
  simulation is illustrated in Figure~\ref{MBHMGAL_IMAGE}. Again here the
  results are reassuring in terms of numerical convergence. If we apply the
  same minimum stellar particle number within the half-mass radius of $80$ as
  we did for our high resolution run (corresponding to a minimum stellar
  half-mass of $10^8 \, M_{\rm \odot}$), the best-fits yield
  slopes of $1.61$ and $1.37$ and normalisations of $-9.08$ and $-6.91$ for
  our low and intermediate resolution runs, respectively. Furthermore, from
  Figure~\ref{MBHMGAL_CONV} it is evident that the massive end of the $M_{\rm
    BH}$ - $M_{\rm bulge}$ relation is not significantly affected by
  resolution effects where for all three runs we reproduce the data well. 
  At the low mass end the simulated relation shifts somewhat to the right for
  higher resolutions, which is caused by the combined effect of stellar masses
  increasing and black hole masses decreasing with higher resolution. We
  finally 
  note that even the main $g-r$ colour trend along the simulated $M_{\rm BH}$ -
  $M_{\rm bulge}$ relation is present at all resolutions, although with
  increasing resolution we find a higher fraction of blue star-forming
  galaxies at the low mass end.

\end{document}